\documentclass[hidelinks,aps,rmp,twocolumn,superscriptaddress,10pt]{revtex4-1}
\usepackage{amsmath,amssymb}
\usepackage{mathrsfs}
\usepackage{epstopdf}
\usepackage{graphicx}
\usepackage{bm,color,bbm}
\usepackage{natbib}
\usepackage{hyperref}
\usepackage{caption}
\usepackage{subcaption}
\usepackage{tabularx}
\newcommand{\be}[1]{\begin{equation}\label{#1}}
\newcommand{\ee}{\end{equation}}
\newcommand{\bea}[1]{\begin{eqnarray}\label{#1}}
\newcommand{\eea}{\end{eqnarray}}
\newcommand{\no}{\nonumber \\}
\newcommand{\Fig}[1]{Fig.(\ref{#1})}
\newcommand{\Eq}[1]{Eq.(\ref{#1})}

\newcommand{\bsub}{\begin{subequations}}
\newcommand{\esub}{\end{subequations}}
\newcommand{\Int}[1]{\int_{0}^{#1 L}\!\!\!\!\!\!\!\!\!\!\!\!dz\,e^{i\xi(\omega)[#1 L - z]} \hat{s}(z,\omega)}
\newcommand{\om}{\omega}

\def\a0{{\alpha_0}}

\def\da0{{\dot{\alpha}_0}}

\def\myoverDefn#1#2{\hbox{\space \raise-2mm\hbox{$\textstyle{#1} \atop \scriptstyle{#2}$} }}
\newcommand{\fracd}[2]{\frac{\displaystyle #1}{\displaystyle #2}}
\def\ha{{\hat{a}}}

\def\hf{{\hat{f}}}
\def\hb{{\hat{b}}}

\def\hs{{\hat{s}}}
\def\om{{\omega}}

\def\t{{\tau}}
\def\ts{{\tau^{*}}}

\def\rs{{\rho^{*}}}

\def\g{{\gamma}}

\def\a{{\alpha}}

\def\vac{\textrm{vac}}
\def\dag{\dagger}
\def\rp{r_{P}}
\def\rp2{r_{p}^{2}}

\def\eithp{e^{i \theta_p(\nu)}}
\def\emithp{e^{-i \theta_p(\nu)}}

\def\kinab{k\in\{a,b\}}
\newcommand{\nn}{\nonumber}

\newcommand{\ket}[1]{ |{#1} \rangle}

\begin{document}

\preprint{}

\title{Introduction to the Absolute Brightness and Number Statistics in Spontaneous Parametric Down-Conversion}


\author{James Schneeloch}
\email{james.schneeloch@gmail.com}
\affiliation{Air Force Research Laboratory, Information Directorate, Rome, New York, 13441, USA}
\author{Samuel H. Knarr}
%\email{sknarr@ur.rochester.edu}
\affiliation{Department of Physics and Astronomy, University of Rochester, Rochester, New York 14627, USA}
\affiliation{Center for Coherence and Quantum Optics, University of Rochester, Rochester, New York 14627, USA}
\author{Daniela F. Bogorin}
%\email{daniela.bogorin.ctr@us.af.mil}
\affiliation{Air Force Research Laboratory, Information Directorate, Rome, New York, 13441, USA}
\author{Mackenzie L. Levangie}
%\email{levangie.m@husky.neu.edu}
\affiliation{Department of Physics, Northeastern University, Boston, Massachusetts, 02115, USA}
\author{Christopher C. Tison}
%\email{christopher.tison.ctr@us.af.mil}
\affiliation{Air Force Research Laboratory, Information Directorate, Rome, New York, 13441, USA}
\author{Rebecca Frank}
\affiliation{Department of Physics, Northeastern University, Boston, Massachusetts, 02115, USA}
\author{Gregory A. Howland}
\affiliation{Air Force Research Laboratory, Information Directorate, Rome, New York, 13441, USA}
\author{Michael L. Fanto}
%\email{fantom1580@gmail.com}
\affiliation{Air Force Research Laboratory, Information Directorate, Rome, New York, 13441, USA}
\affiliation{RIT Integrated Photonics group, Rochester Institute of Technology, Rochester, New York, 14623, USA}
\author{Paul M. Alsing}
%\email{paul.alsing@us.af.mil}
\affiliation{Air Force Research Laboratory, Information Directorate, Rome, New York, 13441, USA}


\date{\today}

\begin{abstract}
As a tutorial, we examine the absolute brightness and number statistics of photon pairs generated in Spontaneous Parametric Down-Conversion (SPDC) from first principles. In doing so, we demonstrate how the diverse implementations of SPDC can be understood through a single common framework, and use this to derive straightforward formulas for the biphoton generation rate (pairs per second) in a variety of different circumstances. In particular, we consider the common cases of both collimated and focused gaussian pump beams in a bulk nonlinear crystal, as well as in nonlinear waveguides and micro-ring resonators. Furthermore, we examine the number statistics of down-converted light using a non-perturbative approximation (the multi-mode squeezed vacuum), to provide quantitative formulas for the relative likelihood of multi-pair production events, and explore how the quantum state of the pump affects the subsequent statistics of the downconverted light. Following this, we consider the limits of the undepleted pump approximation, and conclude by performing experiments to test the effectiveness of our theoretical predictions for the biphoton generation rate in a variety of different sources.
\end{abstract}

%\keywords{}

\maketitle
\tableofcontents

\section{Introduction}
Spontaneous Parametric Down-Conversion (SPDC) is the premier workhorse in quantum optics, both as a source of entangled photon pairs as well as heralded single photons.  As single quantum events, pump photons inside a $\chi^{(2)}$-nonlinear medium interact with the quantum vacuum via this medium to down-convert into signal-idler photon pairs. This process is spontaneous because there is initially no field at the signal or idler frequencies. When there are initial signal and idler fields, the process is known as difference frequency generation, \emph{stimulated} parametric down-conversion, or parametric amplification and is well-treated in classical nonlinear optics \cite{boyd2007nonlinear}. SPDC is a parametric process because the process itself involves no net exchange of energy or momentum between the pump photon and the nonlinear crystal \footnote{Though the crystal absorbs some pump light according to ordinary linear optics, this exchange of energy is an independent interaction not due to SPDC.}. Because of this, we can treat SPDC as the quantum evolution of a closed system (i.e., the electromagnetic field), where the hamiltonian describing the nonlinear interaction determines the state of the field.

In this tutorial, we explore the foundations of SPDC through a comprehensive derivation of the biphoton generation rate for both type-I and type-II phase matching; for both single-mode and multi-mode pump illumination and biphoton collection, and for both bulk crystals and nonlinear waveguides, where some formulas (e.g., for type-I collinear SPDC) are not found elsewhere in the literature. To accomplish this, we develop a general hamiltonian describing all such SPDC processes; show how to specialize it for each situation; and derive the biphoton rates using techniques similar to Fermi's Golden Rule, as discussed in \cite{LingPRA2008}. Furthermore, we discuss the number statistics of the down-converted light (described by the multi-mode squeezed vacuum state \cite{LvovskySqueezed2016}) in order to explore the tradeoff between the number of pairs produced, and the ability to herald \emph{single} photons from coincidence counts due to multi-pair generation events. With the current emphasis of quantum nonlinear optics turning towards chip-scale implementations of quantum information protocols, understanding the factors contributing to the brightness of photon-pair sources is critical for those entering the rich field of quantum optics.

The rest of this reference is laid out as follows. In Section II, we derive the general hamiltonian for SPDC processes and show how biphoton generation rates can be generally obtained from this hamiltonian using first-order time-dependent perturbation theory. In Section III, we calculate the generation rates for SPDC in both bulk and periodically-poled nonlinear crystals, for both collimated and focused pump beams, and for the collection of biphotons in a single transverse (gaussian) mode as well as over all modes. In Section IV, we use a different (non-pertubative) approximation to obtain the number statistics of down-converted light, providing a quantitative description of the (appoximate) multi-mode squeezed vacuum state created by SPDC, and showing how one may optimize both the brightness and heralding efficiency of down-converted light. In Section V, we use a similar approach to examine the brightness of SPDC in waveguides and micro-ring resonators, where pump intensities may be substantially larger than in the bulk crystal regime. In Section VI, we make a digression to consider SPDC with a fully quantum, depletable pump, and examine the effect of the quantum state of the pump on the subsequent intensity of the down-converted light. Finally, we conclude by performing experiments to test the formulas derived for the pair production rate in this tutorial, showing decent agreement relative to experimental design.

\section{Foundation: The Hamiltonian for the SPDC process and Rate Calculation}
The hamiltonian for the electromagnetic field, $\mathcal{H}_{EM}$, is given by its total energy $U_{EM}$ up to a constant offset. This is a common assumption, which is valid when the total energy contains no explicit time dependence, or dissipative terms. To find the total energy, it is easier to work with the rate of change of the energy of the electromagnetic field, $U_{EM}$, since a constant term in the hamiltonian will not alter the equations of motion, and can be neglected\footnote{For a more general method of deriving the hamiltonian starting from the electromagnetic lagrangian, see \cite{HilleryQuantNonOpt1984}}.

As discussed in \cite{JacksonEandM}, the rate of change of the energy of the electromagnetic field $U_{EM}$ is equal and opposite in sign to the rate of work done by the field on electric charges:
\begin{equation}
\frac{d U_{EM}}{dt}=-\int d^{3}r \;\vec{\mathbf{J}}\cdot\vec{\mathbf{E}}.
\end{equation}
Here, the work done is exclusively due to the electric field, since the magnetic field produces force perpendicular to velocity, as shown in the Lorentz force law. Using Maxwell's equations for arbitrary dielectrics, this can be re-expressed purely in terms of fields as:
\begin{equation}
\frac{d U_{EM}}{dt}=\int d^{3}r\; \Big(\vec{\mathbf{H}}\cdot\frac{d \vec{\mathbf{B}}}{dt} +\vec{\mathbf{E}}\cdot\frac{d \vec{\mathbf{D}}}{dt} \Big)
\end{equation}
where $\vec{\mathbf{D}}=\epsilon_{0}\vec{\mathbf{E}}+\vec{\mathbf{P}}$, and $\mu_{0}\vec{\mathbf{H}}=\vec{\mathbf{B}}-\mu_{0}\vec{\mathbf{M}}$.

Here, we make our first three assumptions. First, we assume the material is non-magnetic so that $\mu_{0}\vec{\mathbf{H}}=\vec{\mathbf{B}}$. Second, we assume the frequency spectrum of the light that will be interacting with the material is far enough from any absorption bands (i.e., off-resonance) that the material is approximately lossless. Third, we assume that where the pump light is weak compared to the electric field binding electrons to their atoms, the polarization field $\vec{\mathbf{P}}$ is expressible as a rapidly decaying power series in $\vec{\mathbf{E}}$:
\begin{equation}
P_{i}= \epsilon_{0}\big[\chi^{(1)}_{ij} E_{j} + \chi^{(2)}_{ijk}E_{j}E_{k} + \chi^{(3)}_{ijk\ell}E_{j}E_{k}E_{\ell}+...\big],
\end{equation}
Note that here, $\chi^{(1)}$ and $\chi^{(2)}$ are the first- and second-order optical susceptibility tensors, and we use the Einstein summation convention to simplify notation. Since most crystalline materials respond differently to fields polarized along its different principal axes, the induced polarization will not always point in the same direction as the applied electric field. These assumptions are easily satisfied in most cases in nonlinear optics, as discussed in \cite{boyd2007nonlinear}, and we use them throughout this tutorial. Since we are discussing SPDC, a second order process, we need only expand the polarization to second order in the electric field. Alternatively, we can express the electric field $\vec{\mathbf{E}}$ as a rapidly decaying power series of $\vec{\mathbf{D}}$:
\begin{equation}
E_{i}= \big[\zeta^{(1)}_{ij} D_{j} + \zeta^{(2)}_{ijk}D_{j}D_{k} + \zeta^{(3)}_{ijk\ell}D_{j}D_{k}D_{\ell}+...\big],
\end{equation}
where, for example, $\zeta_{ij}^{(1)}$ is the first-order \emph{inverse} optical susceptibility tensor.

When quantizing the electromagnetic field in matter \cite{HilleryQuantNonOpt1984}, the hamiltonian is best expressed in terms of $\vec{\mathbf{D}}$ and $\vec{\mathbf{B}}$ instead of $\vec{\mathbf{E}}$ and $\vec{\mathbf{B}}$. Indeed, if one were to substitute the quantum operator for the free electric field to create a quantum hamiltonian, the equations of motion obtained are no longer consistent with Maxwell's equations, and lead to nonphysical results. For a straightforward discussion of why this is so, see \cite{QScorrection}.

Expressing $\vec{\mathbf{E}}$ in terms of $\vec{\mathbf{D}}$, we greatly simplify calculating the hamiltonian using the prior assumption of a lossless medium. In this regime, the first and second order contributions have full permutation symmetry, and the total energy rate simplifies to:
\begin{align}
\frac{d U_{EM}}{dt} &= \frac{1}{2}\int d^{3}r\; \frac{d}{dt}\Big(\vec{\mathbf{H}}\cdot\vec{\mathbf{B}} +\vec{\mathbf{D}}\cdot\vec{\mathbf{E}}^{(1)} \Big)+\nn\\
&+\frac{1}{3}\int d^{3}r\; \frac{d}{dt}\Big(\zeta^{(2)}_{ijk}D_{i}D_{j}D_{k} \Big),
\end{align}
which can be integrated to give our hamiltonian for the electromagnetic field in a second order nonlinear dielectric. Here, $\vec{\mathbf{E}}^{(1)}$ is the electric field up to first order in the inverse susceptibility.

The hamiltonian of the electromagnetic field is now expressible as a sum of two terms, one governing the linear-optical response, and one governing the nonlinear response:
\begin{equation}
\mathcal{H}_{EM} = \mathcal{H}_{L} + \mathcal{H}_{NL}
\end{equation}
where,
\begin{equation}
\mathcal{H}_{NL}=\frac{1}{3}\int d^{3}r\;\big(\zeta_{ij\ell}^{(2)}(\vec{r})D_{i}(\vec{r})D_{j}(\vec{r})D_{\ell}(\vec{r})\big).
\end{equation}
Note that here and throughout the paper, we use the interaction picture, where the nonlinear hamiltonian shall be considered a small contribution to the total hamiltonian. In order to obtain the hamiltonian for the quantum electromagnetic field, we use the standard quantization procedure as discussed in \cite{HilleryQuantNonOpt1984,mandel1995optical,DuanQuant1997}. In a medium of index of refraction $n$, the electric displacement field operator $\hat{D}(\vec{r})$ is expressible as a sum over momentum and polarization modes in a rectangular cavity of volume $V$ with dimensions $L_{x}$, $L_{y}$, and $L_{z}$, respectively \footnote{Since the down-converted light can only be created inside the crystal, we can reach the continuum limit ($V\rightarrow\infty$) only in the limit that the crystal is much larger than the wavelength of the created light (i.e., the bulk crystal regime).}. For convenience, $\hat{D}(\vec{r},t)$ is separated into positive and negative frequency components $\hat{D}^{+}(\vec{r},t)+\hat{D}^{-}(\vec{r},t)$, where;
\begin{equation}\label{equantexp}
\hat{D}^{+}(\vec{r},t)=\sum_{\vec{k},s}i\;\sqrt{\frac{\epsilon_{0}n^{2}_{\vec{k}}\hbar\omega_{\vec{k}}}{2 V}}\hat{a}_{\vec{k},s}(t)\vec{\epsilon}_{k,s}e^{i\vec{k}\cdot\vec{r}},
\end{equation}
and $\hat{D}^{-}(\vec{r},t)$ is the hermitian conjugate of $\hat{D}^{+}(\vec{r},t)$. Here, $\hat{a}_{\vec{k},s}$ is the annihilation operator of a photon\footnote{Since we are quantizing the field in matter, the elementary excitations are collective excitations of both the electromagnetic and material degrees of freedom. Even so, they are still regarded as photons.} with momentum $\vec{k}$ and polarization in direction $\vec{\epsilon}_{k,s}$ indexed by $s$ (which can take one of two values for each transverse direction), and $V$ is the quantization volume, which we may take to approach infinity in the continuum limit. With this, the quantum hamiltonian describing linear-optical effects becomes:
\begin{equation}\label{HamLin}
\hat{H}_{L}=\sum_{\vec{k},s}\hbar\;\omega(\vec{k})\Big(\hat{a}^{\dagger}_{\vec{k},s}(t)\hat{a}_{\vec{k},s}(t)+\frac{1}{2}\Big)
\end{equation}
As one can see, the linear hamiltonian cannot be responsible for the creation of photon pairs, as it is only first-order in both the creation and annihilation operators.

The nonlinear quantum hamiltonian $\hat{H}_{NL}$,
\begin{equation}
\hat{H}_{NL}=\frac{1}{3}\int d^{3}r\;\big(\zeta_{ij\ell}^{(2)}(\vec{r})\hat{D}_{i}(\vec{r},t)\hat{D}_{j}(\vec{r},t)\hat{D}_{\ell}(\vec{r},t)\big),
\end{equation}
has a deceptively simple form. With each field operator $\hat{D}(\vec{r},t)$ expressed as  $\hat{D}^{+}(\vec{r},t)+\hat{D}^{-}(\vec{r},t)$, where  $\hat{D}^{+}(\vec{r},t)$ depends only on annihilation operators, and  $\hat{D}^{-}(\vec{r},t)$ on creation operators, the nonlinear hamiltonian is actually a sum over eight distinct terms. Various combinations of these terms correspond to different basic nonlinear-optical processes, but only those processes that conserve energy contribute significantly to the probability-amplitude of down-conversion. For example, the two terms that are third-order in either photon creation or annihilation may be excluded, as their contribution to the probability amplitude of photon pair generation is a rapidly varing phase that becomes negligible even over the small time it takes light to travel through the nonlinear medium. Furthermore, for many nonlinear media, $\zeta^{(2)}$ (or alternatively $\chi^{(2)}$) is only significant for one particular optical process (either by design or happenstance) \footnote{Although not impossible, simultaneous generation of SHG light and SPDC photon pairs in a single nonlinear medium with a single (pump) laser source has yet to be accomplished.}. Even when $\zeta^{(2)}$ is significant for multiple nonlinear optical processes, simultaneously achieving phase-matching (i.e., momentum conservation)  for multiple processes may be prohibitively difficult due to materials having a fixed optical dispersion (i.e., index of refraction as a function of frequency).

In the case of SPDC, either pump photons are destroyed in exchange for signal-idler photon pairs or vise versa, so that $\hat{H}_{NL}$ is well-approximated as:
\begin{equation}\label{hnl}
\hat{H}_{NL}=\frac{1}{3}\int d^{3}r\;\big(\zeta_{ij\ell}^{(2)}(\vec{r})\hat{D}_{i}^{+}(\vec{r},t)\hat{D}_{j}^{-}(\vec{r},t)\hat{D}_{\ell}^{-}(\vec{r},t) + h.c.,\big)
\end{equation}
where $h.c.$ stands for hermitian conjugate. Before we continue, we point out that we have conflated the polarization index $s$ with the displacement field component index $i$. The operator $\hat{D}_{i}^{+}(\vec{r},t)$ is given by the sum in equation \eqref{equantexp}, but where the polarization unit vector $\vec{\epsilon}_{k,s}$ is replaced by its component parallel to the $i$th direction, $\vec{\epsilon}_{k,s}\cdot \vec{x}_{i}$. Throughout this paper, we will be working in the paraxial regime, where the light is propagating primarily along a single direction (i.e., along the optic axis). In this situation, it is a valid approximation to simply replace the displacement field component indices with polarization indices since the component of the displacement field parallel to the optic axis is negligible.

\subsection{Transforming the Hamiltonian into the Hermite-Gauss basis}
The canonical quantization of the electromagnetic field into plane-wave modes with creation operators $\hat{a}^{\dagger}_{\vec{k},s}$ is the first step in the standard quantum treatment of SPDC light. However, it will make subsequent calculations much simpler if we express the transverse momentum components of the field in terms of hermite-gaussian modes, since gaussian pump beams and similar collection modes of down-converted light are valid descriptions of the light generated in SPDC experiments.

In order to do this, we first introduce some notation. Let $\vec{q}$ denote the projection of the momentum $\vec{k}$ onto the transverse plane, so that $\vec{k}=\vec{q} + k_{z}\hat{z}$, and $\hat{z}$ is a unit vector pointing along the optic axis in the direction of propagation. Then, the creation operator $\hat{a}^{\dagger}_{\vec{k},s}$ can be expressed as $\hat{a}^{\dagger}_{(\vec{q},k_{z},s)}$. Since both plane waves and hermite-gaussian wavefunctions form a complete basis in 2-D space, we can express the plane-wave creation operator $\hat{a}^{\dagger}_{(\vec{q},k_{z},s)}$ as a sum over transverse mode creation operators $\hat{a}^{\dagger}_{(\vec{\mu},k_{z},s)}$, where $\vec{\mu}$ is a vector denoting the horizontal and vertical indices of a given hermite-gaussian mode;
\begin{equation}
\hat{a}^{\dagger}_{(\vec{q},k_{z},s)}=\sum_{\vec{\mu}}\tilde{C}_{\vec{q},\vec{\mu}}\;\hat{a}^{\dagger}_{(\vec{\mu},k_{z},s)}.
\end{equation}
Since the boson commutation relation $[\hat{a}_{(\vec{q},k_{z},s)},\hat{a}^{\dagger}_{(\vec{q'},k_{z},s)}] = \delta_{\vec{q},\vec{q'}}$ must be preserved in both representations, it is straightforward to show that:
\begin{equation}
\sum_{\vec{\mu}}\|\tilde{C}_{\vec{q},\vec{\mu}}\|^{2}=1.
\end{equation}
With this established, we can express the displacement field operator $\hat{D}^{-}(\vec{r},t)$ in this new Hermite-Gauss basis.

The transverse spatial dependence of $\hat{D}^{-}(\vec{r},t)$ for a given Hermite-Gauss mode indexed by $\vec{\mu}$ relies on the sum:
\begin{equation}
\sum_{\vec{q}}\tilde{C}_{\vec{\mu},\vec{q}}\;e^{-i\vec{q}\cdot\vec{r}}= \sqrt{L_{x}L_{y}}\; g_{\vec{\mu}}(x,y)
\end{equation}
where we have defined $g_{\vec{\mu}}(x,y)$ to be the normalized\footnote{The hermite-gaussian mode functions are normalized so that the integral over all space of their magnitude square gives unity (as with quantum wavefunctions).} hermite-gaussian wavefunction given by the index $\vec{\mu}$. Here, $\vec{\mu}$ is an ordered pair of non-negative integers corresponding to the horizontal and vertical mode index, respectively. Because the momentum components can only take on values that are integer multiples of $2\pi$ divided by the respective length of the cavity in each direction, this relation is straightforward to check through normalization. For finite size nonlinear crystals, this relation is approximate, but accurate when the hermite-gaussian modes are encompassed by the crystal. Finally, using an element of the paraxial approximation (so that the frequency $\omega$ only depends on $k_{z}$), the displacement field operator becomes:
\begin{equation}
\hat{D}^{-}(\vec{r},t)=-i\!\!\sum_{\vec{\mu},k_{z},s}\!\!\sqrt{\frac{\epsilon_{0}n^{2}_{k_{z}}\!\hbar\omega_{k_{z}}}{2 L_{z}}}\vec{\epsilon}_{k_{z},s}g_{\vec{\mu}}(x,y)e^{-i k_{z} z}e^{i\omega t}\hat{a}^{\dagger}_{\vec{\mu},k_{z},s}
\end{equation}
Here, we point out that $\hat{a}^{\dagger}_{\vec{\mu},k_{z},s}(t)=\hat{a}^{\dagger}_{\vec{\mu},k_{z},s}e^{i\omega t}$.

With the displacement field operators expressed in the Hermite-Gauss basis, we are ready to obtain the nonlinear hamiltonian \eqref{hnl}. In the standard approach for treating SPDC, the pump field is treated as being bright enough that classical electromagnetism is sufficient for its description, and that its intensity is not noticeably diminished due to down-conversion events (also known as the undepeleted pump approximation). We use the classical pump approximation throughout most of this paper, but use a more accurate description when discussing the number statistics of the down-converted light.

\subsubsection{The Classical Pump Field}
Although arbitrary illumination of the nonlinear medium can be expressed as an integral over all frequencies, SPDC occurs only in narrow bands of pump frequencies where phase matching (i.e., momentum conservation) can be achieved due to dispersion \footnote{For example, in degenerate type-I SPDC, momentum conservation $\vec{k}_{p}=2 \vec{k}_{1}$ is achieved when $n_{p}=n_{1}$. This is possible when the dispersion (i.e., dependency of index on frequency) is different for different polarizations in birefringent materials. It is also possible to achieve phase matching if the dispersion is anomalous (e.g., near an absorption peak)\cite{CahillSPDCAnnomalous1989}, but birefringent phase matching is much more straightforward.}. In light of this, we limit ourselves to the ubiquitous case of a monochromatic pump beam with peak magnitude $|D_{p}^{0}|$, frequency $\omega_{p}$, polarization $\vec{\epsilon}_{p}$, and (non-normalized) spatial dependence $f_{p}(\vec{r})$ given by:
\begin{equation}
\vec{D}_{p}(\vec{r},t)=|D_{p}^{0}|\vec{\epsilon}_{p}f_{p}(\vec{r})\;\text{Cos}(\omega_{p}t),
\end{equation}
which can be separated into positive and negative frequency components, giving us:
\begin{equation}
\vec{D}_{p}^{-}(\vec{r},t)=|D_{p}^{0}|\vec{\epsilon}_{p}f_{p}(\vec{r})\frac{e^{i\omega_{p}t}}{2}.
\end{equation}
The (time averaged) pump intensity \footnote{The time averaged pump intensity may be taken as the magnitude of the Poynting vector $|\vec{S}|=\frac{1}{2}|\vec{E}\times\vec{H}^{*}|$, and in our approximations, $|D_{p}^{0}|=\epsilon_{0}n^{2}|E_{p}^{0}|$.} is then:
\begin{equation}
I_{p}=\frac{c}{2\epsilon_{0}n^{3}} \; |D_{p}^{0}|^{2}|f_{p}(\vec{r})|^{2}.
\end{equation}
For later simplification, we factor out the linear phase due to propagating the beam, and get:
\begin{equation}
f_{p}(\vec{r})= \tilde{G}_{p}(\vec{r})e^{-i k_{z} z}
\end{equation}
where $\tilde{G}_{p}(\vec{r})$ is implicitly defined.

Throughout most of this paper, $\tilde{G}_{p}(\vec{r})$ will describe the rest of a gaussian pump beam, so that
\begin{align}
\tilde{G}_{p}(\vec{r})&\equiv\frac{\sigma_{p}}{\sigma(z)}\text{Exp}\Big(-\frac{x^{2}+y^{2}}{4\sigma(z)^{2}}\Big)\text{Exp}\Big(-ik_{z}\frac{x^{2}+y^{2}}{2 R(z)}\Big)\times\nn\\
&\times\text{Exp}\Big(i\;\text{Tan}^{-1}\Big(\frac{z}{z_{R}}\Big)\Big).
\end{align}
Except in Section III-D where we consider SPDC using a focused pump beam, we use the simplifying approximation that the pump beam is collimated, so that we may neglect the Guoy phase and curvature of the phase fronts in our calculations. To condense notation, $\sigma(z)$ is the evolving beam radius (as measured by standard deviation);
\begin{equation}
\sigma(z)\equiv\sigma_{p}\sqrt{1+\Big(\frac{z}{z_{R}}\big)^{2}}
\end{equation}
$R(z)$ is the evolving radius of curvature of the wavefronts:
\begin{equation}
R(z)\equiv z\Big[1+\Big(\frac{z_{R}}{z}\Big)^{2}\Big]
\end{equation}
and $z_{R}$ is the Rayleigh length, such that $\sigma(z_{R})=\sqrt{2}\sigma_{p}$;
\begin{equation}
z_{R}\equiv\frac{4\pi\sigma_{p}^{2}}{\lambda_{p}}.
\end{equation}
As we can see, the first exponential governs the evolving spatial amplitude of the beam; the second exponential describes the propagation and curvature of the phase fronts, while the last exponential describes the Guoy phase.

With the parameters of a gaussian pump beam, the amount of energy per second delivered by such a beam (i.e, its power) is expressed as:
\begin{equation}\label{GaussPower}
P=c\frac{|D_{p}^{0}|^{2}}{n^{3}\epsilon_{0}}\pi\sigma_{p}^{2},
\end{equation}
which equals the mean intensity of the beam times its effective area \footnote{The effective area of a probability distribution (as described by the transverse intensity distribution of light) is the reciprocal of the mean height of the probability density. This is the area that a uniform probability distribution of such a mean height would have to have to be normalized. For a two-dimensional radially symmetric gaussian distribution, the effective area is $4\pi\sigma^{2}$.}.

\subsubsection{Simplifying the nonlinear Hamiltonian}
Incorporating our expressions for the displacement field operators and the classically bright pump field, the nonlinear hamiltonian becomes:
\begin{align}
\hat{H}_{NL}&=\int d^{3}r\Big(\zeta^{(2)}_{eff}(\vec{r})|D_{p}^{0}|\tilde{G}^{*}_{p}(\vec{r})e^{i k_{pz}z}e^{-i\omega_{p}t}\times\nn\\
&\times-i\sum_{\vec{\mu}_{1},k_{1z}}\sqrt{\frac{\epsilon_{0}n^{2}_{1}\hbar\omega_{1}}{2 L_{z}}}g_{\vec{\mu}_{1}}(x,y)e^{-i k_{1z} z}e^{i\omega_{1} t}\hat{a}^{\dagger}_{\vec{\mu}_{1},k_{1z}}\nn\\
&\times-i\sum_{\vec{\mu}_{2},k_{2z}}\sqrt{\frac{\epsilon_{0}n^{2}_{2}\hbar\omega_{2}}{2 L_{z}}}g_{\vec{\mu}_{2}}(x,y)e^{-i k_{2z} z}e^{i\omega_{2} t}\hat{a}^{\dagger}_{\vec{\mu}_{2},k_{2z}}\nn\\
&+h.c. \Big)
\end{align}
Here, we abbreviated $n_{k_{1z}}$ as $n_{1}$, and $\omega_{k_{1z}}$ as $\omega_{1}$. Furthermore, we have already performed the sum over the components of the inverse susceptibility. The additional factor of $6=3!$ comes from the permutation symmetry of the nonlinear susceptibility where the total sum is 6 times the value of each term, where all terms are added together. After simplifying, we find:
\begin{align}
&\hat{H}_{NL}=\frac{\hbar |E_{p}^{0}|}{2 L_{z}}\sum_{\vec{\mu}_{1},k_{1z}}\sum_{\vec{\mu}_{2},k_{2z}}\sqrt{\frac{\omega_{1}\omega_{2}}{n_{1}^{2}n_{2}^{2}}}\times\nn\\
&\times \int d^{3}r\Big(\chi^{(2)}_{eff}(\vec{r})G^{*}_{p}(\vec{r}) g_{\vec{\mu}_{1}}(x,y)g_{\vec{\mu}_{2}}(x,y)e^{-i \Delta k_{z} z}\Big)\times\nn\\
&\times e^{i\Delta\omega t}\hat{a}^{\dagger}_{\vec{\mu}_{1},k_{1z}}\hat{a}^{\dagger}_{\vec{\mu}_{2},k_{2z}}\nn\\
&+h.c.\label{GenHam}
\end{align}
Here we have switched from $\zeta^{(2)}_{eff}$ to the effective nonlinear susceptibility $\chi^{(2)}_{eff}$ using the approximation:
\begin{equation}
-\epsilon_{0}^{2}\;\zeta^{(2)}_{eff}\;n_{p}^{2}n_{1}^{2}n_{2}^{2}\approx \chi^{(2)}_{eff},
\end{equation}
which is satisfied under the same lossless media assumption that allowed us to invoke full permutation symmetry. Since $\chi^{(2)}$ is what is measured experimentally, and tabulated in handbooks of optical materials, the rest of this tutorial will be expressed in terms of the susceptibility, rather than its inverse.

Since the $\chi^{(2)}$ nonlinearity is zero outside the nonlinear medium, the spatial integration is carried over the dimensions of the medium. This hamiltonian describes SPDC in a nonlinear medium of length $L_{z}$, unspecified transverse dimensions (but significantly wider than the pump beam), and unspecified poling when illuminated by a monochromatic pump beam directed along the optic axis. Here, the sum over the polarization indices has already been carried out, giving the effective nonlinearity $\chi_{eff}^{(2)}(\vec{r})$. To simplify notation, we defined $\Delta\omega\equiv\omega(k_{1z})+\omega(k_{2z})-\omega(k_{pz})$, and $\Delta k_{z}\equiv k_{1z}+k_{2z}-k_{pz}$. Note that while the Hermite-Gauss modes of the down-converted light $g_{\vec{\mu}}(x,y)$ are normalized to have unit norm, the pump spatial dependence $\tilde{G}_{p}(\vec{r})$ has a maximum magnitude of unity at $\vec{r}=0$. The peak pump intensity is fixed by the value of the pump field strength $|D_{p}|$.

\subsection{Calculating the biphoton rate from the nonlinear Hamiltonian}
With a general hamiltonian describing most SPDC processes, we could calculate a general rate of biphoton generation using Fermi's golden rule as shown in \cite{LingPRA2008}. Here, we instead show how the direct calculation takes place with first-order time-dependent perturbation theory (from whence Fermi's Golden Rule originates). We take the initial state of the down-converted fields to be the vacuum state, and the final state to be a biphoton with momenta and Hermite-Gauss mode numbers ($k_{1z},\vec{\mu}_{1}$) and ($k_{2z},\vec{\mu}_{2}$),  for the signal and idler photon respectively. The transition probability $P_{k_{1z},\vec{\mu}_{1},k_{2z},\vec{\mu}_{2}}$ is given by:
\begin{align}
P&_{k_{1z},\vec{\mu}_{1},k_{2z},\vec{\mu}_{2}}\equiv|\langle \vec{\mu}_{1}k_{1z},\vec{\mu}_{2}k_{2z}|\Psi(t)\rangle|^{2}\nn\\
&\approx\Big|\langle \vec{\mu}_{1}k_{1z},\vec{\mu}_{2}k_{2z}|\Big(1-\frac{i}{\hbar}\int_{0}^{t}dt'\;\;\hat{H}_{NL}(t')\Big)|0,0\rangle\Big|^{2}
\end{align}
where the expression in parentheses comes from the first-order approximation (a la perturbation theory) of the time propagation operator. Substituting our expression for the nonlinear hamiltonian, we obtain:
\begin{align}
P&_{k_{1z},\vec{\mu}_{1},k_{2z},\vec{\mu}_{2}}=\frac{|E_{p}^{0}|^{2}}{4 L_{z}^{2}}\frac{\omega_{1}\omega_{2}}{n_{1}^{2}n_{2}^{2}}\times\nn\\
&\times \Big|\int d^{3}r\Big(\chi^{(2)}_{eff}(\vec{r})G^{*}_{p}(\vec{r}) g_{\vec{\mu}_{1}}(x,y)g_{\vec{\mu}_{2}}(x,y)e^{-i \Delta k_{z} z}\Big)\Big|^{2}\times\nn\\
&\times \Big|\int_{0}^{t}dt'\;\;e^{i\Delta\omega t'}\Big|^{2}\nn\\
&=W_{k_{1z},\vec{\mu}_{1},k_{2z},\vec{\mu}_{2}}\Big|\int_{0}^{t}dt'\;\;e^{i\Delta\omega t'}\Big|^{2}
\end{align}
where $W_{k_{1z},\vec{\mu}_{1},k_{2z},\vec{\mu}_{2}}$ is defined implicitly to simplify notation. This expression can be further simplified in the limit that $t$ becomes large, and knowing that the magnitude of a complex number is independent of its phase:
\begin{align}
P&_{k_{1z},\vec{\mu}_{1},k_{2z},\vec{\mu}_{2}}=W_{k_{1z},\vec{\mu}_{1},k_{2z},\vec{\mu}_{2}}\Big|t\;\text{Sinc}\big(\frac{\Delta\omega t}{2}\big)\Big|\times\nn\\
&\qquad\qquad\qquad\times\Big|\int_{0}^{t}dt'\;\;e^{i\Delta\omega t'}\Big|\nn\\
&\xrightarrow[\text{large}\;t]{} W_{k_{1z},\vec{\mu}_{1},k_{2z},\vec{\mu}_{2}}\; 2\pi \;\delta(\Delta\omega)\Big|\int_{0}^{t}dt'\;\;e^{i\Delta\omega t'}\Big|
\end{align}
In practice, $t$ need not be arbitrarily large for this limit to apply. Instead, one only needs $t$ to be significantly larger than the inverse of $\Delta\omega$, which is achieved for crystals longer than a few hundredths of a millimeter. The range of values $\Delta\omega$ can take (before this limit is invoked) is known as the phase-matching bandwidth (and is of the order $10^{13}-10^{14}$ for most materials). Although ultimately limited by effective nonlinearity $\chi^{(2)}_{eff}$, the phase-matching bandwidth is primarily determined by the dispersion of the material where the condition $|\Delta k_{z}|<2\pi/L_{z}$ is satisfied. Since the large time limit for $P_{k_{1z},\vec{\mu}_{1},k_{2z},\vec{\mu}_{2}}$ can only be nonzero when $\Delta\omega$ is zero, the second integral is of a constant term, making $P_{k_{1z},\vec{\mu}_{1},k_{2z},\vec{\mu}_{2}}$ linear in time. Where the transition rate $R_{k_{1z},\vec{\mu}_{1},k_{2z},\vec{\mu}_{2}}$ is defined as the time derivative of the transition probability, it levels off to a constant value for large times (e.g., longer than a picosecond):
\begin{equation}
\lim_{t\rightarrow\infty}R_{k_{1z},\vec{\mu}_{1},k_{2z},\vec{\mu}_{2}}=W_{k_{1z},\vec{\mu}_{1},k_{2z},\vec{\mu}_{2}}\;2\pi \;\delta(\Delta\omega)
\end{equation}
Of course, the transition probability cannot increase linearly with time indefinitely; the first-order perturbation approximation breaks down. However, in the undepleted pump approximation, and using times of the order of the time it takes light to pass through the crystal, this approximation is valid. To calculate the total transition rate for down-conversion into a single pair of transverse modes $R_{\vec{\mu}_{1},\vec{\mu}_{2}}$, we must add the transition rates for all values of $k_{1z}$ and $k_{2z}$:
\begin{equation}
R_{\vec{\mu}_{1},\vec{\mu}_{2}}=\sum_{k_{1z},k_{2z}}R_{k_{1z},\vec{\mu}_{1},k_{2z},\vec{\mu}_{2}},
\end{equation}
and we can define $W_{\vec{\mu}_{1},\vec{\mu}_{2}}$ similarly. Because the length of the nonlinear medium $L_{z}$ is much longer than the wavelength of light passing through it, we may approximate the sums over $k_{1z}$ and $k_{2z}$ as integrals over $k_{1z}$ and $k_{2z}$, which in turn, can be expressed as integrals over frequencies $\omega_{1}$ and $\omega_{2}$:
\begin{equation}
\sum_{k_{1z},k_{2z}}\approx\Big(\frac{L_{z}}{2\pi}\Big)^{2}\int dk_{1z}dk_{2z}\approx\Big(\frac{L_{z}}{2\pi}\Big)^{2}\frac{n_{g1}n_{g2}}{c^{2}}\int d\omega_{1}d\omega_{2},
\end{equation}
where $n_{g1}$ $(n_{g2})$ is the group index at the signal (idler) frequency.

With this, the single-mode transition rate $R_{\vec{\mu}_{1},\vec{\mu}_{2}}$ is given by the integral:
\begin{equation}
R_{\vec{\mu}_{1},\vec{\mu}_{2}}=\int d\omega_{1}d\omega_{2}\;W_{k_{1z},\vec{\mu}_{1},k_{2z},\vec{\mu}_{2}}\;\frac{L_{z}^{2}n_{g1}n_{g2}}{2\pi c^{2}}\delta(\Delta\omega),
\end{equation}
where $W_{\vec{\mu}_{1},k_{1z},\vec{\mu}_{2},k_{2z}}$ is readily expressed in terms of $\omega_{1}$ and $\omega_{2}$. The total rate $R$, is then the sum over all transverse modes of the single-mode rates.

\subsubsection{Transition rate vs the rate of generated biphotons}
The transition rate $R$ is taken to be the average number of biphotons per second generated in the nonlinear medium. The reason this is so requires further explanation. The transition rate $R$ is defined as the rate of change of the transition probability. The transition probability $P(t+dt)$ is the probability that the biphoton will be emitted either in the time interval $t\in[0,t]$ or in the interval $t\in[t,t+dt]$. Since these intervals are disjoint, and the transition probability is linear, the quantity $R dt$ is the probability that the biphoton will be emitted in an interval of length dt. Since the state of the signal and idler field in the crystal is once again well-described by the vacuum state as soon as the biphoton exits the crystal, while the pump continues driving transitions, the temporal statistics of biphotons generated in SPDC are well-described as a Poisson point process. In particular, the probability of \emph{not} generating a biphoton in the interval $t\in[0,t+dt]$ is given as the product of the same probability over the interval $t\in[0,t]$, and $(1-R dt)$. This defines a differential equation, allowing one to obtain the exponential distribution for waiting times between biphoton generation events. One can then recursively obtain the probabilities of one, two, or more transitions in an interval of length $T$ from this information as well. For example, the event of two transitions in an interval of length $T$ is broken down into one transition in the interval $[0,t']$, a transition in the interval $[t',t'+dt]$ and no transitions in the interval $[t',T]$.

Similar equations can be developed to describe the probability of detecting $n$ biphotons over time $T$, Indeed, these statistics are described by a Poisson distribution with rate $R$ such that the mean number of biphotons generated over time $T$ is simply $RT$. For a more thorough discussion, see \cite{HayatCoincStat,SheldonProbBook}.

\section{The bulk crystal regime: photon-pair brightness}
Previously, we found a general form for the hamiltonian describing SPDC \eqref{GenHam} in a general bulk nonlinear crystal. All other parameters being fixed by experimental design, the biphoton generation rate depends on the overlap integral $\Phi(\Delta k_{z})$:
\begin{align}\label{Overlap}
&|\Phi(\Delta k_{z})|^{2}\equiv\nn\\
&\equiv \Big|\int d^{3}r\Big(\chi^{(2)}_{eff}(\vec{r})G^{*}_{p}(\vec{r}) g_{\vec{\mu}_{1}}(x,y)g_{\vec{\mu}_{2}}(x,y)e^{-i \Delta k_{z} z}\Big)\Big|^{2}.
\end{align}
The simplest case to solve is that of the collimated gaussian pump beam incident on an isotropic rectangular crystal of dimensions $L_{x}$ by $L_{y}$ by $L_{z}$ centered at the origin of a Cartesian coordinate system with $z$ pointing along the optic axis. If we make the additional assumption that we are collecting the downconverted light into single-mode fibers, then only the photons generated in the zeroth-order hermite-gaussian modes will contribute to the rate of detected events. In this case, $G_{p}(\vec{r})$, $g_{\vec{\mu}_{1}}(x,y)$, and $g_{\vec{\mu}_{2}}(x,y)$ are all gaussian functions, so that $|\Phi(\Delta k_{z})|^{2}$ becomes:
\begin{align}
|\Phi(\Delta k_{z})|^{2}&= \Big(\frac{\chi^{(2)}_{eff}}{2\pi \sigma_{1}^{2}}\Big)^{2}\Big|\int_{-L_{Z}/2}^{L_{Z}/2} dz e^{-i \Delta k_{z} z}\Big|^{2}\times\nn\\
&\times\Big|\int dx dy\Big(\text{Exp}\Big[-(x^{2}+y^{2})\Big(\frac{1}{4\sigma_{p}^{2}}+\frac{2}{4\sigma_{1}^{2}}\Big)\Big]\Big|^{2}
\end{align}
Here, we have let the widths of the hermite-gaussian modes of the signal-idler light be defined as $\sigma_{1}$ in analogy with $\sigma_{p}$ for the pump beam. The value of $\sigma_{1}$ is a free parameter in our definition of the hermite-gaussian basis, but is best set using the mode field diameter of the accepting single-mode fiber, and related collection optics that image the accepting mode to the center of the crystal. To make the limits of the integral over $x$ and $y$ arbitrarily large, it only suffices that the transverse width of the crystal is larger than the dimensions of both the gaussian pump beam and of the signal and idler modes. Even for crystals only a millimeter wide in $x$ and $y$, it is straightforward to have a well-collimated beam whose area is contained within the crystal. With these assumptions, the overlap integral simplifies significantly to:
\begin{equation}\label{GaussOverlap}
|\Phi(\Delta k_{z})|^{2}= \Big(2 \chi^{(2)}_{eff}L_{z}\Big)^{2}\text{Sinc}^{2}\Big(\frac{\Delta k_{z} L_{z}}{2}\Big)\Big|\frac{\sigma_{p}^{2}}{\sigma_{1}^{2}+2\sigma_{p}^{2}}\Big|^{2}
\end{equation}
With this, the total rate for down-conversion from a collimated gaussian pump beam into gaussian signal-idler modes, $R_{SM}$, is readily converted into an integral over the signal and idler frequencies $\omega_{1}$ and $\omega_{2}$:
\begin{align}
R_{SM}&=\int d\omega_{1}d\omega_{2}\frac{|E_{p}^{0}|^{2}(\chi_{eff}^{(2)})^{2}L_{z}^{2}}{2\pi c^{2}}\;\frac{n_{g1}n_{g2}}{n_{1}^{2}n_{2}^{2}}\Big|\frac{\sigma_{p}^{2}}{\sigma_{1}^{2}+2\sigma_{p}^{2}}\Big|^{2}\nn\\
&\times \omega_{1}\omega_{2}\delta(\Delta\omega)\text{Sinc}^{2}\Big(\frac{\Delta k_{z} L_{z}}{2}\Big)
\end{align}
The dependence of $R_{SM}$ on the widths $\sigma_{p}$ and $\sigma_{1}$ is subject to these modes being both well-collimated and contained within the crystal.

To further simplify the total rate $R_{SM}$, we express $\Delta k_{z}$ in terms of the frequencies $\omega_{1}$ and $\omega_{2}$, and integrate over $\omega_{2}$ using the Dirac delta function to find:
\begin{align}\label{SMrate}
R_{SM}&=\int d\omega_{1}\frac{|E_{p}^{0}|^{2}(\chi_{eff}^{(2)})^{2}L_{z}^{2}}{2\pi c^{2}}\frac{n_{g1}n_{g2}}{n_{1}^{2}n_{2}^{2}}\Big|\frac{\sigma_{p}^{2}}{\sigma_{1}^{2}+2\sigma_{p}^{2}}\Big|^{2}\nn\\
&\times \omega_{1}(\omega_{p}-\omega_{1})\text{Sinc}^{2}\Big(\frac{\Delta k_{z} L_{z}}{2}\Big)
\end{align}
where
\begin{equation}
\Delta k_{z}=k(\omega_{1}) + k(\omega_{p}-\omega_{1}) -k(\omega_{p})
\end{equation}
Further simplification requires knowledge of the type of down-conversion being used.

\subsection{Degenerate down-conversion}
Let us consider the case where the crystal is cut and tuned to optimize down-conversion such that the spectra of $\omega_{1}$ and $\omega_{2}$ are both centered at half the pump frequency $\omega_{p}$. Then, the momentum mismatch $\Delta k_{z}$ can be Taylor-expanded \cite{Fedorov2009} about this central frequency so that:
\begin{equation}\label{dkExp}
\Delta k_{z}\approx \Big(\frac{\Delta n_{g}}{c}\Big)\big(\omega_{1}-\frac{\omega_{p}}{2}\big) +\kappa \big(\omega_{1}-\frac{\omega_{p}}{2}\big)^{2}
\end{equation}
where $\kappa$ is the group velocity dispersion constant at half the pump frequency: $|\frac{d^{2}k}{d\omega^{2}}|_{\omega_{p}/2}$, and $\Delta n_{g}$ is the group index mismatich for the signal and idler photons $|n_{g1}-n_{g2}|$ at their central frequencies.

In type-0 and type-I SPDC\footnote{Type-0 SPDC is where the pump, signal, and idler beam all have identical (typically vertical) polarization. In type-I SPDC, the signal and idler polarization are identical, but orthogonal to the pump polarization. In Type-II SPDC, the pump polarization is identical to either the signal or idler polarization, but both signal and idler are mutually orthogonal.}, the group indices of the signal and idler light are identical because their polarizations are identical. Only the second-order contribution to $\Delta k_{z}$ is significant, and we find:
\begin{align}
R_{SM}&\propto\int d\omega_{1} \omega_{1}(\omega_{p}-\omega_{1})\text{Sinc}^{2}\Big(\frac{L_{z}\kappa}{2}\big(\omega_{1}-\frac{\omega_{p}}{2}\big)^{2}\Big)\nn\\
&=\frac{(L_{z}\kappa\omega_{p}^{2}-6)\sqrt{2\pi}}{3\big(L_{z}\kappa\big)^{3/2}}\approx \frac{\omega_{p}^{2}}{3}\sqrt{\frac{2\pi}{L_{z}\kappa}}
\end{align}
The approximation holds well for typical crystal parameters and crystal lengths longer than tenths of a millimeter (as is typical). Here, we have also assumed that the portion of the generation rate formula dependent on the indices of refraction is more or less constant over the bandwidth of the down-converted light, which is reasonable for most nonlinear crystals. Making this final simplification, we arrive at the single-mode rate for degenerate type-0 and type-I SPDC:
\begin{equation}\label{SingleModeType01}
R_{SM}^{t1}=\sqrt{\frac{2}{\pi^{3}}}\frac{2}{3\epsilon_{0}c^{3}}\frac{n_{g1}n_{g2}}{n_{1}^{2}n_{2}^{2}n_{p}}\frac{(d_{eff})^{2}\omega_{p}^{2}}{\sqrt{\kappa}}\Big|\frac{\sigma_{p}^{2}}{\sigma_{1}^{2}+2\sigma_{p}^{2}}\Big|^{2}\frac{P}{\sigma_{p}^{2}}L_{z}^{3/2},
\end{equation}
where $d_{eff}\equiv\chi_{eff}^{(2)}/2$, is the more common convention for defining the effective nonlinear susceptibility, and we substituted the relation for the power of the gaussian pump beam \eqref{GaussPower}.

\subsubsection{Multimode degenerate SPDC}
Although many experiments make use of photon pairs coupled into single-mode fiber, this coupling destroys the transverse spatial correlations and the high-dimensional entanglement in that degree of freedom. In experiments that involve coupling down-converted light into multi-mode fiber, or ones using a large-area photon detector, the relevant rate of biphoton generation is the rate of generation into all transverse hermite-gaussian modes. Ordinarily, the total rate would be the sum of the single-mode rates over all pairs of signal and idler modes \eqref{SMrate}. However, directly evaluating this sum yields non-physical results, as the formula for the single-mode rate is contingent on the paraxial approximation. For a given beam waist, hermite-gaussian beams with sufficiently large transverse momentum (or high mode index) are non-paraxial. Instead, it is much simpler to calculate the relative probability that the biphoton will be emitted into the zeroth order signal and idler gaussian modes, and from this, determine the ratio of the total rate to the single-mode rate. Where the idler mode radius $\sigma_{1}$ defining the Hermite-Gauss basis is a free parameter, we set it equal to the pump radius $\sigma_{p}$ to simplify calculation. For types 0 and 1 \emph{degenerate} collinear down-conversion, the ratio is given by:
\begin{align}
\frac{R_{SM}}{R_{T}}&=|\langle\vec{\mu}_{1}=\vec{0},\vec{\mu}_{2}=\vec{0}|\Psi\rangle|^{2}=\frac{4 a \sigma_{p}^{2}}{\big( \sigma_{p}^{2} + \sqrt{a^{2} + \sigma_{p}^{4}}\big)^{2}}\nn\\
&\approx\frac{L_{z}\lambda_{p}}{4\pi n_{p}\sigma_{p}^{2}},
\end{align}
such that $a=\frac{L_{z}\lambda_{p}}{4\pi n_{p}}$, and,
\begin{equation}
\langle x_{1},y_{1},x_{2},y_{2}|\vec{\mu}_{1},\vec{\mu}_{2}\rangle=g_{\vec{\mu}_{1}}(x_{1},y_{1})g_{\vec{\mu}_{2}}(x_{2},y_{2});
\end{equation}
and the approximation is valid for large pump beam widths and thin crystals.

Deriving the transverse wavefunction of a biphoton generated in collinear SPDC is generally more involved than the case where we also consider only degenerate frequencies \cite{Schneeloch_SPDC_Reference_2016}. Instead, one must integrate the biphoton wavefunction over the frequency spectrum of the down-converted light, and renormalize accordingly, resulting in a substantially broadened wavefunction. However, we may still approximate the accurate biphoton wavefunction as a scaled representation of the biphoton wavefunction in the degenerate frequency case. We scale $a$ by a constant factor $\phi$, and find for type-I SPDC:
\begin{equation}\label{totalType1}
R_{T}^{(t1)}=\frac{32\sqrt{2\pi^{3}}}{27\epsilon_{0}c}\Big(\frac{n_{g1}n_{g2}}{n_{1}^{2}n_{2}^{2}}\Big)\frac{d_{eff}^{2}}{\lambda_{p}^{3}\sqrt{\kappa}}\frac{P\sqrt{L_{z}}}{\phi}.
\end{equation}
Here, we approximate $\phi\approx0.335$ by matching the peaks of the degenerate and more accurate biphoton wavefunction in the same fashion as one can obtain a double-gaussian approximation to the biphoton wavefunction \cite{Schneeloch_SPDC_Reference_2016}. An interesting qualitative point here discussed in other references \cite{suzer2008does} is that although the single-mode brightness increases with focusing (i.e., decreasing $\sigma_{p}$), the overall brightness does not  increase this way, unless the focusing is strong enough that the curvature of the phase fronts of the pump beam affects phase matching.

\subsubsection{Degenerate type-II SPDC}
In type-II SPDC, the signal and idler photons are of orthogonal polarizations, and experience different indices of refraction. In this regime, the linear contribution to $\Delta k_{z}$ about the signal and idler photons' central frequencies \eqref{dkExp} is nonzero, and cannot be ignored. In this case:
\begin{align}
R_{SM}&\propto\int d\omega_{1} \omega_{1}(\omega_{p}-\omega_{1})\times\nn\\
&\times\text{Sinc}^{2}\Big(\frac{L_{z}|n_{g1}-n_{g2}|}{2c}\big(\omega_{1}-\frac{\omega_{p}}{2}\big) + \frac{L_{z}\kappa}{2}\big(\omega_{1}-\frac{\omega_{p}}{2}\big)^{2}\Big)
\end{align}
For most nonlinear optical materials, the quadratic contribution to the argument of the Sinc function is negligible relative to the linear contribution because the group index difference $\Delta n_{g}$ is large enough (of the order $10^{-2}$ or greater for most materials) in comparison to the goup-velocity dispersion $\kappa$. This integral cannot be done analytically, but can be bounded from above. Because the square of the sinc function is a non-negative function, and $\omega_{1}(\omega_{p}-\omega_{1})\leq\omega_{p}^{2}/4$, the rate is bounded above by an integral that can be done analytically. Indeed:
\begin{equation}
R_{SM}\propto \frac{c\pi\omega_{p}^{2}}{2 L_{z}\Delta n_{g}}
\end{equation}
The approximate proportionality is valid, when the width of the sinc function in $\omega_{1}$ is much less than the pump frequency (typically, less than a quarter in most nonlinear media). Consequently, the approximation is an over-estimate (typically by less than seven percent). From this, we can get the single-mode rate for type-II degenerate SPDC:
\begin{equation}\label{type2rate}
R_{SM}^{t2}=\frac{1}{\pi\epsilon_{0}c^{2}}\frac{n_{g1}n_{g2}}{n_{1}^{2}n_{2}^{2}n_{p}}\frac{(d_{eff})^{2}\omega_{p}^{2}}{\Delta n_{g}}\Big|\frac{\sigma_{p}^{2}}{\sigma_{1}^{2}+2\sigma_{p}^{2}}\Big|^{2} \frac{P}{\sigma_{p}^{2}}L_{z}
\end{equation}
Interestingly, one may compare this to the corresponding single-mode rate for collinear type-II SPDC derived in \cite{LingPRA2008}, and see that our formula differs by a near-unity factor of the ratio of the indices of refraction $n_{g1}n_{g2}/n_{1}n_{2}$, amounting to only a 3 percent difference in prediction using their experimental parameters. For a description of the absolute biphoton generation rate into \emph{non-collinear} gaussian modes, such as is useful when using type-II SPDC as a source of polarization-entangled photon pairs, their reference provides an invaluable discussion.

In order to get the total rate for type-II SPDC, one can use the inner product between the zeroth order hermite-gaussian modes, and biphoton wavefunction for type-II SPDC as was done previously \eqref{totalType1} for type-I SPDC. However, the biphoton wavefunction for type-II SPDC is not as straightforward to derive or approximate, due to transverse walk-off between the signal and idler light \footnote{The "walk-off" effect, where the signal and idler light have different mean momenta (though still adding to the pump) is due to the index of refraction in birefringent crystals being dependent on direction of propagation. Because the group velocity depends on the gradient of the frequency with respect to momentum, the group velocity and mean phase velocity may point in different directions.}. For a thorough analysis of the biphoton wavefunction in type-II SPDC, see \cite{walbornSPDC}.

\subsubsection{Degenerate SPDC with narrow frequency filtering}
In certain SPDC experiments where a pair of identical photons is preferable to a pair of highly correlated photons, one can narrowly filter the frequency spectrum of the signal and idler photons so that each is tightly clustered around half the pump frequency. Because the bandwidth of these frequency filters may be some orders of magnitude narrower than the natural bandwidth  of the down-converted light, the rate of biphotons generated passing through a narrowband frequency filter behaves differently than the overall rate of biphoton generation.

In particular, if we include a narrowband frequency filter, the integral over $\omega_{1}$ for the rate \cite{Helt:12} simplifies significantly, since the Sinc function is essentially unity over the passband of the filter. Since the integral no longer depends on the width of the Sinc function, the biphoton rate will not depend on group velocity dispersion $\kappa$ or group index mismatch $\Delta n_{g}$. Moreover, the rate will scale as the square of the crystal length $L_{z}$, as one might expect when the probability amplitude for the SPDC event is obtained by integrating over the volume of the crystal.

\subsection{Non-degenerate SPDC}
By angle and temperature tuning the crystal, it is possible that the signal and idler frequency spectra no longer overlap, having different central frequencies that add up to the pump frequency. The Taylor expansion for $\Delta k_{z}$ is taken with respect to the signal beam's center frequency $\omega_{1(0)}$. In this case:
\begin{equation}\label{dkdeg}
\Delta k_{z}\approx \Big(\frac{\Delta n_{g}}{c}\Big)\big(\omega_{1}-\omega_{1 (0)}\big) +\frac{(\kappa_{1}+\kappa_{2})}{2} \big(\omega_{1}-\omega_{1 (0)}\big)^{2}
\end{equation}
Here, $\kappa_{1}$ and $\kappa_{2}$ are the group velocity dispersion constants at the signal and idler central frequencies, respectively. When the central frequencies are different enough that the group index mismatch $\Delta n_{g}$ is significant (e.g., greater than $10^{-2}$), the rate of biphoton generation is qualitatively identical for both type-I and type-II SPDC.

\subsection{Periodic poling}
Thus far, we have examined the absolute brightness of SPDC in isotropic crystals (i.e., where $\chi^{(2)}$ is a constant throughout the crystal volume). This is a fine regime when perfect phase matching is achievable (that is, where tuning the crystal allows the indices of refraction to be such that $\Delta k_{z}=0$). However, this is not always possible. The general dependence of biphoton brightness on crystal length $L_{z}$ is given by:
\begin{equation}\label{Lzdep}
R_{SM}\propto\int\;d\omega_{1}\omega_{1}(\omega_{p}-\omega_{1})\Bigg|\int_{-\infty}^{\infty} dz\; \bar{\chi}(z) e^{-i\Delta k_{z} z}\Bigg|^{2},
\end{equation}
where for an isotropic crystal $\bar{\chi}(z)$ is unity inside the crystal, and zero outside. When perfect phase matching is not achievable (i.e., when the indices of refraction are not compatible for SPDC at the desired pump and signal/idler frequencies), the magnitude square of the integral over $z$ oscillates with crystal length between zero and $4/\Delta k_{z}^{2}$. The value of $\Delta k_{z}$ is set by the frequencies of the pump, signal, and idler light, and the indices of refraction at their respective frequencies. For a given set of pump, signal, and idler frequencies, imperfect phase matching can be ameliorated by periodically poling the nonlinear crystal. If one switches the poling direction (changing $\bar{\chi}$ from $1$ to $-1$) just as the amplitude is maximum (i.e., when $L_{z}=\pi/\Delta k_{z}$), the amplitude grows further as though it were at a minimum. See Fig.~1 for comparison with and without periodic poling. By switching the poling periodically at these intervals such that the poling period $\Lambda_{\text{pol}}=2\pi/\Delta k_{z}$, one can achieve significant brightness without perfect phase matching. This technique is known as \emph{quasi-phase} matching.

\begin{figure}[t]
\centerline{\includegraphics[width=\columnwidth]{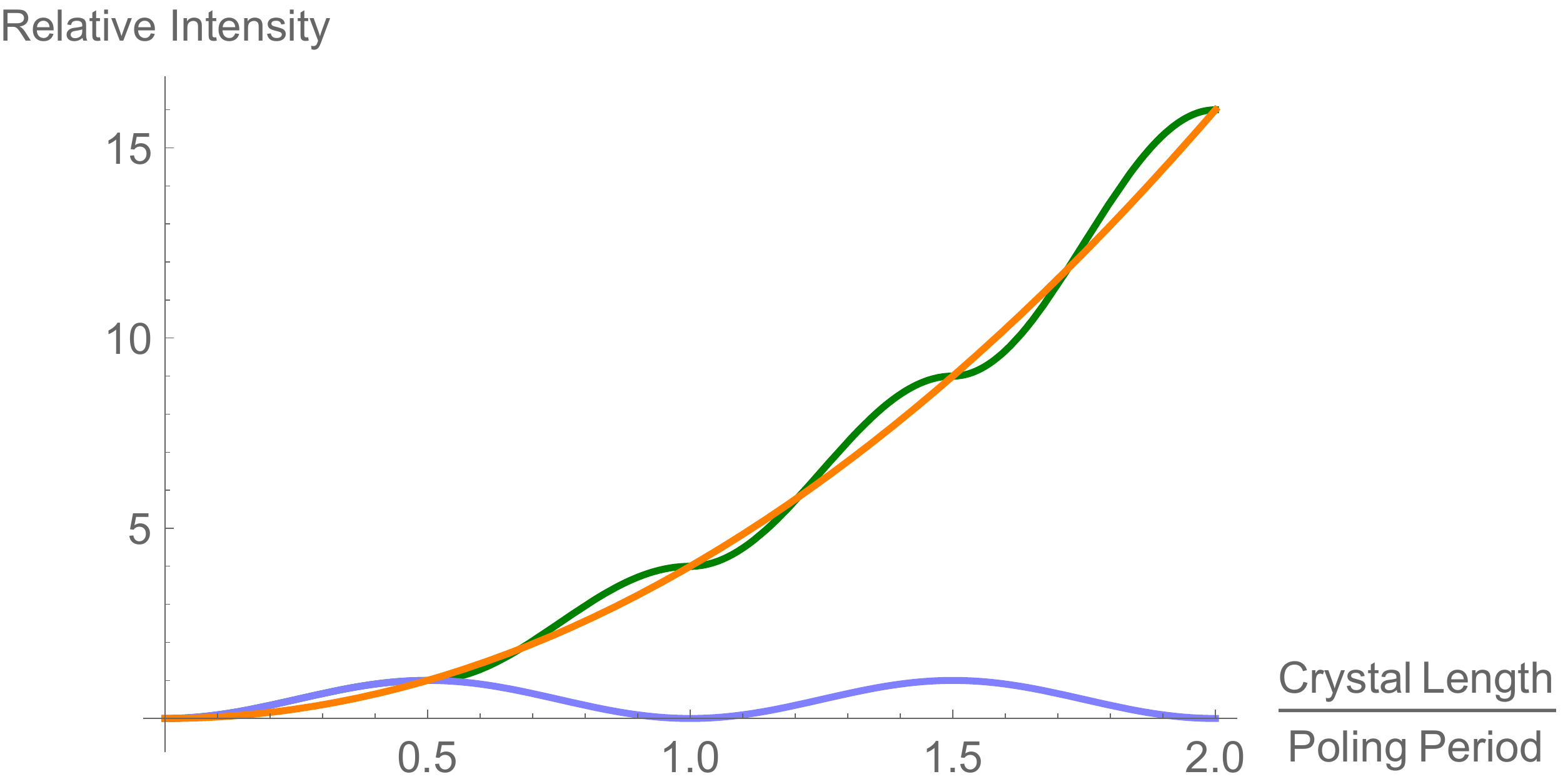}}
\caption{Plot of the relative intensity of down-converted light with quasi-phase matching \eqref{Lzdep} as a function of crystal length measured in units of the poling period $\Lambda$. The sinusoidal blue curve is the relative brightness without periodic poling, while the oscillating ascending green curve gives the relative brightness when the crystal is periodically poled for first-order quasi-phase matching. The parabolic orange curve is the approximate relative brightness with first-order quasi-phase matching \eqref{quasiCorr}.}
\end{figure}

As shown in \cite{boyd2007nonlinear}, the poling profile $\bar{\chi}(z)$ can be broken up into a Fourier series with fundamental momentum $\frac{2\pi}{\Lambda_{\text{pol}}}$:
\begin{equation}
\bar{\chi}(z)=X_{0} +\sum_{n\neq0}X_{n}e^{ik_{n}z}\qquad:\qquad k_{n}=\frac{2\pi}{\Lambda_{\text{pol}}}n
\end{equation}
where $n$ runs from $-\infty$ to $\infty$ excluding zero, $X_{0}=0$, and
\begin{equation}
X_{n}=\text{Sinc}\Big(\frac{n\pi}{2}\Big).
\end{equation}
From this, we see the length dependence \eqref{Lzdep} simplifies to:
\begin{equation}
R_{SM}\propto\int\;d\omega_{1}\omega_{1}(\omega_{p}-\omega_{1})\Bigg|\int_{-\frac{L_{z}}{2}}^{\frac{L_{z}}{2}} \!\!dz\sum_{n}X_{n} e^{-i(\Delta k_{z}-k_{n}) z}\Bigg|^{2},
\end{equation}
In performing this quasi-phase matching, typically only one Fourier component $X_{n}$ will contribute to the brightness because only one value of $k_{n}$ will be close enough to offset $\Delta k_{z}$ to achieve quasi-phase matching. The range of values of $\Delta k_{z}$ over which phase-matching is favorable is approximately $4\pi/L_{z}$, while the shift in $\Delta k_{z}$ between different orders of quasi-phase matching is $4\pi/\Lambda_{\text{pol}}$, which is larger often by multiple orders of magnitude. Since $X_{n}$ decreases with $n$, first-order phase matching (i.e., $n=1$ or $-1$), is most desirable for maximum brightness. The calculation for $R_{SM}$ follows the same steps with periodic poling as with an isotropic crystal. $\Delta k_{z}$ is still Taylor-expanded about the signal and idler central frequencies. The only difference is that the zero-order terms for $\Delta k_{z}$ added to $-k_{m}$ gives zero instead. As such, the single-mode rate of biphoton generation when periodic poling with $n^{\text{th}}$ order quasi-phase matching, $R_{SM}^{PP(n)}$ is multiplied by the factor $X_{n}^{2}$:
\begin{equation}\label{quasiCorr}
R_{SM}^{PP(n)}=\frac{4}{n^{2}\pi^{2}}R_{SM}
\end{equation}
This correction holds for all types of down-conversion, and will work for the multi-mode regime (discussed previously) as well. It is important to note that where published values for $d_{\text{eff}}$ differ between isotropic and periodically poled nonlinear crystals of the same material, these factors of $\frac{2}{n\pi}$ are already included.

Although periodic poling is accomplished by switching the crystal orientation (and therefore the sign of $\chi^{(2)}$) periodically over the length of the crystal, this is not the only fashion in which quasi-phase matching can be achieved. If one instead periodically dopes the crystal, changing its composition periodically over its length, and therefore periodically changing $\chi^{(2)}$, quasi-phase matching may be achieved in precisely the same regimes. Alternatively, in a waveguide, one can produce a sinusoidal variation in the pump intensity by sinusoidally varying the width of the waveguide, which can also be used to achieve quasi-phase matching \cite{rao2017second}. As one final note, poling periods in some materials can be made so small that the fundamental momentum completely offsets the pump momentum. In this regime, it is possible to produce counter-propagating photon pairs \cite{pasiskevicius2008mirrorless, pasiskevicius2012quasi} in SPDC.

\subsection{SPDC with a focused pump beam}
In all the situations considered thus far, the pump beam was considered collimated. However, if one wants to maximize the number of biphotons generated per second that couple into a single-mode fiber, a focused beam offers significant improvement (as discussed previously). In order to see how the single-mode rate changes in the regime of tight focusing, we turn to the work of Bennink \cite{BenninkFocusSPDC}, who treats this situation in detail.

The dependence of the rate of biphoton generation on the spatial aspects of the pump beam is given by the overlap integral
\begin{equation}
R_{SM}\propto \Big|\int d^{3}r\Big(\chi^{(2)}_{eff}(\vec{r})E^{*}_{p}(\vec{r}) E_{1}(\vec{r})E_{2}(\vec{r})\Big)\Big|^{2},
\end{equation}
where in our approximations, $D\approx\epsilon_{0}n^{2}E$. In order to properly treat collinear SPDC into the zeroth-order signal/idler gaussian modes when the pump beam is focused strong enough that its width changes significantly over the length of the crystal, Bennink uses a slightly different expression for the signal/idler spatial modes. Instead of being gaussian in transverse dimensions, and constant along the optic axis (i.e., collimated), Bennink considers the signal/idler fields as focused gaussian beams with their own beam parameters in addition to the pump beam. While a full discussion of his calculations is beyond the scope of this tutorial (and redundant), he finds the joint pair-collection probability, which is proportional to the biphoton generation rate. For type-II SPDC, and non-degenerate type-I SPDC, for near-perfect phase matching, and assuming identical beam focal parameters $\xi$ for the pump, signal and idler modes, one can show:
\begin{equation}
R_{SM}^{t2}\propto \frac{d_{eff}^{2}\omega_{p}^{3}}{\Delta n_{g}}\text{Tan}^{-1}(\xi)P,
\end{equation}
where the (pump) beam focal parameter $\xi$ is defined as the ratio of the crystal length $L_{z}$ divided by twice the Rayleigh range, $z_{R}$. Thus, a small focal parameter indicates a nearly collimated beam. In the limit of a nearly collimated beam, Bennink's formula coincides with the single-mode formula derived previously \eqref{type2rate} up to constant factors.

To date, no calculations have obtained the absolute coincidence rates in the regime of focused pump beams, but Bennink's work captures the salient qualitative behavior of the biphoton generation rate on changing pump focal parameter. In addition, the work of \cite{dixon2014heralding} expands on these results, and shows how one may sacrifice absolute brightness in exchange for a greatly improved heralding efficiency, as is useful in developing SPDC as a source of heralded single photons.

\section{SPDC beyond the first-order approximation: The two-mode squeezed vacuum}
In experimental studies of SPDC, it is only in the case of relatively low pump powers where SPDC is accurately described by first-order perturbation theory. In that approximation, the interaction of the pump beam with the quantum vacuum either produces nothing, or yields a biphoton with low probability. However, the calculation to higher orders of perturbation theory show the down-converted field to be in a superposition of not just the vacuum state and the single biphoton Fock state, but also of multi-biphoton Fock states as well. Although one could perform the perturbation theory calculation to higher orders, it is actually possible in another approximation to solve the Schr\"{o}dinger equation exactly for SPDC \cite{LvovskySqueezed2016,LoMultiSPDC}.

Here we consider the case of a collimated pump beam in the zero-order transverse gaussian mode, coupled to the zero-order signal and idler gaussian modes. In this single-mode approximation, we may solve for the time evolution of the signal and idler creation operators using Heisenberg's equation of motion. This approximation is quite accurate for experiments where the down-converted light is coupled into single-mode fibers, as mentioned previously.

Using the single-mode approximation, the nonlinear hamiltonian \eqref{GenHam} is given by:
\begin{equation}
\hat{H}_{NL}=\hbar\sum_{k_{p},k_{1},k_{2}}\Big(i G_{k_{p},k_{1}k_{2}}\hat{a}_{k_{p}}\hat{a}^{\dagger}_{k_{1}}\hat{a}^{\dagger}_{k_{2}} + h.c.\Big)
\end{equation}
such that
\begin{align}
G_{k_{p},k_{1},k_{2}}&\equiv -\sqrt{\frac{\hbar}{2 L_{z}^{3}\epsilon_{0}}}\sqrt{\frac{\omega(k_{p})\omega(k_{1})\omega(k_{2})}{n^{2}(k_{p})n^{2}(k_{1})n^{2}(k_{2})}}e^{i\Delta\omega t}\nn\\
&\times\int d^{3}r\big(\chi_{eff}^{(2)}(\vec{r}) g^{\ast}_{\vec{\mu}_{p}}(x,y)g_{\vec{\mu}_{1}}(x,y)g_{\vec{\mu}_{2}}(x,y)e^{-i\Delta k z}\big)\label{coupling}
\end{align}
where we let $k_{1}=k_{1z}$ to simplify notation, and $h.c.$ denotes hermitian conjugate. At this point we invoke the approximation that $\Delta\omega\approx 0$ over the time it takes light to propagate through the crystal.

When the pump beam is narrowband enough that its coherence length is longer than the crystal length $L_{z}$ or alternatively that its longitudinal momentum bandwidth $\Delta k_{p}$ is smaller than $2\pi/L_{z}$, we need only consider one value of $k_{p}$ contributing to the general hamiltonian. For typical lasers, this condition is easily satisfied, and makes subsequent calculations much simpler. We make use of this approximation, and let $G_{k_{1},k_{2}}$ be substituted for $G_{k_{p},k_{1},k_{2}}$ to condense notation.

Initially, the signal and idler fields are in the vacuum state. By solving Heisenberg's equations of motion for the annihilation operators of the fields, we can see what the statistics of the signal and idler fields are as the light exits the nonlinear crystal. The evolution of the annihilation operator $\hat{a}_{k_{1}}$ is given by the equation:
\begin{equation}
\frac{d\hat{a}_{k_{1}}}{dt}=\frac{-i}{\hbar}\Big[\hat{a}_{k_{1}},\hat{H}_{NL}\Big].
\end{equation}
Using the boson commutation relation:
\begin{equation}
\Big[\hat{a}_{k_{1}},\hat{a}_{k_{1'}}^{\dagger}\Big]=\delta_{k_{1},k_{1'}}.
\end{equation}
we find that:
\begin{align}
\frac{d\hat{a}_{k_{1}}}{dt}&=-\sum_{k_{2}}G_{k_{1}k_{2}}\hat{a}_{k_{p}}\hat{a}^{\dagger}_{k_{2}}\nn\\
\frac{d\hat{a}_{k_{2}}}{dt}&=-\sum_{k_{1'}}G_{k_{1}k_{2}}\hat{a}_{k_{p}}\hat{a}^{\dagger}_{k_{1}}
\end{align}
and similarly, that
\begin{equation}
\frac{d\hat{a}_{k_{p}}}{dt}=-\sum_{k_{1},k_{2}}G_{k_{1}k_{2}}\hat{a}_{k_{1}}\hat{a}_{k_{2}}
\end{equation}

If we take the undepleted pump approximation, then $\frac{d\hat{a}_{k_{p}}}{dt}\approx 0$, and $\hat{a}_{k_{p}}\hat{a}^{\dagger}_{k_{p}}=\hat{N}_{p}+1\approx\hat{N}_{p}$, and we get a second-order differential equation for the annihilation operator $\hat{a}_{k_{1}}$:
\begin{equation}
\frac{d^{2}\hat{a}_{k_{1}}}{dt^{2}}=\sum_{k_{1'}}(\hat{a}_{kp}\big(GG^{\dagger}\big)_{k_{1}k_{1'}}\hat{a}^{\dagger}_{k_{p}})\hat{a}_{k_{1'}}
\end{equation}
 For all signal modes $\hat{a}_{k_{1}}$, the corresponding linear system of second-order differential equations is expressible with vector notation:
\begin{equation}
\frac{d^{2}\vec{\hat{a}}_{1}}{dt^{2}}=\big((\hat{N}_{p})GG^{\dagger}\big)\cdot\vec{\hat{a}}_{1},
\end{equation}
where $\hat{a}_{k_{1}}$ is a particular component of $\vec{\hat{a}}_{1}$. To solve this system of equations, we can diagonalize $GG^{\dagger}$ and solve for the time evolution of the eigenmodes of the hamiltonian. This calculation greatly simplifies assuming G is hermitian, which it is, under our current approximations. Using this, along with similar equations governing the evolution of $\vec{\hat{a}}_{1}^{\dagger}$, $\vec{\hat{a}}_{2}$, and $\vec{\hat{a}}_{2}^{\dagger}$, one can obtain the solution.
\begin{equation}\label{TimeEvolved}
\vec{\hat{a}}_{1}(t)=\text{cosh}\Big(\sqrt{\hat{N}_{p}}Gt\Big)\cdot\vec{\hat{a}}_{1}-i\text{ sinh}\Big(\sqrt{\hat{N}_{p}}Gt\Big)\cdot\vec{\hat{a}}^{\dagger}_{2}.
\end{equation}

\subsection{The single-mode rate from the two-mode squeezed vacuum}
Having found a formula for the time evolution of the annihilation operators, the number of photon pairs can be calculated by finding the expectation value $\langle \hat{a}_{k_{1}}^{\dagger}\hat{a}_{k_{1}}\rangle$ and summing over all modes $k_{1}$:
\begin{equation}
N_{SM}(t)=\sum_{k_{1}}\langle \hat{a}_{k_{1}}^{\dagger}\hat{a}_{k_{1}}\rangle(t)=\sum_{k_{1}k_{2}}\Big|\text{sinh}\Big(\sqrt{\hat{N}_{p}}Gt\Big)_{k_{1}k_{2}}\Big|^{2},
\end{equation}
so that in the same limits where the first-order approximation is valid:
\begin{equation}
N_{SM}(t)\approx\sum_{k_{1}k_{2}}\Big|G_{k_{1}k_{2}}\Big|^{2}\langle\hat{N}_{p}\rangle t^{2}.
\end{equation}
Here, the mean pump photon number $\langle\hat{N}_{p}\rangle$ will be the average number of pump photons in the nonlinear medium at any given time:
\begin{equation}
\langle\hat{N}_{p}\rangle=\frac{P}{\hbar\omega_{p}}\cdot \frac{L_{z}n_{p}}{c},
\end{equation}
where $P$ is pump power.

Considering the simple case of a collimated gaussian pump beam coupled to a pair of gaussian signal and idler modes, and that the length of the crystal is much larger than the wavelength of the pump light, the only contributions to the sum over $k_{1}$ and $k_{2}$ are those such that $\Delta k_{z}=0$. This is then a sum over one variable, which we may approximate as an integral, and express in terms of frequency. For a given function $f(k_{1},k_{2})$:
\begin{align}
\sum_{k_{1}}&f(k_{1},k_{p}-k_{1})\Big(\frac{2\pi}{L_{z}}\Big)\approx\int dk_{1} f(k_{1},k_{p}-k_{1})\nn\\
&=\int dk_{1} dk_{2} f(k_{1},k_{2})\delta(k_{1}+k_{2}-k_{p})\nn\\
&=\frac{n_{g1}n_{g2}}{c^{2}}\int d\omega_{1}d\omega_{2}f(\omega_{1},\omega_{2})\delta\Big(\frac{n_{1}\omega_{1} + n_{2}\omega_{2}-n_{p}\omega_{p}}{c}\Big)\nn\\
&=\frac{n_{g1}n_{g2}}{n_{1}c}\int d\omega_{1}d\omega_{2}f(\omega_{1},\omega_{2})\delta\Big(\omega_{1}+\frac{n_{2}}{n_{1}}\omega_{2} -\frac{n_{p}}{n_{1}}\omega_{p}\Big)
\end{align}
For type-0 and type-I phase matching, the $\delta$ function simplifies to $\delta(\Delta\omega)$, which gives us:

\begin{align}
&\frac{N_{SM}(t)}{t^2}\approx \frac{2}{\pi^{2}\epsilon_{0}c^{2}}\frac{n_{g1}n_{g2}}{n_{p}n_{1}^{3}n_{2}^{2}}\Bigg|\frac{\sigma_{p}^{2}}{\sigma_{1}^{2} + 2\sigma_{p}^{2}}\Bigg|^{2}\frac{P\; d_{eff}^{2}L_{z}}{\sigma_{p}^{2}}\times\nn\\
&\times\int d\omega_{1}\omega_{1}\big(\omega_{p}-\omega_{1}\big)\text{Sinc}^{2}\Big(\frac{\Delta k_{z}L_{z}}{2}\Big)
\end{align}
These integrals over frequency can be evaluated or approximated with the same methods discussed in the previous section. For type-I degenerate SPDC,
\begin{equation}\label{rsq}
\frac{N_{SM}(t)}{t^2}\approx \sqrt{\frac{2}{\pi^{3}}}\frac{2}{3\epsilon_{0}c^{2}}\frac{n_{g1}n_{g2}}{n_{p}n_{1}^{3}n_{2}^{2}}\frac{d_{eff}^{2}\;\omega_{p}^{2}}{\sqrt{\kappa}}\Bigg|\frac{\sigma_{p}^{2}}{\sigma_{1}^{2} + 2\sigma_{p}^{2}}\Bigg|^{2}\frac{P}{\sigma_{p}^{2}}L_{z}^{1/2}
\end{equation}
For type-II phase matching, the $\delta$ function does not simplify to $\delta(\Delta\omega)$, but the same upper bound approximation may be taken. The value of $N_{SM}(t)$ obtained  will be the same as if one let the $\delta$ function be $\delta(\Delta\omega)$, but with an additional factor of $(2n_{p}-n_{1})/n_{2}$, which is of the order unity.

Finally, to obtain the single-mode \emph{rate}, we point out that $N_{SM}(t)$ is the mean number of biphotons generated as a function of time, for times less than what it takes for a pump photon to travel through the crystal. The rate is the ratio of $N_{SM}(T_{DC})$ over $T_{DC}$, where $T_{DC}$ is the time it takes either the pump or downconverted light to travel the length of the crystal \footnote{Phase matching occurs in degenerate type-0 and type-I SPDC when the indices of refraction of the pump light and the down-converted light are identical.}, which is  $L_{z}n_{1}/c$, giving us:
\begin{equation}
R_{SM}\approx\sqrt{\frac{2}{\pi^{3}}}\frac{2}{3\epsilon_{0}c^{3}}\frac{n_{g1}n_{g2}}{n_{1}^{2}n_{2}^{2}n_{p}}\frac{(d_{eff})^{2}\omega_{p}^{2}}{\sqrt{\kappa}}\Big|\frac{\sigma_{p}^{2}}{\sigma_{1}^{2}+2\sigma_{p}^{2}}\Big|^{2}\frac{P}{\sigma_{p}^{2}}L_{z}^{3/2}
\end{equation}
which agrees precisely with the formula for the rate of generated biphotons we obtained earlier via first-order perturbation theory.

\subsection{The number statistics of the SPDC state}
Previously, we solved for the time evolution of the signal and idler annihilation operators. However, using that relation to obtain the actual quantum state of SPDC light takes one additional step.

If we define $U$ as a unitary transformation diagonalizing the matrix $G$, the same transformation will define eigenmodes of the two-mode squeezing operator.

Let $\Lambda$ be the diagonalized matrix of $G$:
\begin{equation}
\Lambda=UGU^{\dagger}
\end{equation}
Furthermore, let the annihilation operators $\vec{\hat{b}}_{1}$ be defined as $U\cdot\vec{\hat{a}}_{1}$, (i.e., the annihilation operators of the eigenmodes of the SPDC hamiltonian). Then, the linear system of equations for the annihilation operators separates into independent linear equations for the annihilation operators of the eigenmodes:
\begin{equation}
\vec{\hat{b}}_{1}(t)=\text{cosh}(\sqrt{\hat{N}_{p}}\Lambda t)\cdot\vec{\hat{b}}_{1}-i\text{ sinh}(\sqrt{\hat{N}_{p}}\Lambda t)\cdot\vec{\hat{b}}^{\dagger}_{2}.
\end{equation}
The quantum state of SPDC light is obtained from these eigenmodes \cite{LvovskySqueezed2016}, and is a product of multiple two-mode squeezed states (one for each correlated pair of eigenmodes) when the pump beam is in a coherent state. For reference, the two-mode squeezed state between modes 1 and 2 with squeezing amount $r$ is given by:
\begin{equation}
|TMSV\rangle = \frac{1}{\text{cosh}(r)}\sum_{n=0}^{\infty}\text{tanh}^{n}(r)|n\rangle_{1}|n\rangle_{2}
\end{equation}

With the two-mode squeezed vacuum state properly scaled to fit experimental parameters, we can explore what we expect to measure as we increase the intensity of the pump. In a time interval equal to the length of time it takes light to pass through the nonlinear crystal, the state of the field has probabilities to be in a zero biphoton state, a one-biphoton state, a two-biphoton state, and so on. Light whose number statistics obey this exponentially decaying photon number distribution is known as thermal or super-Poissonian light because its variance is larger than its mean. In contrast, coherent light (as from dipole radiation or laser light) has Poissonian number statistics. That said, it may seem surprising that coincidence counting measurements show Poissonian statistics for the down-converted light \cite{PhysRevLett.101.053601}. However, realistic experiments exhibit photodetection across multiple pairs of modes; the emprical number statistics are those of a mixture of multiple exponentially distributed random variables, which is better described with a Poisson distribution.

 In order to serve as a viable source for heralded single photons, the number of higher-order biphoton states generated must be small, relative to the single-biphoton state. Fortunately, the expression for the relative likelihood of higher-order biphoton number states is quite simple:
\begin{equation}
\frac{P(2\text{ or more})}{P(1)}=\text{sinh}^{2}(r)
\end{equation}
When considering the SPDC state as a product of multiple two-mode squeezed vacuum states, the ratio of events of multi-biphoton generation to events of single biphoton generation is straightforward to estimate. First,  the total ratio of multi-biphoton generation events to single biphoton generation events is approximately the mean of the ratios of multi-biphoton to single biphoton events in each pair of modes. We can estimate this as the sum of the ratios over all modes (which happens to equal $N_{SM}(T_{DC})$) times the mean probability over all mode pairs. For type-I SPDC, we find:
\begin{equation}
\frac{P(2\text{ or more})}{P(1)}\approx N_{SM}(T_{DC})*\frac{12}{35}(4-\sqrt{2})c\sqrt{\frac{\kappa\pi}{L_{z}}}
\end{equation}
where again, $\kappa$ is the group velocity dispersion constant for the down-converted light. In order to obtain this formula, we used the large signal-idler correlations to estimate the marginal frequency probability density\footnote{For a good reference detailing the calculation of the joint frequency probability distribution of biphotons in SPDC, see \cite{Mikhailova2008}}, and converted it to momentum to calculate the mean probability as the integral of the square of the probability density times the mode spacing $2\pi/L_{z}$. For typical experimental parameters in bulk, this ratio of multi-biphoton events to single biphoton events is of the order $10^{-8}$ per Watt of pump power. For CW beams of typical intensities, multi-biphoton events would be exceedingly rare. However, using pulsed lasers with a moderate mean power, but small pulse length, it is possible to achieve the high (peak) power levels necessary at the picosecond time scales near $T_{DC}$ (i.e., how long light takes to travel through the crystal). Indeed, when using pulsed SPDC in improved heralded single photon sources, multi-photon events are significant enough to limit the overall system efficiency, so that new strategies (such as in \cite{broomeMultiSPDC}) are being developed to reduce both the number and impact of these events.

\section{SPDC in waveguides and resonators}
Although it is possible to couple entangled light into single-mode fibers, it is also possible to generate SPDC light inside a waveguide made of the appropriate nonlinear material, so that the down-converted light is already propagating in spatial modes easily coupled to fibers physically attached to the nonlinear medium. With the intensity of the pump light being large over the whole length of the waveguide, comparatively large pair generation rates can be achieved in a single spatial mode compared to what has been done in the bulk regime. In this section, we will consider first, the simple case of SPDC in an antireflection (AR) coated nonlinear waveguide, and follow this with the more sophisticated treatment of SPDC in a cavity (e.g., a waveguide without AR coatings) as in a micro-ring resonator. Because the pump light intensity may be much larger inside a cavity, it is possible to increase the efficiency of SPDC, though at the expense of increasing likelihood of multi-biphoton generation events.

\subsection{SPDC in a single-mode waveguide}
In a single-mode waveguide, the rate of biphoton generation in SPDC is particularly simple to calculate. As in a single mode fiber, one pump transverse spatial mode can propagate through the waveguide, in addition to one transverse signal and idler mode. In the bulk crystal regime, we decomposed the down-converted light into hermite-gaussian spatial modes, but we could just as easily decompose them into any basis of modes fitting a particular waveguide. Indeed, we may approximate the spatial modes of the waveguide with hermite-gaussian modes by setting the standard deviations $\sigma_{p}$ and $\sigma_{1}$ as equal to a fourth of the mode field diameters appropriate to those waveguides at the appropriate wavelengths. 

However, because the pump light and down-converted light are a full octave of frequency apart, a waveguide that is single-mode for the down-converted light will be multi-mode at the much shorter pump wavelength. Ordinarily, the multi-mode pump light adds a degree of complication due to modal dispersion \footnote{Modal dispersion is where the group velocity of light in higher-order spatial modes is slower than that of lower-order spatial modes. This is due to the larger transverse component of momentum taking away from the longitudinal component of momentum for an otherwise monochromatic beam.}, which makes phase matching more challenging with each spatial mode experiencing a different effective index of refraction. However, with a graded index profile (as is the case with waveguides produced by diffusing a dopant into a nonlinear medium) this effect can be mitigated, since the range of indices over the spatial modes can be made small. In Section VII, we test our theoretical prediction for type-II SPDC into a single-spatial mode using a Periodically Poled Potassium Titanyl Phosphate (PPKTP) waveguide.

\subsection{SPDC in optical cavities and resonators}
When considering SPDC in optical cavities and resonators, it becomes necessary to accomodate loss (over possibly many round trips) to have even a qualitatively accurate description. This is the case, even when the material is sufficiently lossless to exploit the symmetries of the nonlinear susceptibility for later calculation. While the unitary evolution of a closed quantum system does not permit any loss of energy (by say, absorption), it is straightforward to describe loss as a coupling between modes of an extended quantum system-plus-environment, with a correspondingly extended unitary evolution. In doing this, we remain able to treat SPDC in a lossy medium with our standard nonlinear hamiltonian, but where the signal, idler, and pump creation and annihilation operators experience a continuous series of couplings (theoretically, with generalized beamsplitters (BS)) to scattering modes over the length of the medium, as discussed further in this section.

In our treatment of SPDC in cavities and resonators, we begin with a brief discussion for how the photon creation/annihilation operators evolve when passing through a lossy medium. Following this, we give an abbreviated introduction describing how the modes in a single-bus micro-ring resonator (MRR) are coupled to one another as a prototypical example of an optical cavity (see Fig.~4 for diagram). With this understanding, we then proceed to describe SPDC in a MRR, where the nonlinear medium is the resonator itself. We find the Heisenberg equation of motion for the photon creation/annihillation operators in the lossy MRR, and use the relationship between the fields inside and outside the MRR to obtain the state of the down-converted light in the output bus, where such light can be directed and collected in a variety of experiments. With the state of the exiting SPDC light, we calculate the generation rate of exiting photon pairs, as well as isolated singles due to loss, among other factors, and compare the two to see what factors impact the relative quality (i.e., heralding efficiency) of cavity-based SPDC photon sources. We conclude with a brief discussion on how the time correlations between photon pairs are affected by the MRR. For a thorough discussion of nonlinear optics in micro-ring resonators, we recommend the PhD theses \cite{vernon2017microresonators} and \cite{gentry2018scalable}.

To keep notation simple, we assume a ``particle-in-a-box'' mode expansion vs the more realistic hermite-gaussian decomposition, as discussed above. For simplicity, we will also assume near-perfect phase matching and negligible dispersion. This is a valid approximation when the phase matching bandwidth is much wider than the linewidth of the cavity, and where the optical properties of the material are also essentially constant over this linewidth. With this, we can concentrate on the effects that the passive feedback of the MRR cavity has on photon-pair generation.

\subsubsection{Beam Splitters, propagation loss and cavities}\label{subsec:SPDC:SFWM:preliminaries}
\paragraph{Beam splitters:}
Before discussing cavities, let us  first discuss the simplest of all passive optical elements, the beam splitter (BS) through which fields will enter and exit a cavity. In Fig.~2, we illustrate the standard BS with input modes $\hat{a}_{in}, \hat{b}_{in}$ and output modes $\hat{a}_{out}, \hat{b}_{out}$, related by the unitary matrix $U_{bs}$ with transmission and refection coefficients $\tau, \rho$ such that $|\tau|^2 + |\rho|^2 =1$;

\begin{figure}[h]
\centerline{\includegraphics[width=0.4\columnwidth]{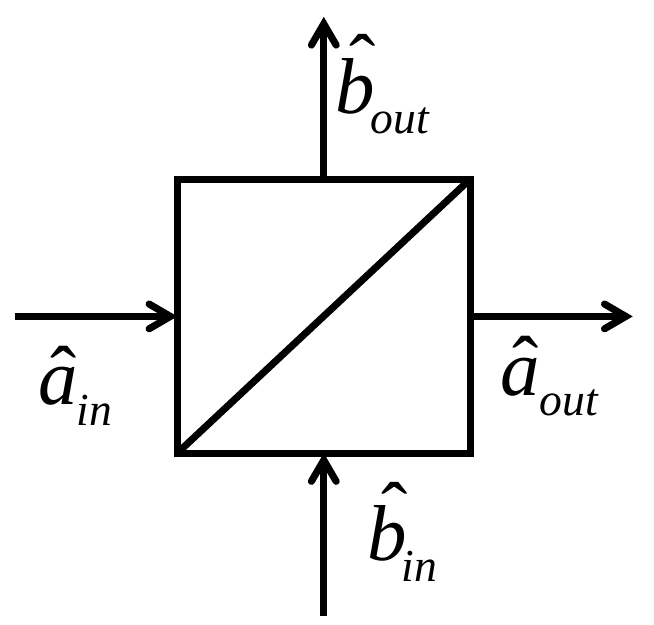}}
\caption{A beam splitter (BS) with input modes $\hat{a}_{in}, \hat{b}_{in}$ and output modes $\hat{a}_{out}, \hat{b}_{out}$.}
\end{figure}

\be{BS:in:out}
\left(
  \begin{array}{c}
    \ha_{out} \\
    \hb_{out} \\
  \end{array}
\right)=
\left(
     \begin{array}{cc}
       \tau & \rho \\
       -\rho^* & \tau^* \\
     \end{array}
\right)\,
\left(
  \begin{array}{c}
    \ha_{in} \\
    \hb_{in} \\
  \end{array}
\right), \; \hat{\vec{a}}_{out} = U_{bs}\, \hat{\vec{a}}_{in}.
 \ee
Typically, one often encounters $\tau$ real with $\rho = i\,\sqrt{1-\tau^2}$. The significance of the unitarity of $U_{bs}$ is that it preserves the commutation relations between fields from input to putput, so that $[\ha_{in},\ha^\dag_{in}]=1\Rightarrow [\ha_{out},\ha^\dag_{out}]=1$, and similarly for the $\hb$ mode. This is just the statement of conservation of probability, i.e. that all signals have been accounted for, and no parts of the signals have been lost.

\paragraph{Loss:}
To incorporate propagation (or scattering) loss in the system, one can use a model
\begin{figure}[h]
\centerline{\includegraphics[width=\columnwidth]{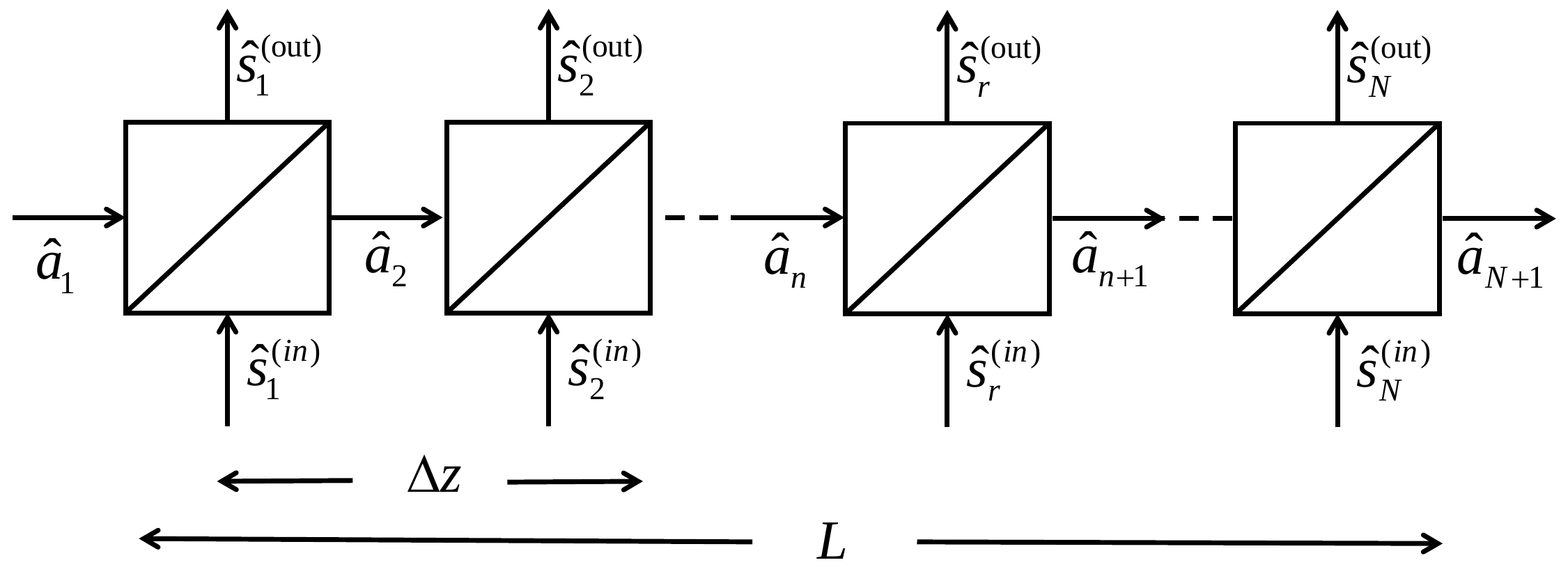}}
\caption{Loudon's propagation loss model based the continuum limit of a series of discrete beam splitters.}
\label{fig:Loudon:BSs}
\end{figure}
developed by Loudon \cite{Loudon:2000, Alsing_Hach:2016} where in the frequency domain
one has
\bsub
\bea{eqn:BSr}
\hat{a}_{r+1}(\omega)     &=&  T(\omega)\,\hat{a}_{r}(\omega) + R(\omega)\,\hat{s}^{(in)}_r(\omega), \label{eqn:BS:ar}\\
\hat{s}^{(out)}_r(\omega) &=&  R(\omega)\,\hat{a}_{r}(\omega) + T(\omega)\,\hat{s}^{(in)}_r(\omega),
\eea
\esub
as illustrated in \Fig{fig:Loudon:BSs}. The attenuated signal (of interest) $\ha_r$ and the scattering sites (unobserved, ``lost" modes) $\hs_r$ satisfy the usual boson commutation relations
$[\hat{a}_{r}(\omega), \hat{a}^\dagger_{r}(\omega')] =
[\hat{s}^{(in, out)}_r(\omega), \hat{s}^{\dagger(in, out)}_r(\omega')] = \delta(\omega-\omega')$.

Successive iteration of  \Eq{eqn:BS:ar} yields,
\be{eqn:ar:discrete}
\hat{a}_{N+1}(\omega) =  T^N(\omega)\,\hat{a}_{1}(\omega) + R(\omega)\,\sum_{r=1}^{N} T^{N-r}(\omega)\,\hat{s}^{(in)}_{r}(\omega).
\ee
In the limit of having an infinite series of beamsplitters with infinitesimal coupling, we obtain the relationship for how loss is treated in a continuous medium. We now take the continuum limit: $N\to\infty$; $\Delta z = L/N\to 0$; and $\sum_{r=1}^N\to (\Delta z)^{-1}\int_0^L dz$. Because an individual BS in this infinite series has infinitesimal coupling (i.e., $|R(\omega)|^2\to 0$), we define the independent attenuation constant $\Gamma(\omega) = |R(\omega)|^2 / \Delta z$.
Then, using $|T(\omega)|^2 + |R(\omega)|^2=1$ we have,
\be{eqn:TN}
|T(\omega)|^{2N} = (1-|R(\omega)|^2)^N = (1-\Gamma(\omega)L/N)^N \to e^{-\Gamma(\omega) L},
\ee
for which we define,
\bea{eqn:T}
T(\omega) &\equiv& e^{i\xi(\omega) \Delta z}= e^{i\,n(\omega) (\omega/c)-\frac{1}{2} \Gamma(\omega)\,\Delta z},\\
\xi(\omega) &\equiv& \beta(\omega) + i \Gamma(\omega)/2, \\
\beta(\omega) &\equiv& n(\omega) (\omega/c).
\eea
In \Eq{eqn:T} we have chosen the phase of $T(\omega)$ to incorporate the free propagation constant (i.e., wavenumber) $\beta(\omega) \equiv n(\omega) (\omega/c)$ through a
medium of index of refraction $n(\omega)$. In addition, we have defined the complex propagation constant as $\xi(\omega) \equiv \beta(\omega) + i \Gamma(\omega)/2$.

To complete our treatment of loss in a continuous medium, we use $(N-r)\Delta z = L-z$, and convert from discrete to continuous modes
%through the identification,
to obtain Loudon's expression for an attenuated traveling beam \cite{Loudon:2000}:
\be{aL:schneeloch}
\hat{a}_L(\omega) = e^{i\xi(\omega)L} \, \hat{a}_0(\omega) + i \sqrt{\Gamma(\omega)}\,\int_{0}^{L} dz \,e^{i\xi(\omega)(L-z)}\,\hat{s}(z,\omega).
\ee
For convenience, we have introduced the shorthand notation
for the input field at $z=0\,$ ($\ha_1$ in \Fig{fig:Loudon:BSs}),  as $\hat{a}_0(\omega)=\hat{a}(z=0,\omega)$ and
for the output field at $z=L\,$ as $\hat{a}_L(\omega)$.
%Note that since $\hat{s}(z,\omega)$ are internal noise operators, and $\hat{a}_0(\omega)$ is the input field before any interactions with the scattering centers, these operators commute, $[\hat{a}_0(\omega),\hat{s}(z',\omega')]= [\hat{a}_0(\omega),\hat{s}^\dagger(z',\omega')]= 0$.
An explicit computation \cite{Alsing_Hach:2016} shows that $[\hat{a}_L(\omega), \hat{a}^\dagger_L(\omega')]=\delta(\omega-\omega')$; %using the identity $i[\xi(\omega)-\xi^*(\omega')] = -\Gamma(\omega)$. 
the expression for the attenuated traveling wave $\hat{a}_L(\omega)$ 
%in \eqref{aL:schneeloch} 
explicitly preserves the output field commutation relations.

To connect our expressions to alternative treatments of lossy media, we can rewrite \Eq{aL:schneeloch} in a \textit{Langevin} form \cite{Walls_Milburn:1994,Scully_Zubairy:1997,Orszag:2000} as,
\bsub
\bea{aL:f:schneeloch}
\hat{a}_L(\omega) &=&   e^{i\xi(\omega)L} \, \hat{a}_0(\omega) + i \sqrt{1-e^{-\Gamma(\omega) L}}\, \hat{f}(\omega), \label{defn:aL:f}\\
\hat{f}(\omega) &\equiv& \sqrt{\frac{\Gamma(\omega)}{1-e^{-\Gamma(\omega) L}}} \,\int_{0}^{L} dz \, e^{i\xi(\omega)(L-z)}\,\hat{s}(z,\omega), \qquad \label{defn:f:s}
\eea
\esub
where the Langevin noise operators $\hat{f}(\omega)$ satisfy the commutation relations,
\be{}
[\hat{f}(\omega), \hat{f}^\dagger(\omega')] = \delta(\omega-\omega').
\ee

%Note, that in the absence of loss (i.e., $\Gamma=0$), \Eq{aL:f:schneeloch} reduces to the
%un-attenuated free propagating field expression $\hat{a}_L(\omega) =   e^{i\beta(\omega)L} \, \hat{a}_0(\omega)$, which is unitary
%since $|e^{i\beta(\omega)L}|=1$. 
One could deduce  \Eq{defn:aL:f} by phenomenologically introducing loss as
$\hat{a}_L(\omega) \sim   e^{[i\beta(\omega) -\Gamma(\omega)/2] L} \, \hat{a}_0(\omega)$, assuming that $\hat{a}_L(\omega)$ takes the form of $\hat{a}_L(\omega) = \mathcal{A}\,\hat{a}_0(\omega) + \mathcal{B}\,\hat{f}(\omega)$, 
%with $[\hat{f}(\omega), \hat{f}^\dagger(\omega')] = \delta(\omega-\omega')$,
and \textit{requiring} by quantum mechanics that
$[\hat{a}_L(\omega), \hat{a}_L^\dagger(\omega')] = \delta(\omega-\omega')$. 
%This implies that $|\mathcal{B}| = \sqrt{1-|\mathcal{A}|^2}$ with freedom to choose the phase of $\mathcal{B}$. 
This deduction is the essence of the Langevin approach, where the inclusion of loss requires the introduction of
additional noise operators $\hat{f}(\omega)$ to ensure that the quantum-mechanical commutation relations
are preserved.
This is also an embodiment of the \textit{fluctuation-dissipation theorem} \cite{Mandel_Wolf:1995}. What is not obtained from this procedure is the actual from
of $\hat{f}(\omega)$ as given by \Eq{defn:f:s}.
%The above derivation of $\hat{a}_L(\omega)$ by Loudon preserves the commutation relations
%$[\hat{a}_L(\omega), \hat{a}^\dagger_L(\omega')] = \delta(\omega-\omega')$ by explicit construction.

%=====================
Alternatively, one can treat loss in an optical medium as the Hamiltonian evolution of an extended quantum system. If we consider the total hamiltonian as the sum of the system hamiltonian $(\hat{H}_{sys}=\hat{H}_{L})$ (see \eqref{HamLin}), an environment hamiltonian of free photons $\hat{H}_{env}=\int_{-\infty}^{\infty} d\omega\,\hbar\,\omega\,\hat{e}^\dagger(\omega)\,\hat{e}(\omega)$, and a coupling interaction between the two; $\hat{H}_{int}= i\,\hbar\int_{-\infty}^{\infty} d\omega\, \kappa(\omega)\,\left(\hat{e}^\dagger(\omega)\, \hat{a}(\omega) - \hat{e}(\omega)\,\hat{a}^\dagger(\omega)\right)$ the Heisenberg equation of motion for this system in a reference frame rotating with respect to the central frequency of the light approximates to the Heisenberg-Langevin Equation \cite{Walls_Milburn:1994,Orszag:2000}:
\be{eqn:aindot:ain:b0}
\dot{\hat{a}}(t) = -\frac{i}{\hbar}\,[\hat{a}, \hat{H}_{sys}] -\frac{\g_a}{2}\,\hat{a}(t) + \sqrt{\g_a}\,\hat{f}_a(t) ,
\ee
where $\gamma_{a}=\Gamma_{a}c/n_{ga}$, is the attenuation constant in time. The solution to this equation also yields \Eq{aL:f:schneeloch}. Here, it is also understood that $\hat{a}(t)$ is the time evolution of a \emph{single mode} of the electromagnetic field $\hat{a}(\omega)$ in a lossy medium. In the lossless case (i.e., $\gamma_{a}=0$), the cavity mode evolves unitarily under the system hamiltonian $\hat{H}_{sys}$. When loss is present, the mode is damped by the operator loss term $-(\gamma_{a}/2)\; \hat{a}$, but the total evolution remains unitary; it is preserved due to the additional noise term $\sqrt{\gamma_{a}}\hat{f}$.

In the section where we specifically tackle the problem of SPDC in a lossy cavity, we use a Heisenberg-Langevin equation similar to \Eq{eqn:aindot:ain:b0}, but where $\hat{H}_{sys}$ includes both $\hat{H}_{L}$ and $\hat{H}_{NL}$. Moreover, we use a rotating frame of reference so that the total time derviative of the propagating mode, $\dot{\hat{a}}$, is given as $(\partial_{t} + (c/n_{g})\partial_{z})\hat{a}$. Once the equation of motion is solved to find $\hat{a}$ as a function of position in the MRR, this expression is incorporated into the interaction-picture hamiltonian to find the state of the down-converted light.

\paragraph{Cavities and MRR:}
We now use the above results to examine the output mode $\ha_{out}$ of a cavity subject to an input mode driving field $\ha_{in}$, with the internal cavity mode $\ha$. Without loss of generality, we take the cavity to be a Micro-Ring Resonator (MRR) as illustrated in Fig.~4, which also corresponds to a Fabry-Perot cavity with one input/output semitransparent mirror, and one fully reflecting mirror.

\begin{figure}[h]
\centerline{\includegraphics[width=0.6\columnwidth]{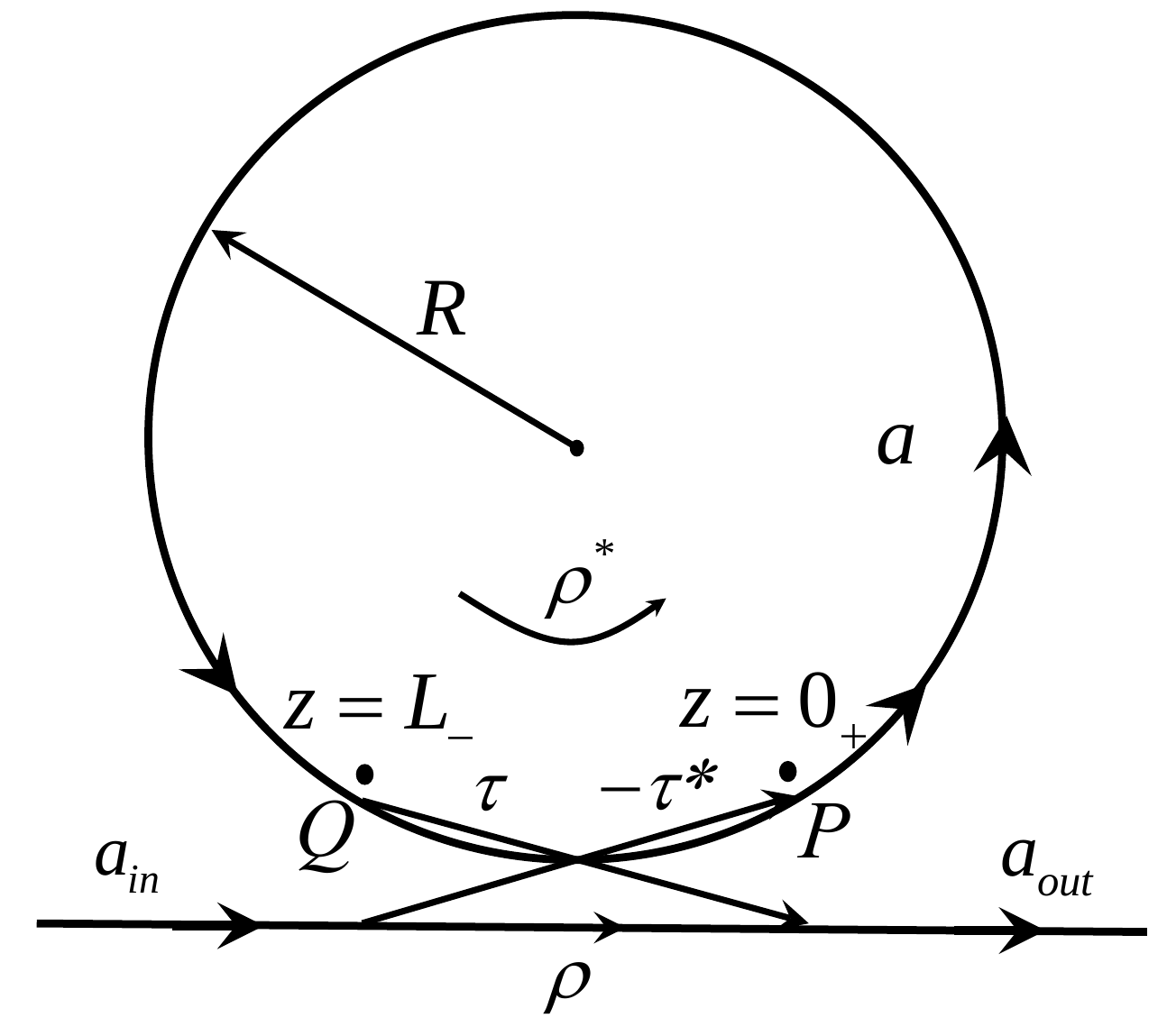}}
\caption{A single bus (all-through) micro-ring resonator (MRR) of length $L=2\pi R$ with cavity field $\hat{a}$, coupled to a waveguide bus with input field $\ha_{in}$ and output field $\ha_{out}$. The constants $\rho$ and $\tau$ are the self-coupling and cross-coupling coefficients, respectively, of the bus to the MRR. The value $z=0_+$ is the point $P$ just inside the MRR that cross-couples to the input field $\ha_{in}$, and $z=L_-$ is the point $Q$ after one round trip in the MRR that cross-couples to the output field $\ha_{out}$.}
\end{figure}

%In Fig.~4, the parameters $\rho,\,\tau$ are the beam-splitter-like self-coupling (reflection) and cross-coupling (transmission) coefficients, respectively, of the bus to the MRR such that $|\tau|^2 + |\rho|^2 =1$. $z=0_{+}$ is the point $P$ just inside the MRR that cross-couples to the input field $a_{in}$, and $z=L_{-}$ is the point $Q$ after one round trip in the MRR that cross-couples to the output field $a_{out}$.

In analogy with a classical field derivation \cite{Alsing_Hach:2016}, the output mode $\ha_{out}$ is a function of the sum over all possible trajectories from the input mode $\ha_{in}$, as it makes an arbitrary number (including zero) of circulations around the cavity:
\bsub
\bea{pma:deriv}
\ha_{out} = &\rho&\,\ha_{in} \label{pma:deriv:1} \\
% 0
&+& \!\big(-\ts\big)_{a_{in}\to a_{0}}\big(\hat{a}_{L}\big)_{a_{0}\to a_{L}}\big(\t\big)_{a_{L}\to a_{out}}  \label{pma:deriv:2} \\
% (rho^*)^0
&+& \!\big(-\ts\big)_{a_{in}\to a_{0}}\big(\rs \hat{a}_{2L}\big)_{a_{0}\to a_{2L}}\big(\t\big)_{a_{2L}\to a_{out}}  \label{pma:deriv:3} \\
% (rho^*)^2
&+& \!\big(-\ts\big)_{a_{in}\to a_{0}}\big((\rs)^{2}\hat{a}_{3L}\big)_{a_{0}\to a_{3L}}\big(\t\big)_{a_{3L}\to a_{out}}\label{pma:deriv:4} \\
&+& \ldots,\no
= &\rho&\,\ha_{in} - |\t|^2\,\sum_{n=0}^\infty (\rs)^n\,\ha_{(n+1)L}, \label{pma:deriv:5} \\
= &\Big(& \rho - \alpha\,e^{i\theta}\,|\t|^2\,\sum_{n=0}^\infty (\rs\alpha\,e^{i\theta})^n\Big)\,\ha_{in} \no
&{}&  -i |\t|^2\,\sqrt{\Gamma}\sum_{n=0}^\infty (\rs)^n\,\Int{(n+1)},\qquad\;   \label{pma:deriv:6} \\
=
&\bigg(&
\frac{\rho-\alpha\,e^{i\theta}}{1-\rs\,\alpha\,e^{i\theta}}
\bigg)\,\ha_{in} \no
&{}&  -i |\t|^2\,\sqrt{\Gamma}\sum_{n=0}^\infty (\rs)^n\,\Int{(n+1)}. \label{pma:deriv:7}
\eea
\esub
First, the output photon can arrive directly from the input bus by ``reflection'' off the MRR (as described in \Eq{pma:deriv:1}). Next, (as written diagramatically in \Eq{pma:deriv:2}), the photon can couple into the MRR, acquiring factor $-\ts$, evolve through one circulation (circumference $L$ of the resonator) as described in \Eq{aL:schneeloch}, and couple out of the resonator acquiring factor $\t$. Successive paths involve multiple circulations within the resonator, acquiring additional factors of $\rs$ from self-coupling (i.e., ``reflection'') after each circulation. To simplify notation, we have used the definition $e^{i\xi L}  \equiv \alpha\,e^{i\theta}$ defining  $\alpha =e^{-\frac{1}{2}\Gamma L}$ to be the internal loss factor in one circulation of the resonator, and $\theta \equiv \beta L$ to be the phase gained in free propagation over the same distance.

As derived in \cite{Alsing_Hach:2016}, an explicit calculation of the output field commutation relation yields,
\be{c:comm:pma}
[\ha_{out}(\om),\ha_{out}^\dagger(\om')]=\delta(\om-\om').
\ee
This preservation of unitarity allows us to write
\bsub
\bea{aout:AH:v3}
\ha_{out}(\om) &=& G_{out,in}(\om)\,\ha_{in} + H_{out,in}(\om)\,\hf_a(\om),\qquad \\
|H_{out,in}(\om)| &=& \sqrt{1 - |G_{out,in}(\om)|^2},
\eea
\esub
where $G_{out,in}(\omega)$ is the coefficient preceding $\hat{a}_{in}$ in \Eq{pma:deriv:7}, whose magnitude is always less than or equal to unity.  This defines the Langevin quantum noise operator $\hf_a(\om)$ from the unitary requirement of the preservation of the free field output commutator. Interestingly, $G_{out,in}(\omega)$ is identical in form to the classical transmission coefficient \cite{Yariv:2000}, as would be expected. It is important to note that in treating loss in a MRR, we implicitly assumed the medium is istropic. However, as shown in \cite{Alsing_Hach:2016} this assumption can be relaxed and the commutation relations \Eq{c:comm:pma} still hold for multiple, piecewise defined propagation wavevectors and losses along the ring resonator of circumference $L$.

\subsubsection{Biphoton generation within the MRR}\label{subsec:preliminaries_biphoton:generation}
For biphoton generation arising from either the $\chi^{(2)}$ process of Spontaneous Parametric Down-Conversion (SPDC), or the $\chi^{(3)}$ process of Spontaneous Four-Wave Mixing (SFWM), Alsing and Hach \cite{Alsing_Hach:2017a} consider a signal mode $\ha$, and an idler mode $\hb$ circulating within the MRR, and here, we do the same.

In the non-depleted pump approximation, one can arrive at the hamiltonian:
\begin{align}
\hat{H}_{NL}=&\int dz d\omega_{1}d\omega_{2}\; g(\omega_{p}\omega_{1}\omega_{2})e^{-i\Delta k_{z} z}e^{i\Delta\omega t}\times\nn\\
&\times\Big(\alpha(z,\omega_{p})\hat{a}^{\dagger}(z,\omega_{1})\hat{b}^{\dagger}(z,\omega_{2})\Big) + h.c.
\end{align}
where for SPDC:
\begin{equation}
g_{spdc}=-i\Big(\frac{\chi_{eff}^{(2)}}{4\pi c \sqrt{L}}\Big)\sqrt{\frac{n_{g1}n_{g2}}{n_{1}^{2}n_{2}^{2}n_{p}^{2}}}\sqrt{\frac{(\hbar\omega_{p})^{3}}{2\epsilon_{0}}}\Phi_{xy}^{SPDC}
\end{equation}
and for SFWM\footnote{The expression for $g_{sfwm}$ uses the additional (though common) assumption of a $\chi^{(3)}$-nonlinear medium with no $\chi^{(2)}$ nonlinearity, such as any material with a centro-symmetric structure (e.g., amorphous solids, liquids, gases, and any crystal whose unit cell is indentical under reflection). Under this assumption, we have the approximation: $\chi^{(3)}_{eff}\approx -\epsilon_{0}^{3}n_{p}^{4}n_{1}^{2}n_{2}^{2}\zeta^{(3)}_{eff}$.}:
\begin{equation}
g_{sfwm}=-\Big(\frac{3 \chi_{eff}^{(3)}}{4\pi c  L}\Big)\sqrt{\frac{n_{g1}n_{g2}}{n_{1}^{2}n_{2}^{2}n_{p}^{4}}}\Big(\frac{(\hbar\omega_{p})^{2}}{\epsilon_{0}}\Big)\Phi_{xy}^{SFWM}.
\end{equation}
In more accurate treatments of SPDC and SFWM in a MRR, the mode functions $g_{\vec{\mu}}(\vec{r})$ would be calculated given the geometry of the material, and how the index of refraction varies spatially (e.g., step-index vs graded index). Here, we define the spatial overlap integrals $\Phi_{xy}^{SPDC}$ and $\Phi_{xy}^{SFWM}$ as:
\begin{align}
\Phi_{xy}^{SPDC}&=\int \!\!dx dy \;g_{\vec{\mu_p}}^{*}(x,y)g_{\vec{\mu_1}}(x,y)g_{\vec{\mu_2}}(x,y)\\
\Phi_{xy}^{SFWM}&=\int \!\!dx dy \;g_{\vec{\mu_{p1}}}^{*}(x,y) g_{\vec{\mu_{p2}}}^{*}(x,y)g_{\vec{\mu_1}}(x,y)g_{\vec{\mu_2}}(x,y)
\end{align}
In the ``particle-in-a-box'' mode basis, the coupling constants are given by:
\begin{equation}
g_{spdc}=-i\Big(\frac{32 \chi_{eff}^{(2)}}{9\pi^{3}c }\Big)\sqrt{\frac{n_{g1}n_{g2}}{n_{1}^{2}n_{2}^{2}n_{p}^{2}}}\sqrt{\frac{(\hbar\omega_{p})^{3}}{2\epsilon_{0}V_{ring}}}
\end{equation}
and
\begin{equation}
g_{sfwm}=-\Big(\frac{27 \chi_{eff}^{(3)}}{16\pi c}\Big)\sqrt{\frac{n_{g1}n_{g2}}{n_{1}^{2}n_{2}^{2}n_{p}^{4}}}\Big(\frac{(\hbar\omega_{p})^{2}}{\epsilon_{0}V_{ring}}\Big).
\end{equation}

In order to obtain this approximate hamiltonian, we have used the lowest-order plane-wave cavity modes (i.e., ``particle-in-a-box'' modes) instead of the hermite-gaussian modes to describe $g_{\vec{\mu}}(x,y)$, and integrated over both transverse dimensions. We let $L=2\pi R$, the circumference of the ring, essentially treating the ring as a conformal mapping of a rectangular nonlinear waveguide\footnote{For a treatment of photon-pair generation in a MRR that does not rely on this conformal approximation, see \cite{Camacho:2012}.}. With this, we also let $V_{ring}\equiv L_{x}L_{y}L$ using the dimensions of the deformed rectangular medium. Furthermore, where the pump is undepleted and in a coherent state, we have replaced the pump annihilation operator $\hat{a}_{p}$ with its corresponding coherent state amplitude $\alpha_{p}$. We have taken the same steps used before to express the hamiltonian as an integral over frequency, and we make the approximation that $\sqrt{\omega_{1}\omega_{2}}$ is approximately equal to the corresponding square root product of their central values. In SFWM, we let $\alpha(\omega_{p},z)$ represent the \emph{square} of the pump coherent state amplitude. For the rest of this section, we will focus on SPDC, but it is instructive to be aware that besides issues related to different phase matching, dependence on pump intensity, and the much smaller value of $\chi^{(3)}_{eff}$ relative to $\chi^{(2)}_{eff}$, the physics of photon pair generation in a cavity is very similar for both SPDC and SFWM.

For further simplification, and to arrive at the essential aspects of SPDC in a MRR, we first use the simplifying approximation of near-perfect phase matching, so that $e^{-i\Delta k_{z}z}\approx 1$.  Next, we  use the approximation of interaction times long enough to enforce energy conservation so that $e^{i\Delta \omega t}\rightarrow (\sqrt{2\pi}/T_{DC})\delta(\Delta\omega)$. This judicious substitution allows us to abbreviate the calculations done to calculate the biphoton rate in first-order perturbation theory as in previous sections. The interaction time, $T_{DC}$ is the round-trip time of light at the signal/idler frequencies. With these substitutions, the hamiltonian simplifies to:
\begin{align}
\hat{H}_{NL}=&\int dz \;d\omega_{1}\;\frac{\sqrt{2\pi}}{T_{DC}}g(\omega_{p},\omega_{1},\Omega_{p}-\omega_{1})\times\\
&\times\Big(\alpha(z,\Omega_{p})\hat{a}^{\dagger}(z,\omega_{1})\hat{b}^{\dagger}(z,\Omega_{p}-\omega_{1})\Big) + h.c.\nn,
\end{align}
where $\Omega_p = \om_p\,(\text{alt.~}2\,\om_p)$ for SPDC (alt.~SFWM) such that the signal frequency is at $\Omega_p/2 + \nu$ and the idler frequency is at $\Omega_p/2 - \nu$. Note that for later convenience, we define
$(\hbar \,g)\equiv(\sqrt{2\,\pi}/T_{DC})\,g_{spdc (sfwm)}$.To simplify the hamiltonian even further, we shift to a reference frame rotating at the central frequency $\Omega_p/2$. Then, in the following, the frequency $\nu$ represents an offset from $\Omega_p/2$, so that $\ha(\Omega_p/2+\nu)\rightarrow\ha(\nu)$ and $\hb(\Omega_p/2-\nu)\rightarrow\hb(-\nu)$. We will further use the common quantum-optical shorthand notation $\hb^\dag(\nu)\equiv [\hb(-\nu)]^\dag$ \cite{Orszag:2000}.
Thus, in the non-depleted pump approximation, we obtain the hamiltonian:
\bea{H:NL}
 \!\!\!\!\!\!\!\!\!\!\hat{H}_{NL} &=&  \!\int\!dz\,d\nu\,\hbar\,g\!
 \left(\!\alpha_p\, \ha^\dag(z,\nu)\,\hb^\dag(z,\nu) \!\right) + h.c.\,, \quad
\eea
%%\esub
%where  $\alpha_p$ is the complex c-number (constant) amplitude (squared amplitude) for the pump for spdc (sfwm), and the transverse dimensions have been integrated over, giving a Hamiltonian density as a function of $z$, the distance traveled in the resonator.
where we will take $\alpha_{p}\equiv\alpha_p(z,\Omega_p/2) =$ constant throughout the MRR.

As was discussed previously, the signal and idler modes satisfy the Heisenberg-Langevin equation of motion in the frequency domain (using $\partial_t\,\ha(t) = -i\,\nu\,\ha(t)$) \cite{Raymer:2013,Alsing_Hach:2017a,Alsing_Hach:2017b}, where this time, $\hat{H}_{NL}$ is included in $\hat{H}_{sys}$. In the rotating reference frame, the equations for the signal and idler modes are given by:
\bsub
\bea{a:b:EoM}
\Big(-i\,\nu + \frac{c}{n_{ga}}\,\partial_z\Big)\, \ha(z,\nu) &=&  -i\,g\,L\,\alpha_p(z,\Omega_{p}/2)\,\hb^\dag(z,\nu) \no
-\;\frac{\g'_a}{2}\,\ha(z,\nu) &+& \alpha_{polz}\,\hf_a(z,\nu),\;\qquad \label{a:EoM}\\
\Big(-i\,\nu + \frac{c}{n_{gb}}\,\partial_z\Big)\, \hb^\dag(z,\nu) &=&   i\,g\,L\,\alpha^*_p(z,\Omega_{p}/2)\,\ha(z,\nu) \no
-\;\frac{\g'_b}{2}\,\hb^\dag(z,\nu) &+& \alpha_{polz}\,\hf_b^{\dagger}(z,\nu). \;\;\qquad\label{b:EoM}
\eea
\esub
where $\g'_k$  is the internal propagation loss for mode $\kinab$, and $\hat{f}_k$ are corresponding Langevin noise operators added to preserve the canonical form of the output commutators. The constant $\alpha_{polz}$ is a Langevin coupling constant to the scattered modes required to preserve the unitary evolution of the fields in the lossy MRR.

By expressing the relations between the input, cavity, and output fields in terms of matrices, we greatly simplify the subsequent algebra used to find the state of the output fields. In particular, the input-output boundary condtions are given by:
\bsub
\bea{BCs:a:b}
\!\!\!\!\!\left(\!\!
  \begin{array}{c}
    \ha_{0_+} \\
    \hb_{0_{+}}^{\dagger} \\
  \end{array}
\!\!\right) \!&=&\!
\left(\!\!
     \begin{array}{cc}
       -\tau_a^*\!\! & 0 \\
       0 &\!\! -\tau_b \\
     \end{array}
\!\!\right)\!
\left(\!\!
  \begin{array}{c}
    \ha_{in} \\
    \hb_{in}^{\dagger}\\
  \end{array}
\!\!\right)\!+\!\left(\!\!
     \begin{array}{cc}
       \rho_a^* & 0 \\
       0 & \rho_b \\
     \end{array}
\!\!\right)\!
\left(\!\!
  \begin{array}{c}
    \ha_{L_{-}} \\
    \hb_{L_{-}}^{\dagger}\\
  \end{array}
\!\!\right)_,  \label{BCs:a}\\
\left(\!\!
  \begin{array}{c}
    \ha_{out} \\
    \hb_{out}^{\dagger} \\
  \end{array}
\!\!\right)\! &=&\!
\left(\!\!
     \begin{array}{cc}
       \tau_a & 0 \\
       0 & \tau_b^{*} \\
     \end{array}
\!\!\right)\!
\left(\!\!
  \begin{array}{c}
    \ha_{L_{-}} \\
    \hb_{L_{-}}^{\dagger}\\
  \end{array}
\!\!\right)\!+\!\left(\!\!
     \begin{array}{cc}
       \rho_a & 0 \\
       0 & \rho_b^{*} \\
     \end{array}
\!\!\right)\!
\left(\!\!
  \begin{array}{c}
    \ha_{in} \\
    \hb_{in}^{\dagger}\\
  \end{array}
\!\!\right)_,\label{BCs:b}
\eea
\esub
Defining vector and matrix notation implicitly, the boundary conditions \eqref{BCs:a}\eqref{BCs:b} may be written in simplified form:
\begin{subequations}
\begin{eqnarray}
\vec{\hat{a}}_{0_{+}}&=-\mathbf{X}\cdot \vec{\hat{a}}_{in} + \mathbf{T}\cdot \vec{\hat{a}}_{L_{-}}\\
\vec{\hat{a}}_{out}&=\;\mathbf{T}^{*}\cdot \vec{\hat{a}}_{in} \;+\; \mathbf{X}^{*}\cdot \vec{\hat{a}}_{L_{-}}.
\end{eqnarray}
\end{subequations}

The equations \Eq{a:EoM} and \Eq{b:EoM} in matrix notation are given by:
\begin{align}
\partial_{z}\vec{\hat{a}}(z,\nu)=\mathbf{M}\cdot \vec{\hat{a}}(z,\nu) +\frac{n_{g}\alpha_{polz}}{c}\vec{\hat{f}}(z,\nu)
\end{align}
where
\begin{align}
\mathbf{M}\equiv \!\left(\!\!
     \begin{array}{cc}
       i\fracd{n_{g}\nu}{c}-\fracd{\Gamma_{a}}{2} & - \fracd{n_{g}}{c}|g| L |\alpha_{p}|e^{i\theta_{p}} \\
        - \fracd{n_{g}}{c}|g| L |\alpha_{p}|e^{-i\theta_{p}} &  i\fracd{n_{g}\nu}{c}-\fracd{\Gamma_{b}}{2} \\
     \end{array}
\!\!\right).
\end{align}
The solutions of \Eq{a:EoM} and \Eq{b:EoM} are then:
\begin{equation}\label{PropSolution}
\vec{\hat{a}}_{L_-}=e^{\mathbf{M}L}\cdot \vec{\hat{a}}_{0_{+}} +\frac{n_{g}\alpha_{polz}}{c}\int_{0}^{L}\!\!dz\; e^{\mathbf{M}(L-z)}\cdot \vec{\hat{f}}(z).
\end{equation}
Although the solution requires taking the matrix exponential, we use the approximation of equal loss ($\Gamma_{a}=\Gamma_{b}=\Gamma$), and equal group index for signal and idler (as in type-I SPDC) to obtain the solution:
\begin{align}
\left(\!\!
  \begin{array}{c}
    \ha_{L_{-}} \\
    \hb_{L_{-}}^{\dagger} \\
  \end{array}
\!\!\right)\! &\approx\!\alpha e^{i\theta}
\left(\!\!
     \begin{array}{cc}
        \text{cosh}(r) \!&\! - e^{i\theta_{p}}\text{sinh}(r) \\
       - e^{-i\theta_{p}}\text{sinh}(r) \!&\! \text{cosh}(r) \\
     \end{array}
\!\!\right)\!\!
\left(\!\!
  \begin{array}{c}
    \ha_{0_{+}} \\
    \hb_{0_{+}}^{\dagger}\\
  \end{array}
\!\!\!\right)+\nn\\
&\;\;\;\;+\left(\!\!
     \begin{array}{cc}
       B_{11} \!\!&\!\! \!\!0 \\
       \!\!0 \!&\! B_{22} \\
     \end{array}
\!\!\right)\!
\!
\left(\!\!
  \begin{array}{c}
    \hf_{a} \\
    \hf_{b}^{\dagger}\\
  \end{array}
\!\!\right)_,
\end{align}
or in vector notation:
\begin{equation}
\vec{\hat{a}}_{L_-}=\mathbf{R}\cdot \vec{\hat{a}}_{0_+} + \mathbf{B}\cdot \vec{\hat{f}},
\end{equation}
where $\mathbf{R}$ and $\mathbf{B}$ are defined implicitly. The coefficients $B_{11}=\sqrt{1 - \alpha^{2}}$, and $B_{22}=\sqrt{1-\alpha^{2}}$, where $\alpha = e^{-\Gamma L/2}$. These coefficients are determined by requiring preservation of the commutation relations $[\hat{a}_{L-},\hat{a}_{L-}^{\dagger}]=[\hat{b}_{L-},\hat{b}_{L-}^{\dagger}]=[\hat{a}_{0+},\hat{a}_{0+}^{\dagger}]=[\hat{b}_{0+},\hat{b}_{0+}^{\dagger}]$. Here, we have used the notation: $r=|g| L |\alpha_{p}|T_{DC}$, and $\alpha_{p}=|\alpha_{p}|e^{i\theta_{p}}$, and $\theta_{p}=(1/2)\Omega_{p} T_{DC}$, and $\Gamma = \gamma n_{g}/c$.  It is interesting to point out, that in the limit of zero loss, the squeezing transformation is essentially identical to that derived in the previous section, though now expressed in terms of length instead of time. 

In addition, we can consider many circulations within the resonator to examine the net relationship between gain and loss. While $\Gamma/2$ represents the amplitude loss per unit length in the resonator, the quantity $r/L$ represents the amplitude gain per unit length due to SPDC. Incorporating out-coupling loss $|\rho|^{2}$ into the total loss per round trip, we find that in order to have a net exponential gain of SPDC light, the pump intensity must be high enough that $r$ exceeds the threshold:
\begin{equation}\label{OPO}
r_{thresh}\geq \frac{\Gamma L}{2} + \ln\Big(\frac{1}{|\rho|}\Big)
\end{equation}
This is also known as the threshold for optical parametric oscillation, where intensities of down-converted light may be bright enough to be comparable to the pump. When using SPDC as a source of heralded single photons, we operate well below this threshold, because multi-biphoton events would overwhelm the photon pair statistics at such high intensities. For typical MRR parameters, $r_{thresh}$ corresponds to input pump powers of the order 1-10 milliwatts, though higher-Q resonators will lower this threshold further. These approximations are liberal and numerous, as an accurate result requires knowing what the actual spatial modes of the waveguide are, what the effective index of refraction of the propagating spatial modes are, and how much of the pump power in the MRR is in the lowest order spatial mode. In particular, this is important because an MRR that is single-mode at the down-converted wavelength will be multi-mode at the pump wavelength. Due to conservation of momentum, only the lowest-order pump mode in such an MRR can drive photon pair generation if it is single-mode at the down-converted wavelength.

\subsubsection{The output two-photon signal-idler state}\label{sec:2photon_state:outside}
To obtain the state of the down-converted light outside the resonator, we may use the matrix expressions in \Eq{BCs:a} and \Eq{BCs:b},  to express the output fields $\vec{\hat{a}}_{0_{+}}$ in terms of $\vec{\hat{a}}_{out}$ and $\vec{\hat{f}}$. To do this, we can express the output field operator as a sum over the possible number of circulations in the MRR, as was done  previously in relating $\vec{\hat{a}}_{out}$ to $\vec{\hat{a}}_{in}$. In this case, there are no photons in the input field at the frequency of the down-converted light, as the down-converted light is being generated within the MRR. When relating $\vec{\hat{a}}_{0_{+}}$ to $\vec{\hat{a}}_{out}$, we find:
\begin{equation}\label{aOutInSPDC}
\vec{\hat{a}}_{0_{+}}(\nu) = \mathbf{D}(\nu)\cdot\vec{\hat{a}}_{out}(\nu) + \mathbf{J}(\nu)\cdot\vec{\hat{f}}(\nu)
\end{equation}
where
\begin{subequations}
\begin{eqnarray}
\mathbf{D}&=&\Big(\mathbf{R} -\mathbf{T}^{*}\Big)^{-1}\mathbf{X}\\
\mathbf{J}&=&-\Big(\mathbf{R} -\mathbf{T}^{*}\Big)^{-1}\mathbf{B}
\end{eqnarray}
\end{subequations}
Assuming the parameters are the same for $a$ and $b$, these matrices have relatively simple expressions:
\begin{align}\label{OutInRelations}
\mathbf{D}&=\tau\left(
  \begin{array}{cc}
    \fracd{(\text{Cosh}(r) (\alpha e^{i\theta}) - \rho)}{\mathcal{D}} & \fracd{e^{i\theta_{p}}(\alpha e^{i\theta})\text{Sinh}(r)}{\mathcal{D}}  \\
   \fracd{e^{-i \theta_{p}}(\alpha e^{i\theta})\text{Sinh}(r)}{\mathcal{D}}  & \fracd{(\text{Cosh}(r) (\alpha e^{i\theta}) - \rho)}{\mathcal{D}} \\
  \end{array}
\right)
\\
\mathbf{J}&=-\;\fracd{\sqrt{1-\alpha^{2}}}{\tau}\;\mathbf{D}
\end{align}
where 
\begin{equation}
\mathcal{D}\equiv (\alpha e^{i\theta})^{2} + \rho^{2} -2(\alpha e^{i\theta})\rho\; \text{Cosh}(r),
\end{equation}
and the dependence on $\nu$ is given by $\theta=\nu T_{DC}$. 

For later convenience, we also define the notation:
\begin{subequations}
\begin{eqnarray}
\mathbf{D}(\nu)&=\left(\begin{array}{cc}
    D_{aa}(\nu) & D_{ab}(\nu) \\
    D_{ba}(\nu) &  D_{bb}(\nu) \\
  \end{array}\right),\\
  \mathbf{J}(\nu)&=\left(\begin{array}{cc}
    J_{aa}(\nu) & J_{ab}(\nu) \\
    J_{ba}(\nu) &  J_{bb}(\nu) \\
  \end{array}\right).
\end{eqnarray}
\end{subequations}

For weak, but classically bright pump fields, the state of the down-converted fields is well approximated to first order in $\hat{H}_{NL}$, and given by:
\begin{widetext}
 \bsub
 \bea{H_SPDC:state:ab}
 \ket{\Psi(T_{DC})}_{ab} &=& e^{-i/\hbar\,\hat{H}^{NL}\,T_{DC}}\, \ket{\Psi}_{in} \approx \left(1 -\frac{i}{\hbar}\,\hat{H}^{NL}\,T_{DC}\right)\ket{\vac} \\
 &=&
 \left[1 - \int_{-\infty}^{\infty}\,d\nu\,|g|T_{DC}\int_{0}^{L}\,dz\,|\alpha_{p}|\left( \eithp\,\hat{a}^\dag(z,\nu)\,\hat{b}^\dag(z,\nu) + \emithp\,\hat{a}(z,\nu)\,\hat{b}(z,\nu) \right)\right]\ket{\vac} \\
 &\approx&
 \left[1 - \int_{-\infty}^{\infty}\,d\nu\,r_{ab}(\nu) \,\left( \eithp\,\hat{a}^\dag(0_+,\nu)\,\hat{b}^\dag(0_+,\nu) + \emithp\,\hat{a}(0_+,\nu)\,\hat{b}(0_+,\nu) \right)\right]\ket{\vac}, \qquad\quad 
 \label{H_SPDC:state:ab:integral}
 \eea
 \esub
 \end{widetext}
 where $\hat{H}^{NL}$ is given in \Eq{H:NL}, and for simplicity: $r_{ab}(\nu) \equiv |g|\,|\alpha_p|\,L\,T_{DC}$. Note that in this estimation of the quantum state, we use the interaction picture, where the state evolves according to $\hat{H}_{NL}$, while the creation and annihilation operators evolve according to $\hat{H}_{L}$ and the interaction hamiltonian accounting for loss. In this case, we treat the evolution of the operators as in the Heisenberg-Langevin equation \eqref{PropSolution}, but without the contribution of $\hat{H}_{NL}$, effectively setting $r=0$. In this picture, we can relax the assumption of near-perfect phase matching, so that $r_{ab}(\nu)$ acquires an additional factor of $Sinc(\Delta k_{z} L/2)$ after integrating over $z$. The integration over $z$ is approximated under the assumption that the damping over $z$ is slow enough that the exponential damping can be approximated to first order (i.e., linearly) from $0$ to $L_{-}$. For $r=0$, the output relation matrices $\mathbf{D}$ and $\mathbf{J}$ are greatly simplified to:
\begin{subequations}
\begin{eqnarray}
\mathbf{D}(\nu,r=0)&=\fracd{\tau}{\alpha e^{i\theta} - \rho}\left(\begin{array}{cc}
    1 & 0 \\
    0 &  1 \\
  \end{array}\right),\\
  \mathbf{J}(\nu,r=0)&=\fracd{-\sqrt{1-\alpha^{2}}}{\alpha e^{i\theta} - \rho}\left(\begin{array}{cc}
    1 & 0 \\
    0 & 1 \\
  \end{array}\right).
\end{eqnarray}
\end{subequations}
Although these matrices blow up in the limit of critical coupling (i.e., $\rho\rightarrow\alpha$) where $r=0$, first-order perturbation theory is no longer accurate in such regimes. When calculating expectation values using the solutions to the Heisenberg-Langevin equation (where $r>0$) to get a more accurate estimate, the number of generated biphotons exiting the resonator is maximum at critical coupling, but finite.

Now that the state of the SPDC light inside the resonator has a straightforward form, the output state $\ket{\Psi}_{out}$ is obtained from the internal state $\ket{\Psi(T_{ab})}_{ab}$ as the Heisenberg operators $\vec{\hat{a}}_{0_{+}}$ evolve through the resonator and couple out, becoming $\mathbf{X}^{*}\mathbf{R}\;\vec{\hat{a}}_{0_{+}}$. The scattered light given by creation operator $\vec{\hat{f}}$ has exited the system, and does not enter into the Heisenberg propagation of the down-converted light from inside to outside the resonator. Then, using our expression for $\vec{\hat{a}}_{0_{+}}$ in terms of $\vec{\hat{a}}_{out}$ and $\vec{\hat{f}}$. The output state of the fields has a straightforward expression (with $\nu$ argument suppressed to save space):

\bea{Psi:out:SPDC}
 \!\!\!\!\!\!\ket{\Psi}_{out} &=& |\text{vac}\rangle - \int_{-\infty}^{\infty} d\nu \;\alpha_{a}^{*}\alpha_{b}^{*}\tau_{a}^{*}\tau_{b}^{*}r_{ab}\;\text{Sinc}\Big(\frac{\Delta k_{z}L}{2}\Big)\times\nn\\
 &{}&\times\,\Big[\big( e^{i\theta_{p}}D^{*}_{aa}D_{bb} + e^{-i\theta_{p}}D_{ab}D_{ba}^{*}\big)\hat{a}_{out}^{\dagger}\hat{b}_{out}^{\dagger}\nn\\
 &{}&\;\; +\big( e^{i\theta_{p}}D^{*}_{aa}J_{bb} + e^{-i\theta_{p}}D_{ba}^{*}J_{ab}\big)\hat{a}_{out}^{\dagger}\hat{f}_{b}^{\dagger}\nn\\
 &{}&\;\; +\big( e^{i\theta_{p}}D_{bb}J^{*}_{aa} + e^{-i\theta_{p}}D_{ab}J^{*}_{ba}\big)\hat{f}_{a}^{\dagger}\hat{b}_{out}^{\dagger}\nn\\
 &{}&\;\; +\big( e^{i\theta_{p}}J^{*}_{aa}J_{bb} + e^{-i\theta_{p}}J_{ab}J^{*}_{ba}\big)\hat{f}_{a}^{\dagger}\hat{f}_{b}^{\dagger}\Big]|\text{vac}\rangle,\nn\\
\eea
where annihilation operators acting on the vacuum state yield a null result. In the previous matrix expressions, we let $\tau_{a}=\tau_{b}=\tau$, and let $\tau$ be real to simplify notation. Interestingly, the phase of $\tau$ can be incorporated as a contribution to the phase $e^{i\theta_{p}}$ because although  the previous expression contains terms associated to $e^{i\theta_{p}}$ and $e^{-i\theta_{p}}$, closer examination of the coefficients associated to these terms reveals a global phase dependence of $e^{i\theta_{p}}$.

With the state of the down-converted light exiting the  resonator $|\Psi\rangle_{out}$ known, we see that it is readily decomposed into four elements. The amplitude for biphoton production precedes $\hat{a}_{out}^{\dagger}\hat{b}_{out}^{\dagger}$, while the amplitude for a signal photon with scattered idler precedes $\hat{a}_{out}^{\dagger}\hat{f}_{b}^{\dagger}$. The corresponding amplutides for scattered idlers, and both scattered photons are staightforward as well.

\begin{figure}[t]
\centerline{\includegraphics[width=\columnwidth]{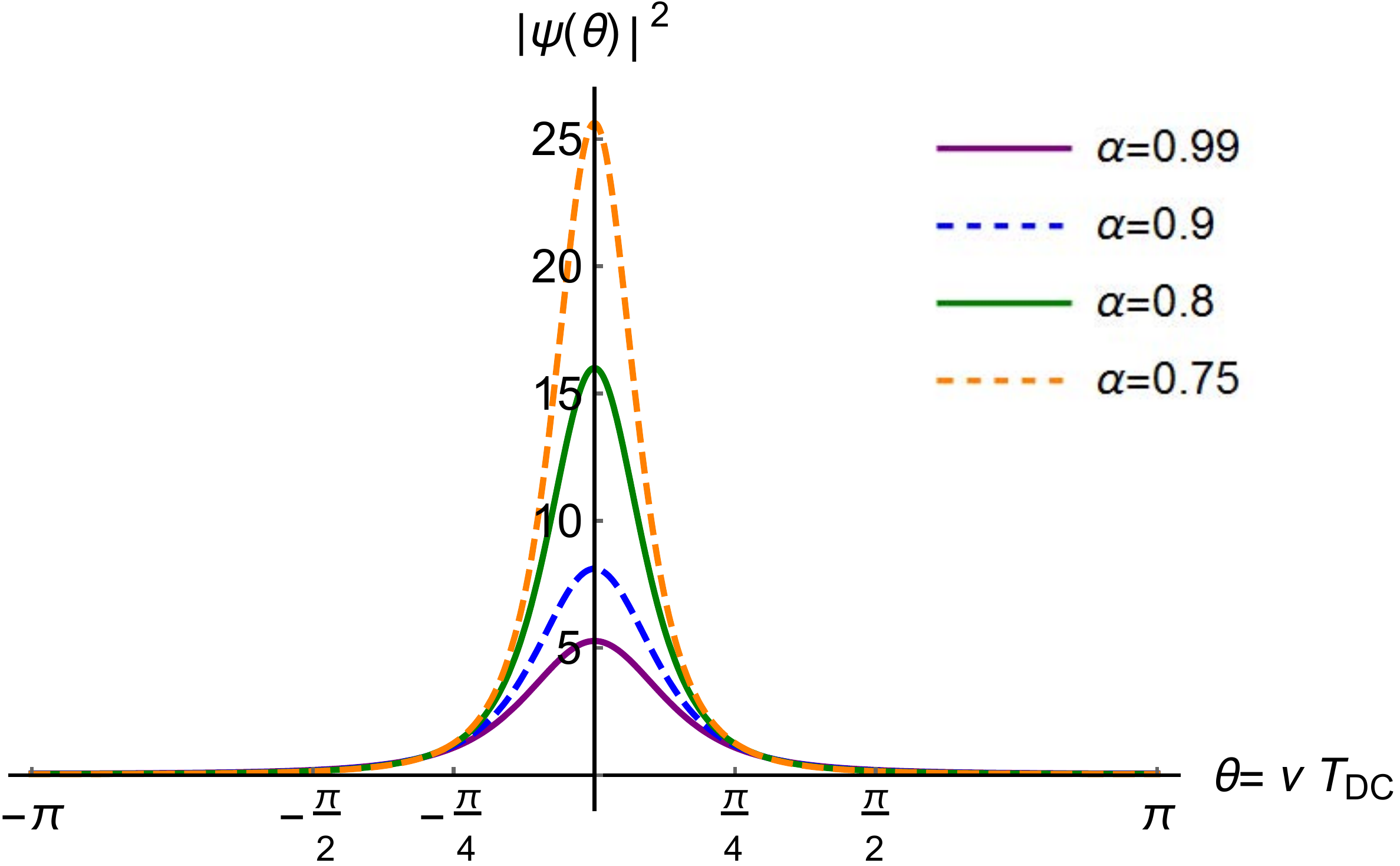}}
\caption{Plot of $|\psi(\theta)|^{2}$ as a function of $\theta=\nu T_{DC}$, capturing the frequency dependence of a single pair of signal/idler resonances in a single-bus MRR. Here, we have assumed $\rho=0.5$. The FWHM of the resonance is approximately $|\alpha-\rho|/\sqrt{2\alpha\rho}$ with a peak height of $(1-\rho^{2})^{4}/(1-\rho/\alpha)^{4}$, so long as the coupling is non-critical (i.e., $(\alpha-\rho)\gg r$).}
\end{figure}

\subsubsection{Rate and heralding efficiency of biphotons exiting cavity}
As was discussed previously for bulk crystals, the rate of biphotons coupling out of the resonator is given by the probability for the existence of the biphoton from $|\Psi\rangle_{out}$, divided by the round-trip time $T_{DC}$, where $|\Psi\rangle_{out}$ is obtained fom $|\Psi(T_{DC})\rangle_{ab}$. As a function of $\nu$, the biphoton rate per unit frequency $\mathcal{R}_{ab}(\nu)$ is given by:
\begin{equation}
\mathcal{R}_{ab}(\nu)=\frac{2\pi L^{2}}{\hbar^{2}T_{DC}}|g_{spdc}|^{2}|\alpha_{p}|^{2}\,|\psi_{ab}(\nu)|^{2}\text{Sinc}^{2}\Big(\frac{\Delta k_{z}L}{2}\Big)
\end{equation}
where
\begin{align}
\psi_{ab}(\nu)&\equiv\alpha_{a}^{*}\alpha_{b}^{*}\tau_{a}^{*}\tau_{b}^{*}\times\nn\\
&\times(e^{i\theta_{p}}D^{*}_{aa}(\nu)D_{bb}(\nu) + e^{-i\theta_{p}}D_{ab}(\nu)D_{ba}^{*}(\nu));
\end{align}
$P$ is the input pump power (in the bus) and $B$ is the cavity buildup factor at the pump wavelength, approximately equal to the Finesse $\mathcal{F}$ divided by $\pi/2$\footnote{Where optical cavities are also often rated by their $Q$ ``quality'' factor, it is useful to know that in the low loss limit (and at the pump wavelength), $B\approx \frac{2\lambda_{p}}{n_{p}L\pi}Q$.}. In Fig.~5, we've plotted $|\psi_{ab}(\nu)|^{2}$ to examine the shape of the spectrum of down-converted light when the cavity linewidth is much narrower than the phase matching bandwidth. Where we have assumed strict energy conservation, this spectrum represents a subset of the detected biphotons, i.e., the spectrum of the signal light over one linewidth of the cavity. When reflected about $\nu=0$, this is the idler spectrum. With the rate $\mathcal{R}_{ab}(\nu)$ known, we integrate over the area of a single resonance to obtain the coincidence rate due to emission into a single pair of frequency peaks $R_{ab}^{(peak)}$, and find:
\begin{equation}
R_{ab}^{(peak)}\approx \frac{8192}{81 \pi^{4}\epsilon_{0}c^{2}}\;\frac{n_{g1}n_{g2}}{n_{1}^{2}n_{2}^{2}n_{p}^{2}} \;\;\frac{d_{eff}^{2}\omega_{p}^{2}}{L_{x}L_{y}}L(1-\rho^{4})\,B\,P,
\end{equation}
where the approximation assumes $\alpha\approx 1$ for the integration of $|\psi(\nu)|^{2}$, and we are sufficiently far from critical coupling that $|\psi(\nu)|^{2}$ is not significantly altered when assuming $r\approx 0$. In the limit of zero self-coupling ($\rho\rightarrow 0$) the down-converted light can only make one round trip around the MRR, and the formula becomes identical to the single-mode rate in the bulk crystal (i.e., waveguide) regime. The pump buildup factor $B$ approaches unity, $|\psi_{ab}(\nu)|^{2}$ grows wider than the phase-matching bandwidth of the light so that it is near unity over the bandwidth of the sinc function, and we must carry out the same phase-matching integrals as in previous sections. In this same limit, we see that the effect of loss is that $R_{ab}$ scales as $|\alpha|^{4}$, or by two factors of the power loss; one for the signal photon and one for the idler photon.

It is interesting to point out that here, the total rate $R_{ab}^{(peak)}$ scales linearly with $L$, even though the narrow frequency filtering of the MRR would suggest a quadratic dependence. This is due to the linewidth of the MRR itself depending on $L$, where longer resonators have a corresponding narrower linewidth.

As an example of the utility of this formula, consider the following. Let us assume Type-I SPDC in a MRR of Aluminum Nitride with radius $30\mu$m, with transverse horizontal and vertical thicknesses of $1.0\mu$m and $0.3\mu$m, respectively. The effective nonlinearity $d_{eff}\approx 4.7 pm/V$. Let the quality factor at the pump wavelength be $10^{4}$ which gives a buildup factor $B$ of about 12.3. Let the pump wavelength $\lambda_{p}=775$nm. We will let $n_{g1}=n_{g2}=2.19$ and $n_{1}=n_{2}=2.16$ and $n_{p}=2.14$. With these parameters, we obtain an astonishingly high rate $R_{ab}^{(peak)}$ of approximately $3.0\times10^{7}$ pairs per second per mW of pump power between correlated resonances in the cavity. In the limit of no self coupling ($\rho\rightarrow0$) and no cavity buildup $B\rightarrow 1$, $|\psi(\nu)|^{2}\approx 1$ over all frequency so that an accurate treatment must explicitly consider phase-matching (i.e., $e^{i\Delta k_{z}}\not\approx 1$), and an accurate treatment is well described in the bulk crystal regime. While the ideality of our approximations  (including our choice of basis modes) makes it unrealistic that this formula provides an accurate estimate of the number of exiting photon pairs per second, it does illustrate the potential single-bus MRRs have as a bright source of photon pairs via SPDC.

In more practical implementations of SPDC in micro-ring resonators, a dual-bus configuration may be used so that one waveguide may be dedicated to coupling in/out pump light, and the other for outcoupling SPDC photon pairs. Alternatively, in type-II SPDC, the coupling between bus and MRR can be strongly polarization dependent, and it may be possible to well-separate the signal and idler photons from one another instead of tolerating the reduction in coincidences relative to singles that comes with separation with a non-polarizing beamsplitter.

In order to gauge the utility of the photon pairs exiting the resonator, it is not enough to simply know the photon pair rate. Because of loss in the resonator among other places, the number of signal photons without matching idlers exiting the resonator is significant enough, that its dependence on experimental parameters is important to know. Although usually discussed in the context of spatial correlations, here, we shall define the resonator heralding efficiency $\eta_{R}$ to be the ratio of the signal photon rate coming from exiting photon pairs (equal to the photon pair rate discussed previously), divided by the sum of this rate and the rate of signal photons exiting the resonator, where the idler has been lost. Where common factors in the ratio cancel out, we find:
\begin{equation}
\eta_{R}(\nu)\approx\fracd{|D_{bb}(\nu)|^{2}}{|D_{bb}(\nu)|^{2}+|J_{bb}(\nu)|^{2}},
\end{equation}
where we take the same approximations for calculating the individual rates as before. In this case, we find the heralding efficiency is nearly constant over the FSR of the resonator, and arrive at the approximation:
\begin{equation}
\eta_{R}\approx\fracd{1-\rho^{2}}{2- \rho^{2} - \alpha^{2}}
\end{equation}
For the single-bus MRR studied here, we see a tradeoff between enhancing either brightness or heralding efficiency due to the parameters of the resonator. While lower intrinsic loss (i.e., $\alpha\rightarrow\approx 1$) is an absolute improvement, increasing the self-coupling $\rho$ only increases brightness at the expense of lowering heralding efficiency. Indeed, $\eta_{R}$ is maximized in the limit of no self coupling (i..e, where $\rho\rightarrow 0$), and only approaches $50$ percent at critical coupling. In the limit of strong self-coupling, where $\rho\rightarrow 1$ for constant loss $\alpha$, the heralding efficiency decreases towards zero, since it becomes progressively more and more likely that a photon in the resonator will be scattered out as loss rather than couple into the output bus.

\subsubsection{Time correlations of biphotons exiting cavity}
In order to accurately treat the time correlations between the signal and idler photons exiting the cavity, it is necessary to include phase matching. Energy conservation allows us to say $\psi_{ab}(\nu)\approx \psi_{ab}(\nu_{-}/\sqrt{2})$, where $\nu_{-}\equiv (\nu_{1}-\nu_{2})/\sqrt{2}= (\omega_{1}-\omega_{2})/\sqrt{2}$. We are interested in this model only for discussion of the behavior of the time correlations between biphotons exiting a micro-ring resonator. For type-I SPDC, the overall phase matching function is given by $\text{Sinc}(L\kappa \nu_{-}/\sqrt{8})\psi_{ab}(\nu_{-}/\sqrt{2})$, which can be broken up into three different terms. The sinc function is a broad envelope function multiplying $\psi_{ab}(\nu_{-}/\sqrt{2})$, and $\psi_{ab}(\nu_{-}/\sqrt{2})$ is well approximated as the convolution of a Lorentzian ``tine'' of FWHM $|\alpha-\rho|\sqrt{8}/(T_{DC}\sqrt{\alpha \rho})$, convolved with a Dirac comb with spacing $2\sqrt{2}\pi/T_{DC}$ (see Fig.~6a for diagram of $|\psi(\nu_{-})|^{2}$). Because of the simplicity of our expression, we can readily take the inverse Fourier transform to examine the time correlations. Using the convolution theorem to our advantage, we see that in time (as in Fig.~6b), the amplitude of $t_{-}$ has a similar breakdown to the corresponding function of $\nu_{-}$. The ``envelope'' in time is given by the inverse transform of the ``tine'' function in frequency, and the tine function in time is given by the inverse transform of the envelope function in frequency. The spacing of the comb in $t_{-}$ is given by $T_{DC}/(2\sqrt{2}\pi)$.

\begin{figure}[t]
\centerline{\includegraphics[width=\columnwidth]{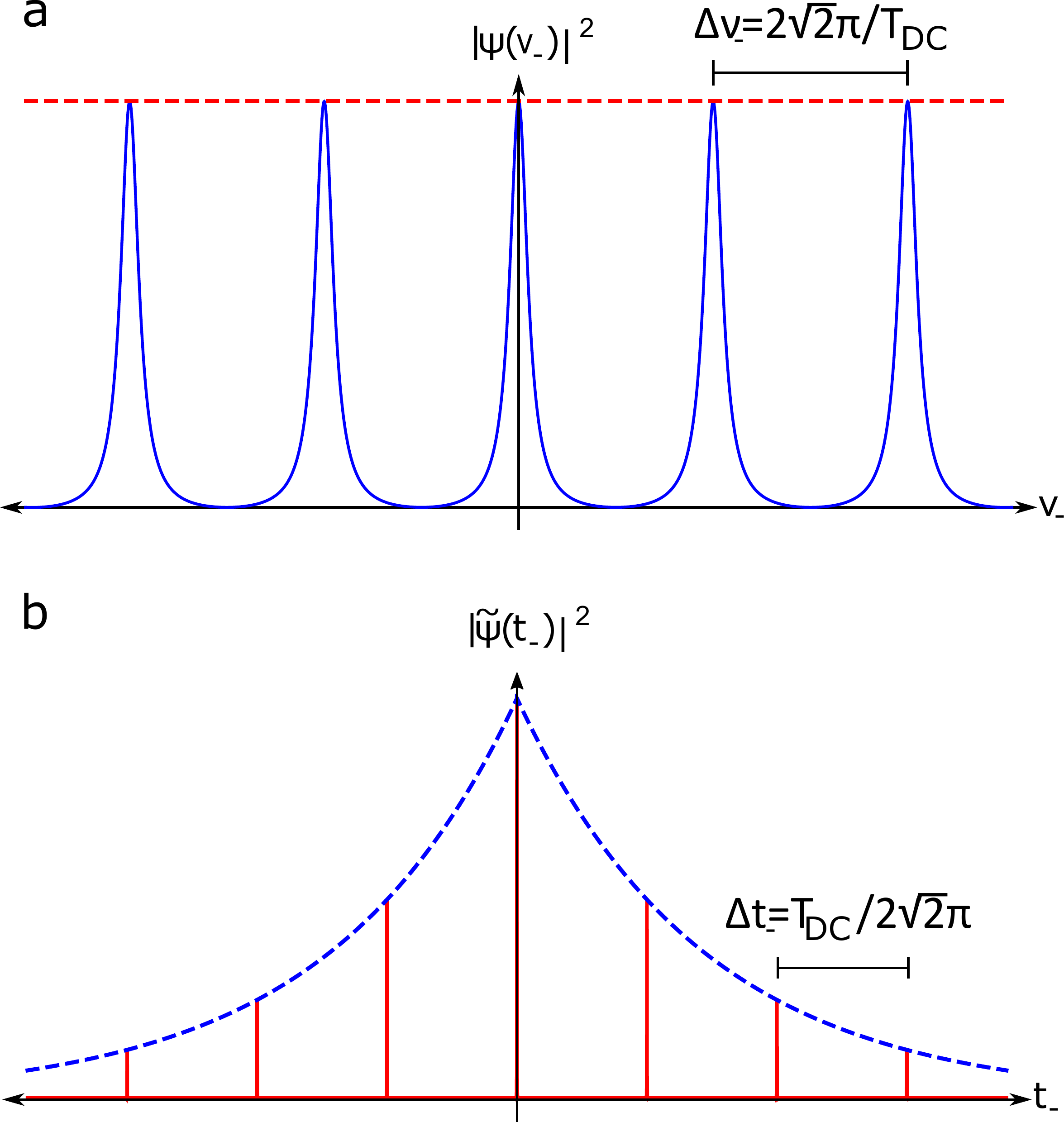}}
\caption{Plots showing reciprocal scaling of phase-matching (red) and resonance widths (blue). a) Plot of $|\psi_{ab}(\nu_{-})|^{2}$, where the phase-matching ``sinc-like'' function (red) is much wider than the resonance widths, and serves as an envelope function for the frequency difference spectrum. b) Plot of $|\tilde{\psi}_{ab}(t_{-})|^{2}$. When transforming from frequency to time, the resonance widths in time come from the inverse transform of the phase-matching envelope in frequency, while the inverse transform of the resonance peaks in frequency becomes the exponential envelope function in time.}
\end{figure}

In the time domain, the inverse transformed Lorenzian is an exponential spike with decay constant in $t_{-}$ of $|\alpha-\rho|\sqrt{2}/(T_{DC}\sqrt{\alpha \rho})$, which serves as an envelope for a comb of inverse-transformed sinc resonances, with spacing equal to $T_{DC}/(2\sqrt{2}\pi)$. The exact shape of the ``inverse-transformed sinc resonances'' is determined by the type of phase matching, as discussed in previous sections. Where $T_{DC}$ is on the order of a few picoseconds, experimental measurements of the time correlations by coincidence counting are not yet capable of resolving individual peaks, but may have sufficient range to capture the breadth of these time correlations. Indeed, the number of tines in $t_{-}$ until the exponential envelope decays to $1/e$ of its peak value is directly proportional to the finesse of the resonator at the down-conversion frequency \footnote{The decay constant in number of tines is equal to $\mathcal{F}/8\pi^{2}$, where $\mathcal{F}$ is the finesse of the resonator at the given wavelength.}. As an example, when $T_{DC}$ is of the order of 2 picoseconds, coincidence counting setups with range of 20 nanoseconds will adequately capture the time correlations in single-bus micro ring resonators with a finesse of the order $10^{3}$.

\section{SPDC with pump depletion}
Throughout this paper, we have considered SPDC in the regime where the pump illumination is bright enough to be treated classically, but not so bright that multi-biphoton creation events become significant. We later explored a more fully quantum treatment of SPDC light \eqref{TimeEvolved}, but only in the undepleted pump approximation. In this section, we will consider SPDC  in the regime of longer interaction times, where the pump light may be significantly depleted in exchange for bright intensities of the down-converted fields. We limit ourselves to the case of a simple waveguide, where a single pump mode is coupled to a single pair of signal and idler modes, and do not consider loss  due either to absorption or coupling with other modes. In the regime where the pump is undepeleted, we will conclude by discussing how the number of generated biphotons is affected when using different quantum states of pump light as the source (e.g., Fock states).

When the pump light is dim enough that a fully quantum descripton of the pump is necessary, it is also wise to consider when it is no longer possible to invoke the undepleted pump approximation. In this section, we show how the mean number of downconverted photon pairs changes with time when the pump can be depleted, and how in the limit of small times, we obtain the same result as in the undepeleted pump approximation.

The simplest hamiltonian describing SPDC from a single pump mode to a single pair of signal and idler modes is given by:
\begin{equation}
H_{NL}=i\hbar \Big( g \;\hat{a}_{p}\hat{a}^{\dagger}_{1}\hat{a}^{\dagger}_{2} - g^{\ast}\;\hat{a}_{p}^{\dagger}\hat{a}_{1}\hat{a}_{2}\Big)
\end{equation}
where $g$ is the coupling constant between the pump mode, and the signal-idler mode pair as seen in \eqref{coupling}, albeit without incorporating a static pump power. Using the Heisenberg equation of motion, and the commutator algebra for the creation and annihilation operators for each of the three modes, we can obtain a differential equation for the photon number operator $\hat{N}_{1}\equiv\hat{a}_{1}^{\dagger}\hat{a}_{1}$.
\begin{equation}
\frac{d^{2}\hat{N}_{1}}{d t^{2}}=2|g|^{2}\big(\hat{N}_{p}\big(\hat{N}_{1}+\hat{N}_{2}+1\big)    -\hat{N}_{1}\hat{N}_{2}\big).
\end{equation}
Since the hamiltonian also guarantees that:
\begin{equation}
\frac{d\hat{N}_{1}}{dt}=\frac{d\hat{N}_{2}}{dt} = - \;\frac{d\hat{N}_{p}}{dt},
\end{equation}
the initial vacuum state of the down-converted fields also guarantees that $\langle\hat{N}_{1}\rangle=\langle\hat{N}_{2}\rangle$, so that $\hat{N}_{1}(t)$ is described by the simpler equation:
\begin{equation}\label{depletPumpRaw}
\frac{d^{2}\hat{N}_{1}}{d t^{2}}=2|g|^{2}\big(\hat{N}_{p} + 2\hat{N}_{1}\hat{N}_{p} - \hat{N}_{1}^{2}\big).
\end{equation}
From this, one can obtain a (semi-classical) differential equation describing the expectation value $\langle\hat{N}_{1}(t)\rangle$ using the simplifying assumptions that the signal and idler fields are in the vacuum state at time $t=0$, and that the thermal statistics of the photon pairs described by the two-mode squeezed vacuum state for a coherent state pump follow the law for the geometric distribution: $\langle\hat{N}_{1}^{2}\rangle=2\langle\hat{N}_{1}\rangle^{2} + \langle\hat{N}_{1}\rangle$:
\begin{equation}\label{depletPump}
\frac{d^{2}N_{1}}{d t^{2}}=2|g|^{2}\big(N_{p}^{(0)}(2 N_{1}+1)-N_{1}(6N_{1}+4)\big).
\end{equation}
Here, $N_{p}^{(0)}$ is the initial mean number of pump photons in the medium, which may be given by the (instantaneous) pump power, multiplied by the time it takes light to move through the crystal, and divided by the energy of a pump photon. In addition, we used the fact from our initial conditions, that $\langle\hat{N}_{1}\rangle=N_{p}^{(0)}-\langle\hat{N}_{p}\rangle$. For simplicity, we let $N_{1}\equiv \langle\hat{N}_{1}(t)\rangle$. A fully quantum treatment will account for the departure from a coherent state pump, as the down-converted photon pairs are later up-converted again in the reverse process, altering the pump statistics. For a fully quantum treatment, in which the complete number statistics of the pump, signal, and idler light are considered, see \cite{nation2010trilinear,alsing2015parametric}.

Although the depleted pump equation \eqref{depletPump} is nonlinear, it is integrable using techniques similar to those used to solve the ordinary nonlinear pendulum. In doing so, we obtain an implicit solution in the form of an integral:
\begin{equation}
\int_{0}^{N_{1}(t)}\frac{dx}{\sqrt{-2x^{3}+(N_{p}^{(0)}+2)x^2 + N_{p}^{(0)}x}}=2|g|t
\end{equation}
For typical experimental parameters, $N_{p}^{(0)}\gg2$; we may omit the correction of $2$ to the quadratic term in the integrand. Even so, this integral cannot be expressed in terms of elementary functions, though certain definite integrals have straightforward expressions. In particular, the time to maximum depletion $T_{D}$ can be found by taking the integral from zero to approximately $N_{p}^{(0)}/2$, and solving for the time $t$ The approximation becomes exact in the lmit of large $N_{p}^{(0)}$\footnote{The maximum (critical) value of $N_{1}$ expanded to first nontrivial order is $\frac{N_{p}^{(0)}}{2} + \frac{1}{2}+ \mathcal{O}((N_{p}^{(0)})^{-1})$.}. Values of $N_{1}$ larger than the critical value make the integrand imaginary, so that $50$ percent pump depletion is the maximum amount allowed in this coherent state model. Alternative derivations of the maximum power conversion efficiency in SPDC, for this simple setup, also exhibit this approximate theoretical limit \cite{BreitenbachSPDCEfficiency}, though more sophisticated experiments using optical cavities give different values. The solution (simplified assuming $N_{p}^{(0)}\>>1$) can be expressed in terms of elliptic integrals:
\begin{equation}
T_{D}\approx-\frac{\sqrt{2}\Big(\mathcal{F}\Big[i \;\text{csch}^{-1}\Big(\sqrt{\frac{N_{p}^{(0)}}{2}}\Big),-\frac{N_{p}^{(0)}}{2}\Big]-i\mathcal{K}\Big[\frac{N_{p}^{(0)}}{2}\Big]\Big)}{|g|\sqrt{-N_{p}^{(0)} + \sqrt{N_{p}^{(0)}(N_{p}^{(0)}+8)}}},
\end{equation}
where $\mathcal{F}(a,b)$ and $\mathcal{K}(a)$ are incomplete and complete elliptic integrals of the first kind, respectively. Here, $|g|^{2}$ takes the value (using our Hermite-Gauss quantization basis):
\begin{equation}
|g|^{2}=\frac{8\hbar\;\pi^{2} c^{3}d_{eff}^{2}}{\epsilon_{0} n_{1}^{2}n_{2}^{2}n_{p}^{2}L_{z}\lambda_{p}^{3}\sigma_{p}^{2}}\Bigg|\frac{\sigma_{p}^{2}}{\sigma_{1}^{2} + 2\sigma_{p}^{2}}\Bigg|^{2},
\end{equation}
which, for typical experimental values is of the order $10^{6}$. Note that the absence of a Sinc function in this expression is due to our approximation of a single pump mode coupled to a single pair of signal and idler modes, where the Sinc function can be taken to be unity. For typical pump wavelengths and crystal lengths, $N_{p}^{0}$ is of the order $10^{4}$, and $T_{D}$ is of the order $10^{-5}$ seconds\footnote{For a 1mW pump, with 404nm wavelength, incident on a BiBO crystal 3mm long, and a pump radius $\sigma_{p}$ of 0.4mm, $|g|^{2}$ is about $8.136\times10^{6}$, $N_{p}^{(0)}$ is about $3.71\times10^{4}$, and $T_{D}$ is about $1.147\times10^{-5}$ seconds.}. At much larger pump powers, where multi-biphoton events become significant, near the optical damage threshold of the crystal, the depletion time can be less than a nanosecond. Here it is important to point out, that our model applies only for times less than the coherence time of the pump light, and less than the time it takes light to travel through the crystal. Instead of using unreasonably long nonlinear media, one could instead keep pump light in the crystal for microsecond-scale times with an optical cavity with a finesse\footnote{For small round trip losses, the finesse is approximately $2\pi$ divided by the fraction of light lost in one round trip.} in excess of $5\times10^{6}$, though an accurate description of this requires us to treat SPDC in a cavity, as seen in Section $5b$.

Of particular interest is the case of small times, where the cubic term can be neglected in the integrand. In this approximation, the integral has the form of an hyperbolic arcsine, which leads to the solution:
\begin{equation}
N_{1}(t)\approx \text{sinh}^{2}\Big(\sqrt{N_{p}^{(0)}}|g|t\Big)\approx N_{p}^{(0)}|g|^{2}t^{2},
\end{equation}
which is in agreement with the undepleted pump approximation where $N_{p}^{(0)}$ is the mean number of pump photons in the crystal at any given time, given pump power and crystal length. See Fig.~7 for a side by-side comparison of the different approximations for $N_{1}(t)$. In the picosecond time scales light takes to travel through nonlinear crystals, there is no meaningful distinction between these approximations, and the simplest one will suffice.
\begin{figure}[t]
\centerline{\includegraphics[width=\columnwidth]{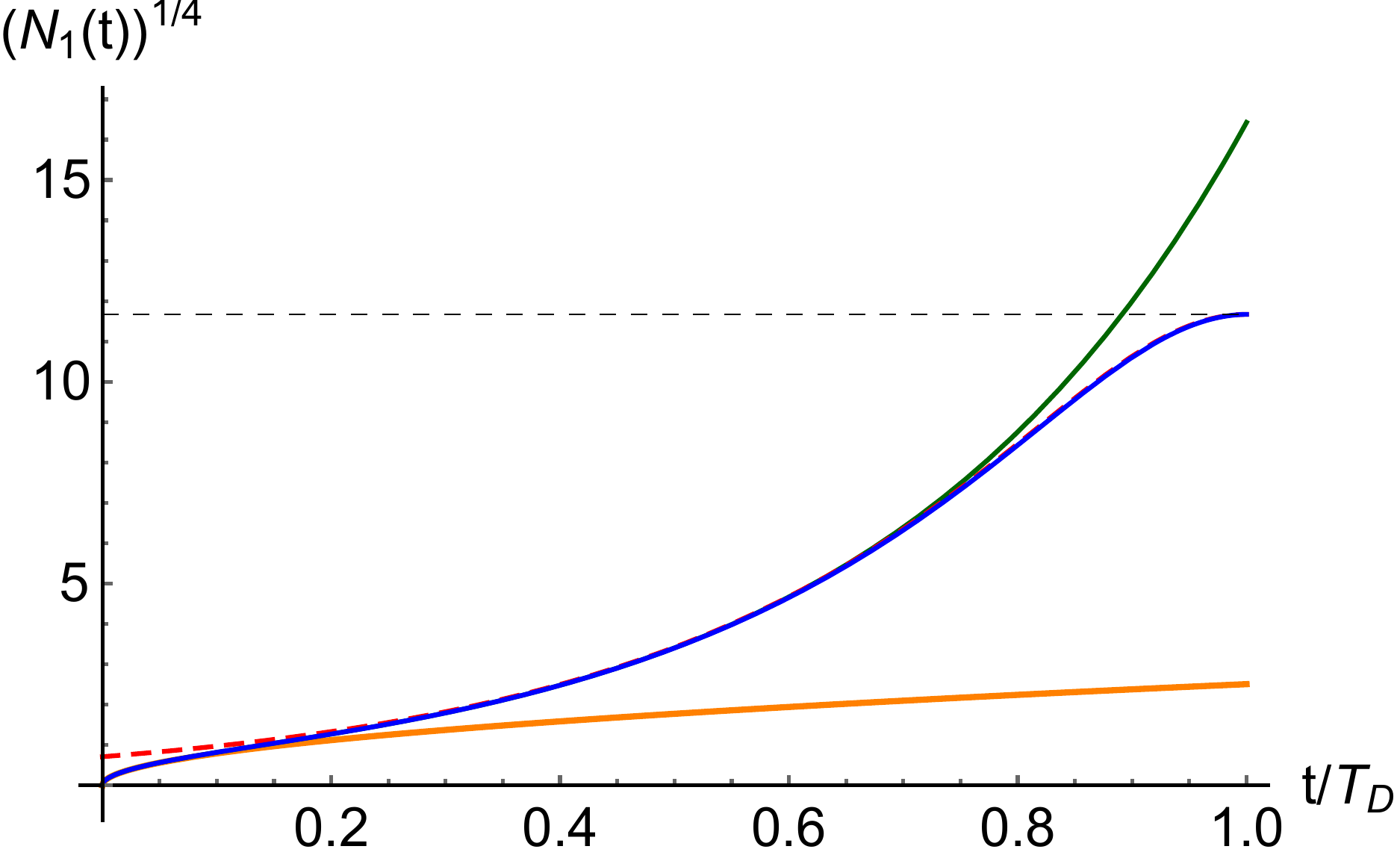}}
\caption{Plots of the number of signal photons as a function of time, for a coherent state pump, scaled with respect to $T_{D}$. The blue curve gives the exact solution obtained from numerically solving the differential equation \eqref{depletPump}. The green curve gives the hyperbolic sine approximation, which rapidly diverges for times beyond $T_{D}$. The shallow orange curve gives the first-order approximation to $N_{1}(t)$, which agrees within ten percent for times less than  $T_{D}/12$. The red dotted curve gives the approximation as a hyperbolic secant, which only differs noticeably from the exact numerical solution for times less than $T_{D}/5$. The fourth root of $N_{1}$ is taken to allow better visual comparison of the extreme variation in the approximations at large time values.}
\end{figure}

In the limit of times on par with $T_{D}$, the differential equation \eqref{depletPump} is such that the constant term contribution to the second derivative may be neglected, and the approximate solution has the form of the square of the hyperbolic secant:
\begin{equation}
N_{1}(t\sim T_{D})\approx \frac{N_{p}^{(0)}}{2}\text{sech}^{2}\Big(\sqrt{N_{p}^{(0)}|g|^{2}}(t-T_{D})\Big)
\end{equation}
Plotting this in Fig.~2 shows no significant departure from the exact numerical solution for times larger than $T_{D}/2$, indicating a valid approximation. Indeed, using the Hyperbolic sine approximation for times less than $T_{D}/2$, and the hyperbolic secant approximation for times greater than $T_{D}/2$, yields a maximum percent error of $0.7$ percent for times between $0$ and $T_{D}$.

\subsection{SPDC with different quantum pump statistics}
Regardless of the initial quantum state of the pump, we can use the differential equation for $\hat{N}_{1}(t)$ \eqref{depletPumpRaw} to  find the rate of photon pair generation. For a given quantum state of the field $\hat{\rho}$, the mean number of photon pairs, also given by $\langle\hat{N}_{1}(t)\rangle$ is:
\begin{equation}
N_{SM}(t)=\text{Tr}[\hat{\rho}\;\hat{N}_{1}(t)]
\end{equation}

Because the signal and idler fields are initially in the vacuum state, for times before significant pump depletion, this simplifies to:
\begin{equation}
N_{SM}(t)=\text{Tr}[\hat{\rho}\;\text{sinh}^{2}(\sqrt{\hat{N}_{p}}\;g t)]
\end{equation}
At smaller pump powers or smaller times, this simplifies further to:
\begin{equation}
N_{SM}(t)\approx\text{Tr}[\hat{\rho}\;(\hat{N}_{p})\;g^{2} t^{2}]=\langle\hat{N}_{p}\rangle\;g^{2} t^{2}.
\end{equation}
Therefore, at small times, and pump powers, the average number of generated biphotons depends only on the mean pump power, regardless of whether it is in a coherent state, Fock state, or any other state. For higher pump powers, where this approximation no longer applies, there is some qualitative difference between the efficiency of SPDC with different pump photon statistics and same mean pump power.

If we take the trace in the photon number basis, and let $P(n_{p})$ be the probability of measuring $n_{p}$ photons at time $t$, then the number of generated biphotons $N_{SM}(t)$ is expressible as:
\begin{equation}
N_{SM}(t)=\sum_{n_{p}=0}^{\infty}P(n_{p})\;\text{sinh}^{2}(\sqrt{n_{p}}\;g t)
\end{equation}
Since the function $f(x)=\text{sinh}(\sqrt{x})^{2}$ is a convex, monotonically increasing function\footnote{A convex function is a function with non-negative second derivative i.e., "concave-up").} of $x$ for all positive values of $x$, the mean value of the function $\langle f(x)\rangle$ is larger than the function of the corresponding mean value of $x$, $f(\langle x\rangle)$. Consequently, pump beams with larger fluctuations of photon number will have larger biphoton generation efficiency solely by the virtue of there being probable events of larger photon number. Whether this is due to power instability in the pump, or a fundamental difference in the quantum number statistics of the pump, the overall effect on biphoton generation rate will remain the same. Even so, comparing the mean number of biphotons generated for a Fock state pump, a coherent state pump, and a thermal state pump with same mean photon number yields an inconsequential discrepency. Even at pump intensities approaching the damage threshhold of many nonlinear materials (e.g., $1 MW/mm^{2}$), the estimated difference in $N_{SM}(T_{DC})$  between a Fock pump, a coherent pump, and a thermal pump is less than one percent.

When entering the regime of significant pump depletion and long interaction times, the efficiency of SPDC can vary significantly. Although we showed earlier that coherent state pumps incident on simple nonlinear media have a maximum down-conversion efficiency of approximately $50$ percent, it has been shown \cite{MurphySPDCUnitEfficiency} that a 1-photon Fock state pump can have 100 percent down-conversion efficiency, while $n$-photon Fock states up to $n=50$ have maximum efficiencies above 77 percent.

Efficiency aside, it is a very interesting question how the quantum state of the down-converted fields changes with the quantum state of the pump. The two-mode squeezed vacuum state for SPDC light assumes a coherent state pump, but the state of the down-converted fields for a Fock state pump, or a thermal state pump will differ greatly. The nature of the down-converted fields as a function of exotic quantum pump states remains a rich field for further development.

\section{Comparisons with experiment}
In order to compare theoretical biphoton generation rates with experimental data, we create a simple model accommodating loss and various efficiencies throughout the experiment. Let us consider the following setup (See Fig. 8). Here, we will assume $N$ biphotons per second are separated into the signal and idler arms, eventually arriving at the respective arm's single-photon detector. In addition, we assume non-number resolving detectors, so that a biphoton hitting one detector registers as a single count. Here we define the coupling efficiencies into the collection modes as $C_{1}$. $C_{2}$ and $C_{12}$ for signals, idlers, and coincidences, repsctively. When we use a non-polarizing beamsplitter, we define the beamsplitter efficiencies as $\beta_{1}$, $\beta_{2}$ and $\beta_{12}$. Independent losses in the signal and idler channel due to, e.g., scattering, detector efficiency, and absorption, are given by the efficiencies $E_{1}$ and $E_{2}$. For a 50/50 beamsplitter, $\beta_{1}=\beta_{2}=3/4$, since three out of four times, at least one photon of the pair will exit a given output mode of the beamsplitter. Furthermore, $\beta_{12}=1/2$ since half of the time, both photons exit the same port. When coupling down-converted light into a single-mode fiber, the coupling efficiencies $C_{1}$ and $C_{2}$ are given as equal to $C$, while the coincidence coupling efficiency $C_{12}=\eta C$, where $\eta$ is the heralding efficiency.

For the experiments using type-0 and type-I SPDC, the down-converted light was separated with a 50/50 beamsplitter. In this situation $N_{1}$ and $N_{2}$ are related to the raw rate $N$ and coincidence count rate $N_{12}$ in the following way:
\begin{align}
N_{1}&=N\cdot E_{1} C \beta_{1} + \Phi_{1}\\
N_{2}&=N\cdot E_{2} C \beta_{2} + \Phi_{2}\\
N_{12}&= N\cdot E_{1} E_{2} \;\eta \;C\beta_{12} + A_{12}
\end{align}
When the photon pairs can be completely separated, such as by polarization in type-II SPDC, the relative beamsplitter losses $(\beta_{1}, \beta_{2},\beta_{12})$ can all be set equal to unity, with indpendent losses already being captured by $E_{1}$ and $E_{2}$. Here, $\Phi_{1}$ (alt. $\Phi_{2}$) is the count rate due to uncorrelated photons such as external noise, dark counts, and uncorrelated fluorescence stimulated by the pump. Finally, $A_{12}$ is the count rate of accidental coincidences due to a variety of sources, but nonetheless detectable. Using straightforward algebra, one can show that when the biphotons are separated by a 50/50 beamsplitter, the number of biphotons generated at the source $N$ is given by:
\begin{equation}
N^{(50/50)}=\frac{(N_{1}-\Phi_{1})(N_{2}-\Phi_{2})}{(N_{12}-A_{12})}\Bigg(\frac{\beta_{12}}{\beta_{1}\beta_{2}}\Bigg)\frac{\eta}{C},
\end{equation}
where the fraction of beamsplitter efficiencies for a 50/50 BS is $8/9$. The most challenging aspect of applying this formula in general is to obtain the coupling efficiency $C$, and heralding efficiency $\eta$. When the beamsplitter is asymmetric, so that fraction $\gamma_{t}$ of the light is transmitted, and fraction $\gamma_{r}$ is reflected (and normalized so that $\gamma_{t}+\gamma_{r}=1$), one finds:
\begin{align}
\beta_{1}&=\gamma_{t}^{2} + 2 \gamma_{t}\gamma_{r}\nn\\
\beta_{2}&=\gamma_{r}^{2} + 2\gamma_{t}\gamma_{r}\nn\\
\beta_{12}&=2\gamma_{t}\gamma_{r}
\end{align}
To test the validity of the generation rate formulas derived earlier in this paper, we performed three simple experiments.

\begin{figure}[t]
\centerline{\includegraphics[width=\columnwidth]{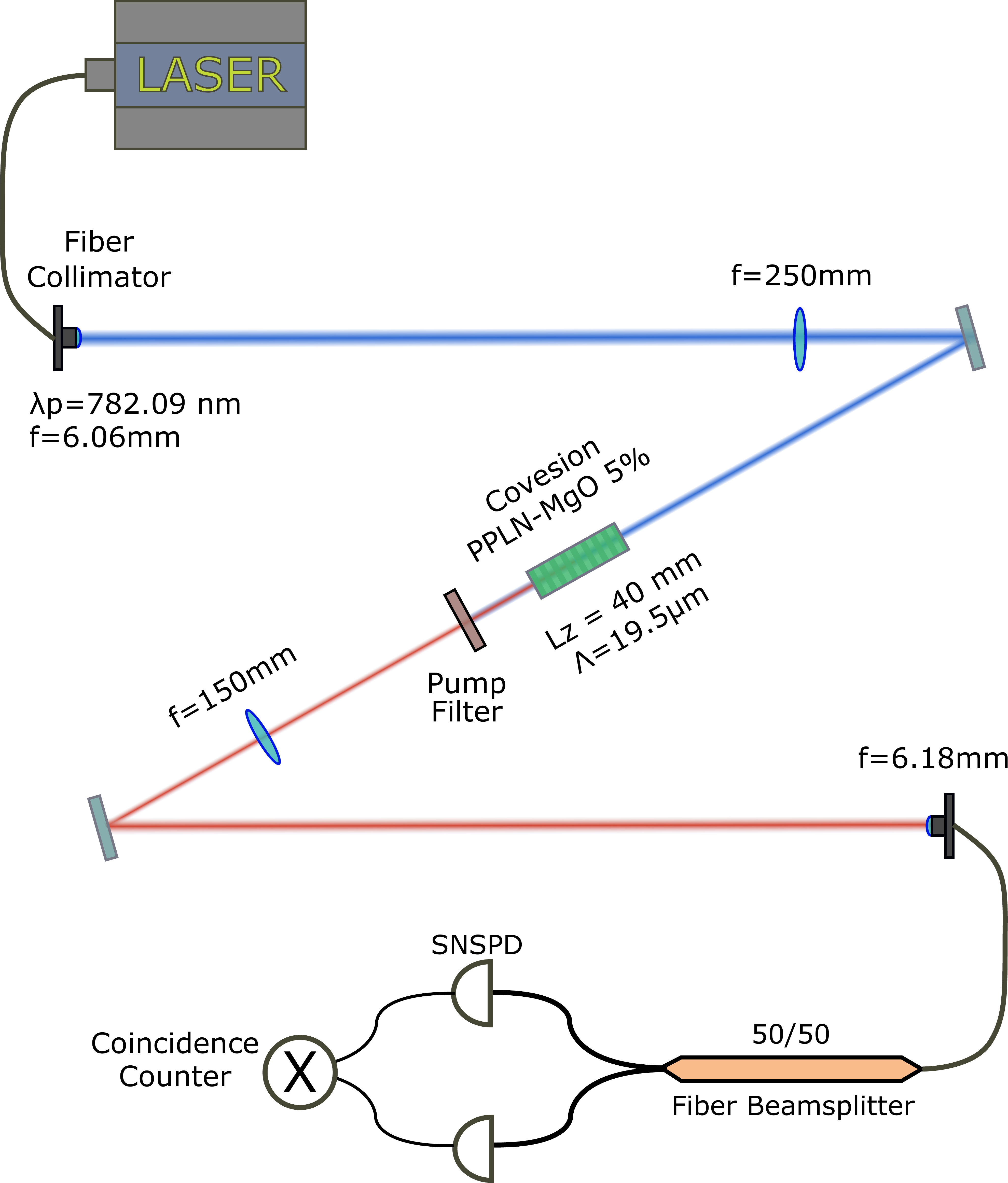}}\label{diagram}
\caption{Diagram of experiment used to obtain coincidence count rate for type-0 SPDC. The pump light exiting a single mode fiber is focused to a given spot size at the center of the nonlinear crystal (NLC), and is subsequently filtered out. The downconverted light is collimated, and collected into a single mode fiber, and split by a 50/50 fiber beamsplitter (BS), and sent to Superconducting Nanowire Single-Photon Detectors (SNSPDs). The coincidence counter records time intervals between detection events on each detector. The experiment allows us to directly measure the single-mode rate $R_{SM}$ with optics determing $\sigma_{p}$ and $\sigma_{1}$ relative to the mode field diameters of the input and output fibers.}
\end{figure}

\subsection{Type-0 SPDC in PPLN crystal coupled to single-mode fiber}
The first experiment (Fig.~8) tests the single-mode rate for degenerate type-0 SPDC with a periodically poled nonlinear crystal. We used a 40mm Periodically Poled Lithium Niobate (PPLN) crystal manufactured by Covesion with a $1$mm (transverse) width, and $19.5\mu$m poling period, temperature tuned to $107.2^{\circ}$C for degenerate SPDC from $782.09$nm to $1564.18$nm. Our pump laser was an OBIS laser with measured wavelength of $782.09$ nm and bandwidth of approximately $0.01$ nm. The pump laser light was directed into the crystal through a single mode fiber, triplet fiber collimator, and focusing lens to obtain a well-approximated gaussian beam with spot size $\sigma_{p}=52.6\pm2\mu$m at the center of the crystal. Using corresponding collection optics for the downconverted light, we obtain a mode-matched down-converted beam radius of $\sigma_{1}=55.1\pm2\mu$m also at the center of the crystal. Using the Sellmeyer equations for Lithium Niobate, and published values for $d_{\text{eff}}$  \cite{gayer2008temperature}, we obtained the necessary phase and group indices of refraction, as well as the group velocity dispersion constant $\kappa$.

To simplify the initial alignment of our setup, we input 1564.18 nm light into the back end of the experiment, and coupled the Second Harmonic Generation (SHG) light into the fiber that would later be connected to the pump laser. Since the exit fiber tip is in an image plane of the center of the crystal, the down-converted light is spatially correlated at the fiber tip, and we let the coupling loss through the exiting fiber collimator to be the correlated efficiency C. Since the down-converted light was too dim to be seen in free space with ordinary power meters, we estimated C using laser light at 1564nm shining through an experiment with identical focusing optics and found C to be approximately $(0.807\pm0.025)$ though the coupling to an ideal mode-matched gaussian beam may be higher. We estimate the heralding effiency $\eta\approx (0.862\pm0.022)$ with our experimental beam parameters, and the formula for the Heralding efficiency for SPDC with focused gaussian beams in \cite{dixon2014heralding}. Per milliwatt of pump power per second, we measured singles rates of $16.00\pm0.21$ million and $17.99\pm0.22$ million for the signal and idler detectors, and background noise levels $\Phi_{1}\approx 0.05\times10^{6}$ and $\Phi_{2}\approx 0.06\times10^{6}$. We measured a coincidence count rate of $2.93\pm0.05$ million with accidentals rate $A_{12}\approx0.02\times 10^{6}$, giving coincidence to singles ratios of $16.1$ and $18.1$ percent, respectively, which in turn gives us a raw pair generation rate $N$ of $(95.63\pm2.71)$ million pairs per second per mW of pump power.

With our experimental parameters, our formula \eqref{SingleModeType01} predicts a rate of $(94.86\pm10.89)\times 10^{6}$ coincidence counts per second per mW of pump power. The raw pair generation rate obtained from our experiment was approximately $(95.63\pm2.71)\times 10^{6}$ per second per mW of pump power, differing from our prediction by less than $1\%$, or $0.1$ standard deviations. The relatively large uncertainty in the theoretical prediction is due to the propagation of uncertainties of multiple variables. The individually large $5\%$ uncertainty in $d_{\text{eff}}$ is due to imperfections between different manufacturing process of otherwise identical crystals. To have such a small disagreement between theory and experiment is  subject to multiple caveats, namely, that the true coupling efficiency is unmeasured. Because we only measure the maximum coupling in a parallel experiment, we can only assume that this represents the coupling efficiency in the experiment if it too is optimally coupled. Though much effort was devoted to maximizing the coupling of the down-converted light into the single mode fiber, it is likely that the experimental coupling efficiency is less than $0.805$ by possibly as much as $10-20$ percent, which would then increase our estimate of $N$ by $10-20$ percent, significantly exceeding the theoretical value.

\begin{figure}[t]
\centerline{\includegraphics[width=\columnwidth]{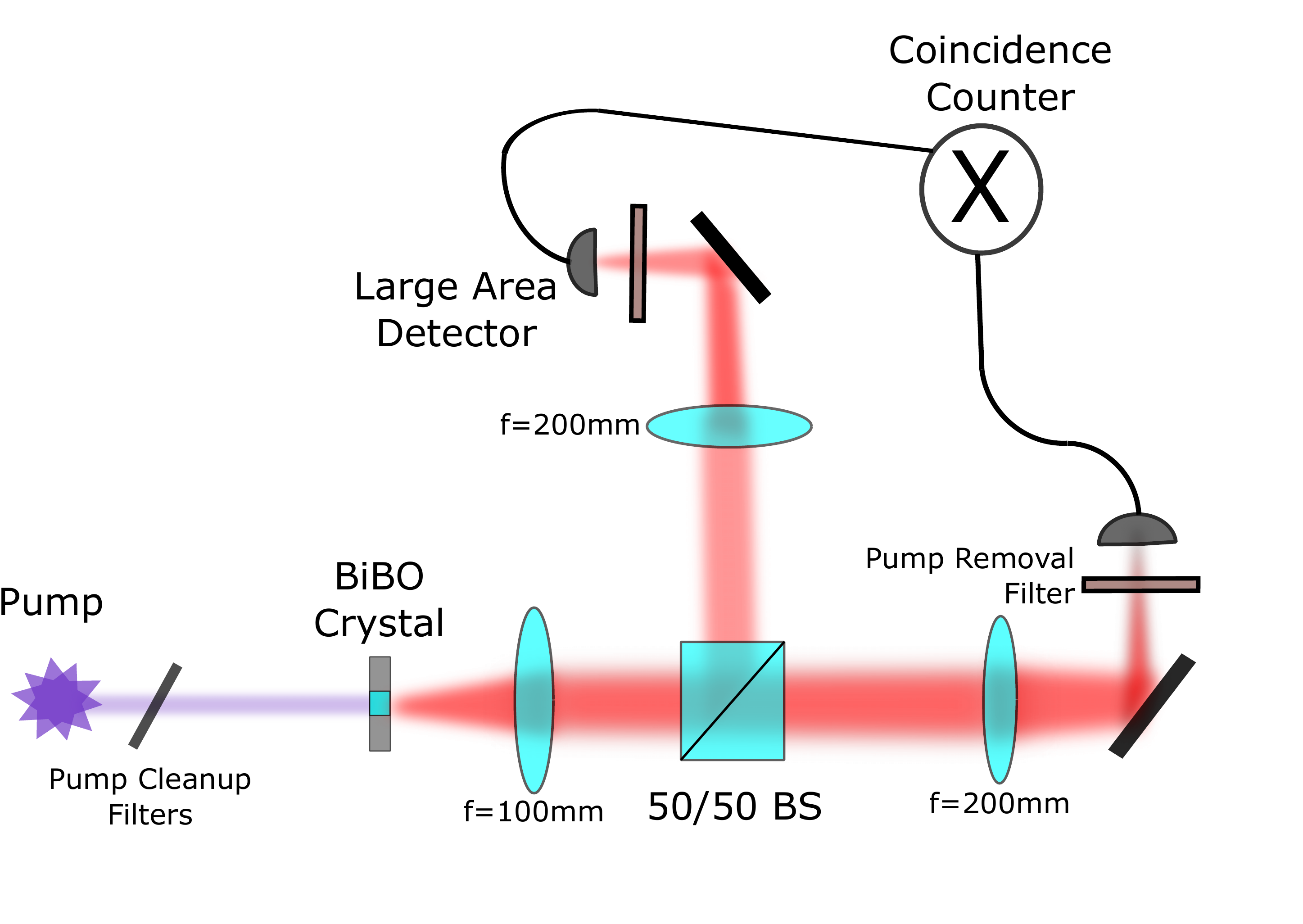}}\label{LargeAreaSetup}
\caption{Diagram of experiment used to obtain total coincidence count rate for type-I SPDC. The pump light is directed through a nonlinear crystal, and is subsequently filtered out. The downconverted light is split by a 50/50 beamsplitter and is focused onto Large area Single photon detectors. The experiment allows us to directly measure the total rate $R_{T}$, though the relation between $R_{SM}$ and $R_{T}$ is determined by the overlap of the total biphoton spatial amplitude with the zero-order gaussian modes used to compute $R_{SM}$.}
\end{figure}

\subsection{Type-I SPDC in BiBO crystal incident on large area single-photon detectors}
The next experiment we performed tests our formula for the total biphoton generation rate for collinear type-I SPDC in an isotropic crystal \eqref{totalType1}. Here, we used a $1$ mm crystal of Bismuth Barium Borate (BiBO) manufactured by Newlight Photonics. We used a $405$ nm OBIS laser to produce down-converted photon pairs centered at $810$nm. We separated the photons with a 50/50 beamsplitter, and focused the light onto large-area single photon detectors. Because we are sampling over all modes, extracting the raw pair generation rate $N$ from the singles and coincidences is simpler; we can set $\eta$ and $C$ equal to unity. Given our experimental parameters, we predict a pair generation rate of $(53.87\pm10.87)\times10^{6}$ per second per mW of pump power. The experiment measured singles rates per mW of pump power of $(6.16\pm0.05)\times10^{5}$ and $(6.02\pm0.05)\times10^{5}$ per second, with respective background rates of $(6.04\pm0.15)\times10^{4}$ and $(6.40\pm0.10)\times10^{4}$ per second. We recorded a coincidence rate of $(2.71\pm0.06)\times10^{3}$ per second and an accidentals rate of $(4.39\pm2.79)$ per second. From these statistics, we obtain a raw pair generation rate of $(64.68\pm1.69)$ million pairs per second per mW of pump power, exceeding our theoretical prediction by 20 percent, though this is still within the large range of uncertainty due to limited knowledge of the biphoton wavefunction, among other factors.

\begin{figure}[t]
\centerline{\includegraphics[width=\columnwidth]{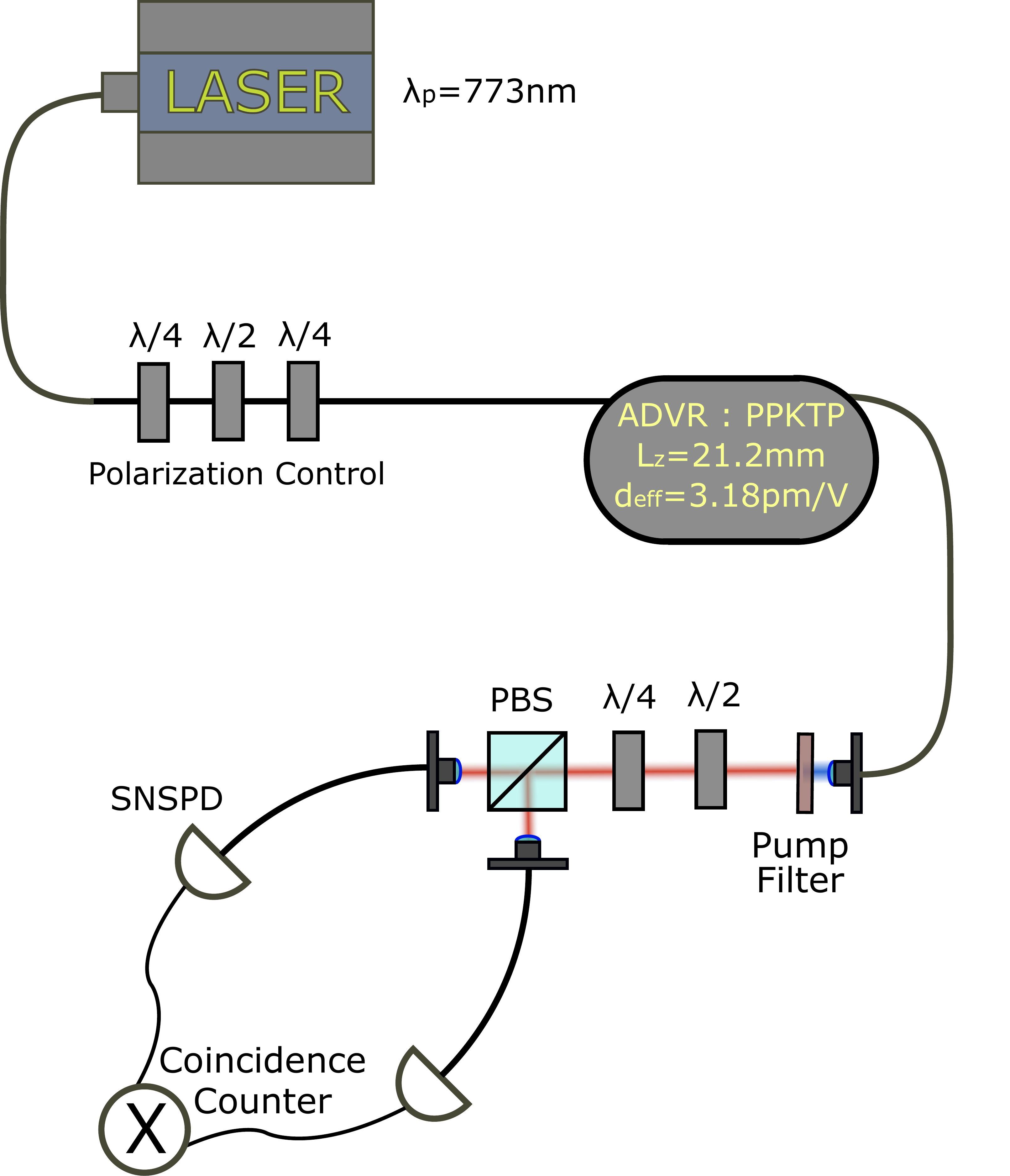}}\label{PPKTPsetup}
\caption{Diagram of experiment used to obtain coincidence count rate for type-II SPDC in a periodically poled, single-mode waveguide. The pump light is directed through an optical fiber coupled to a nonlinear-optical waveguide, and is later filtered out. Because this is type-II SPDC, the downconverted light is split efficiently with a Polarizing beamsplitter (PBS) and is directed to a pair of Supercondicing nanowire single-photon detectors (SNSPDs), from which coincidence counts are recorded.}
\end{figure}

\subsection{Type-II SPDC in single-mode PPKTP waveguide}
For our third experiment, we used a waveguide of Periodically-Poled Potassium Titanyl Phosphate (PPKTP) manufactured by AdvR, for type-II SPDC from 773nm to 1546nm poled for first-order quasi-phase matching. This experiment was pumped with a Newport NewFocus tunable laser centered at 773nm. Here, we separated the signal and idler photons completely with a polarizing beamsplitter. Moreover, we may set $C=1$ since both pair-generation, and collection occur in a single optical mode. The waveguide we used was 21.2mm long, with values for $\sigma_{p}$ and $\sigma_{1}$ being $(0.875\pm0.125)\mu$m, and $(1.875\pm0.125)\mu$m, respectively. Per mW of pump power, we measured singles rates of $(3.71\pm0.05)\times10^{6}/s$ and $(4.51\pm,0.05)\times10^{6}/$s, with a coincidence count rate of $(4.71\pm0.07)\times10^{5}/$s. From these rates, we obtain a raw pair generation rate of approximately $(35.5\pm0.8)\times 10^{6}/$s per mW of pump power.

Using our single-mode formula for Type-II SPDC in a periodically poled medium and the given experimental parameters, we estimate a rate of $(23.58\pm5.60)\times10^{6}$ per second per mW of pump power. This differs from the experimental rate by as much as 33 percent, but due to asymetries in the eigenmodes of the waveguide \cite{fiorentino2007spontaneous, Shukhin2015SPDC}, a simple gaussian mode of equal widths in both transverse dimensions cannot be assumed to be what couples into the exit fiber. Indeed, given the  rubidium doping needed to create the waveguide, the waveguide itself has different effective widths in each transverse dimension. Assuming a 30 percent difference between the different transverse widths of the eigenmodes is reasonable (see diagram in \cite{fiorentino2007spontaneous}), and is sufficient to produce a theoretical preciction that agrees well with experimental data. The theoretical estimate is also subject to the relatively large uncertainties in the pump and signal/idler radii inside the waveguide (of approximately $0.18\mu$m), whose value is generally more difficult to determine than in step-index optical fibers. Moreover, the waveguide is small enough that modal dispersion may noticeably change the effective index of refraction in comparison to bulk media. In addition, there is a rather large ($\approx10\%$) uncertainty in $d_{\text{eff}}$, which varies significantly between different PPKTP crystals, likely due to thermal stress patterns in the manufacturing process. For our theoretical prediction, we used the $d_{24}$ coefficient responsible for type-II SPDC given in \cite{fiorentino2007spontaneous} (so that $d_{eff}=d_{24}$ not counting quasi-phase matching factors), which treats SPDC in a PPKTP waveguide. Where they list $d_{24}=3.92$ pm/V for SPDC for a 405nm pump, we use Miller's rule \footnote{Miller's rule is the approximation that the second order susceptibility $\chi^{(2)}_{eff}(\omega_{p},\omega_{1},\omega_{2})$ is proportional to the product of the first-order susceptibilities $\chi^{(1)}(\omega_{p})\chi^{(1)}(\omega_{1})\chi^{(1)}(\omega_{2})$. For transparent media with negligible absorption, $\chi^{(1)}(\omega)\approx n(\omega)^{2}-1$.} to obtain $d_{24}\approx 3.18$ pm/V for SPDC with a 773nm pump. To describe our waveguide adequately, it is single-mode at the down-conversion wavelength, but it is multi-mode at the pump wavelength. The mode field diameter at the pump wavelength is given as the diameter of the light entering the crystal from a single mode fiber fused to the waveguide, which is not the diameter of the TEM00 mode accepted by the waveguide.

\section{Conclusion}
In this tutorial, we have shown the essential factors contributing to the absolute photon-pair generation rate via Spontaneous Parametric Down-Conversion by deriving this rate from first principles. We began with deriving a general hamiltonian for SPDC processes, and simplified it for the popular cases of bulk crystals, single-mode waveguides and for generation in micro-ring resonators as a prototypical example of cavity-enhanced SPDC, and for its importance in integrated photonics. We examined the effect of focusing the pump beam, and of using periodically poled crystals. We discussed how to describe the field without perturbation theory via the two-mode squeezed vacuum state, and the behavior of SPDC when the pump light can be depleted. We investigated the number statistics of down-converted light and developed useful guidelines for optimizing the coincidence to accidentals ratio, important in using SPDC as a heralded single photon source, and in loophole-free quantum secure communication. Most importantly, we compared our theoretical predictions with experimental data, and find that to the extent that the theoretical approximations resemble the reality of the experiment, the agreement improves correspondingly.

We gratefully acknowledge support from the National Research Council Research Associate Programs, and funding from the OSD ARAP QSEP program, as well as insightful discussions with Dr. A. Matthew Smith. In addition, SHK acknowledges support from the Air Force Office of Scientific Research Grant FA9550-16-1-0359 and from Northrop Grumman Grant 058264-002.  Any opinions, findings and conclusions or recommendations expressed in this material are those of the author(s) and do not necessarily reflect the views of AFRL.
\newline
\newline
\newline
\newline

\appendix
\section{Tables of experimental results and parameters}
\begin{widetext}
\centering
 \begin{tabular}{|c|c|c|c|}
 \hline
 \multicolumn{4}{|c|}{Table of experimental parameters and results}\\
 \hline
  & Type-0, SM in PPLN & Type-I, MM in BiBO & Type-II, SM, in PPKTP \\
 \hline
 $\lambda_{p}$ & $782.09\pm 0.1$ nm & $405.0\pm 1.0$ nm & $773.0\pm1.0$ nm \\
 $d_{eff}$ & $23.95\pm1.20$ pm/V & $3.70\pm0.18$ pm/V & $3.18\pm0.32$ pm/V \\
 $L_{z}$ & $40.0\pm0.001$ mm & $1.0\pm0.001$ mm & $21.2\pm 0.01$ mm \\
 \hline
 $\sigma_{p}$ & $52.6\pm 2.0 \mu$m & N/A & $0.875\pm0.125\mu$m \\
  $\sigma_{1}$ & $55.1\pm2.0 \mu$m & N/A & $1.875\pm0.125\mu$m \\
  \hline
   $n_{1}$ & $2.155\pm0.001$ & $1.822\pm0.001$ & $1.736\pm0.002$ \\
    $n_{2}$ & $2.155\pm0.001$ & $1.822\pm0.001$ & $1.783\pm0.002$ \\
     $n_{p}$ & $2.195\pm0.001$ & $1.822\pm0.001$ & $1.759\pm0.002$ \\ \hline
      $n_{g1}$ & $2.200\pm0.001$ & $1.866\pm0.001$ & $1.765\pm0.002$ \\
       $n_{g2}$ & $2.200\pm0.001$ & $1.866\pm0.001$ & $1.815\pm0.002$ \\ \hline
        $\kappa$ & $96.75\pm0.2\times 10^{-27} s^{2}$/m & $160.9\pm0.2 \times 10^{-27}s^{2}$/m & N/A \\
        \hline
         $R_{th}$ & $94.86\pm10.89\times 10^{6}$/s/mW & $53.87\pm10.87 \times 10^{6}$/s/mW & $23.58\pm5.60 \times 10^{6}$/s/mW \\
          $R_{exp}$ & $95.63\pm2.71\times 10^{6}$/s/mW & $64.68 \pm 1.69 \times 10^{6}$/s/mW & $35.5\pm0.8 \times 10^{6}$/s/mW \\
 \hline
 \end{tabular}
 \\
 TABLE 1: Here, $R_{th}$ and $R_{exp}$ are the theoretically predicted and experimentally determined pair generation rates.
 \newline
 \begin{center}
 \begin{tabular}{|c|c|}
 \hline
 \multicolumn{2}{|c|}{Table of generation rate formulas for different types of SPDC}\\
 \hline
 Type  &  Formula\\
 \hline
 \rule{0pt}{15pt}Type-0/I, SM & $\sqrt{\frac{2}{\pi^{3}}}\frac{2}{3\epsilon_{0}c^{3}}\frac{n_{g1}n_{g2}}{n_{1}^{2}n_{2}^{2}n_{p}}\frac{(d_{eff})^{2}\omega_{p}^{2}}{\sqrt{\kappa}}\Big|\frac{\sigma_{p}^{2}}{\sigma_{1}^{2}+2\sigma_{p}^{2}}\Big|^{2}\frac{P}{\sigma_{p}^{2}}L_{z}^{3/2}$\\
 \hline
  \rule{0pt}{15pt}Type-0/I, MM & $\frac{32\sqrt{2\pi^{3}}}{27\epsilon_{0}c}\Big(\frac{n_{g1}n_{g2}}{n_{1}^{2}n_{2}^{2}}\Big)\frac{d_{eff}^{2}}{\lambda_{p}^{3}\sqrt{\kappa}}\frac{P\sqrt{L_{z}}}{\phi}$ \\
 \hline
 \rule{0pt}{15pt} Type-II, SM & $\frac{1}{\pi\epsilon_{0}c^{2}}\frac{n_{g1}n_{g2}}{n_{1}^{2}n_{2}^{2}n_{p}}\frac{(d_{eff})^{2}\omega_{p}^{2}}{\Delta n_{g}}\Big|\frac{\sigma_{p}^{2}}{\sigma_{1}^{2}+2\sigma_{p}^{2}}\Big|^{2} \frac{P}{\sigma_{p}^{2}}L_{z}$\\
 \hline
 \end{tabular}
 \\
 TABLE 2: Here, $\phi$ is approximately 0.335 and $\Delta n_{g} = |n_{g1}-n_{g2}|$.
 \\
\end{center}
 \end{widetext}

\bibliography{EPRbib15}

\end{document}